\documentclass[a4paper,fleqn,usenatbib]{mnras}

\usepackage[T1]{fontenc}
\usepackage{ae,aecompl}
\UseRawInputEncoding

\usepackage{graphicx}	
\usepackage{amsmath}	
\usepackage{amssymb}	
\usepackage{multicol, wrapfig}
\usepackage{pdflscape}
\usepackage{float}
\usepackage{subfloat}
\usepackage{newtxtext,newtxmath}

\title[The most variable VVV sources]{The most variable VVV sources: eruptive protostars, dipping giants in the
Nuclear Disc and others}

\author[P. W. Lucas et al.]{
P. W. Lucas$^{1}$\thanks{E-mail: p.w.lucas@herts.ac.uk (PWL)},
L. C. Smith$^{2}$, Z. Guo$^{3,4,1,5}$, C. Contreras Pe\~{n}a$^{6,7}$, D. Minniti$^{8,9,10}$, 
\newauthor N. Miller$^1$, J. Alonso-Garc\'{i}a$^{11,12}$, M. Catelan$^{13,12}$, J. Borissova$^{3,12}$, R. K. Saito$^{10}$,
\newauthor R. Kurtev$^{3,12}$, M. G. Navarro$^{14}$, C. Morris$^1$, H. Muthu$^1$, D. Froebrich$^{15}$, 
\newauthor V. D. Ivanov$^{16}$, A. Bayo$^{16}$, A. Caratti o Garatti$^{17}$ and J. L. Sanders$^{18}$\\
$^{1}$Centre for Astrophysics, University of Hertfordshire, College Lane, Hatfield, AL10 9AB, UK\\
$^{2}$Institute of Astronomy, University of Cambridge, Madingley Road, Cambridge, CB3 0HA, UK\\
$^{3}$Instituto de F{\'i}sica y Astronom{\'i}a, Universidad de Valpara{\'i}so, ave. Gran Breta{\~n}a, 1111, Casilla 5030, Valpara{\'i}so, Chile\\
$^{4}$N\'ucleo Milenio de Formaci\'on Planetaria (NPF), ave. Gran Breta{\~n}a, 1111, Casilla 5030, Valpara{\'i}so, Chile\\
$^{5}$Departamento de F{\'i}sica, Universidad Tecnic{\'a} Federico Santa Mar{\'i}a, Avenida Espa{\~n}a 1680, Valpara{\'i}so, Chile\\
$^{6}$Department of Physics and Astronomy, Seoul National University, 1 Gwanak-ro, Gwanak-gu, Seoul 08826, Republic of Korea\\
$^{7}$Research Institute of Basic Sciences, Seoul National University, Seoul 08826, Republic of Korea\\
$^{8}$Departamento de Ciencias Fisicas, Universidad Andres Bello, Republica 220, 8320000 Santiago, Chile\\
$^{9}$Vatican Observatory, Castelgandolfo, V00120, Vatican City State, Italy\\
$^{10}$Departamento de F{\'i}sica, Universidade Federal de Santa Catarina, Trindade 88040-900, Florianopol{\'i}s, SC, Brazil\\
$^{11}$Centro de Astronomía (CITEVA), Universidad de Antofagasta, Av. Angamos 601, 1240000 Antofagasta, Chile\\
$^{12}$Millennium Institute of Astrophysics, Av. Vicuna Mackenna 4860, 782-0436, Macul, Santiago, Chile\\
$^{13}$Instituto de Astrof{\'i}sica, Pontificia Universidad Cat{\'o}lica de Chile, Av. Vicu{\~n}a Mackenna 4860, 7820436 Macul, Santiago, Chile\\
$^{14}$INAF - Osservatorio Astronomico di Roma, Via di Frascati 33, I-00078, Monte Porzio Catone, Roma, Italy\\
$^{15}$Centre for Astrophysics and Planetary Science, University of Kent, Canterbury CT2 7NH, UK\\
$^{16}$European Southern Observatory, Karl Schwarzschildstr. 2, D-85748 Garching bei MŸnchen, Germany\\
$^{17}$INAF - Osservatorio Astronomico di Capodimonte, salita Moiariello 16, 80131 Napoli, Italy\\
$^{18}$Department of Physics and Astronomy, University College London, London WC1E 6BT, UK}

\date{Accepted XXX. Received YYY; in original form ZZZ}

\pubyear{2023}

\begin{document}
\label{firstpage}
\pagerange{\pageref{firstpage}--\pageref{lastpage}}
\maketitle

\begin{abstract}
We have performed a comprehensive search of a VISTA Variables in the Via Lactea (VVV) database of 
9.5~yr light curves for variable sources with $\Delta K_s \ge 4$~mag, aiming to provide a large sample of high 
amplitude eruptive young stellar objects (YSOs) and detect unusual or new types of infrared variable source. 
We find 222 variable or transient sources in the Galactic bulge and disc, most of which are new discoveries. The sample mainly 
comprises novae, YSOs, microlensing events, Long Period Variable stars (LPVs) and a few rare or unclassified sources.
Additionally, we report the discovery of a significant population of aperiodic late-type giant stars suffering deep extinction events,
strongly clustered in the Nuclear Disc of the Milky Way. We suggest that these are metal-rich stars in which radiatively 
driven mass loss has been enhanced by super-solar metallicity.
Among the YSOs, 32/40 appear to be undergoing episodic accretion. Long-lasting YSO eruptions have a typical rise time of 
$\sim$2~yr, somewhat slower than the 6--12 month timescale seen in the few historical events observed on the rise.
The outburst durations are usually at least 5~yr, somewhat longer than many lower amplitude VVV events detected previously. 
The light curves are diverse in nature, suggesting that multiple types of disc instability may occur. 
Eight long-duration extinction events are seen wherein the YSO dims for a year or more, attributable to inner disc 
structure.  One binary YSO in NGC~6530 displays periodic extinction events ($P$=59~days) similar to KH~15D.
\end{abstract}




\section{Introduction}

Time domain astrophysics has led to the discovery of several new types of variable star in recent years. Mid-infrared studies have found transients known as ``eSPecially Red 
Intermediate Luminosity Transient Events (SPRITES)" in nearby galaxies, having luminosities between classical novae and supernovae \citep{kasliwal17}. 
In optical studies of variable stars, the abundant OGLE Small Amplitude Red Giants (OSARGs, \citealt{soszynski04, soszynski13}) and the rarer Blue Large Amplitude Pulsators (BLAPs, 
\citealt{pietrukowicz17}) have been found by the Optical Gravitational Lensing Experiment \citep{udalski03}. Circumstellar matter is the cause of other new types of variable star. 
For example, ASAS~J140748$-$3945.7 (J1407) hosts a transiting giant ring system bound to the substellar companion of a pre-main sequence star \citep{mamajek12}. 
Unpredictable dips and slow variations in brightness are also seen in main sequence stars and giant stars \citep[e.g.][]{boyajian16, boyajian18, schmidt19, rodriguez16, smith21}, 
attributed to occultation by irregular distributions of circumstellar matter in the main sequence stars or a circumsecondary disc in the case of the giants.

The VISTA Variables in the Via Lactea (VVV) survey \citep{minniti10} is the first major near infrared variability survey of the Milky Way. In view of the above
list of discoveries, we might reasonably hope that it will discover new types of infrared variable source. Due to the Wien displacement law, it is perhaps likely
that the physical mechanisms behind any such infrared discoveries would involve circumstellar matter. The most numerous high amplitude variable stars
discovered by VVV thus far are YSOs with circumstellar accretion discs, see \citet{cp17a,cp17b}, hereafter CP17a and CP17b and
\citet{medina18,teixeira18,guo20,guo21,guo22}. These generally have 
amplitudes $\Delta K_s \le 4$~mag in these previous works (measured as $K_{s,max}-K_{s,min}$) and a diverse range of light curve types. Short term 
variability on timescales of a few days is  common at $1 < \Delta K_s < 1.5$. At higher amplitudes, longer timescale variability predominates: mostly cases of 
eruptive variability but also cases of variable extinction, sometimes in the same star \citep[see e.g.][]{kospal11}. 
CP17a found that eruptive variability is dominated by optically faint class I and flat spectrum YSOs, with very few examples of class II YSOs being observed.
Variability in embedded YSOs is more easily studied in the near infrared, the mid-infrared \citep[e.g.][]{park21, cp23} or the submillimetre waveband \citep{lee21}.

Most past studies of YSO outbursts have focussed on the most dramatic events, either long duration events (FUors) with optical amplitudes up to 6~mag or more
(e.g. FU Ori, V1057 Cyg, V1515 Cyg, V346 Nor and PTF14jg \citealt{herbig77, graham85, kospal20, hillenbrand19}) or shorter duration events (EXors) with 
amplitudes up to 5 mag such as EX Lup and V1118~Ori \citep{herbig08}. In recent years, the {\it Gaia} Alerts data \citep{kostrzewa20, hodgkin21} have 
enabled more systematic searches that have discovered two new FUors, Gaia~17bpi \citep{hillenbrand18} and Gaia~18dvy \citep{szegedi-elek20} amongst other
things, complementing the more general {\it Gaia} classification of YSOs and other variable stars \citep{eyer23, rimoldini23, marton23}.
VVV and its extension, VVVX (\url{vvvsurvey.org}) offer the chance to establish a large sample of events that characterises the outburst onset better than in most 
previous (often accidental) discoveries and lacks any significant bias against lower amplitude events.
The great majority of outbursts have lower amplitude, especially in the near infrared. E.g. CP17a listed 106 candidates with $1<\Delta K_s<4$, with 
mean $\Delta K_s=1.72$. 

Outburst durations do not always fit neatly
into the original FUor (multi-decade) and EXor (8-16 month) categories \citep{herbig77, herbig08}. Spectroscopically confirmed VVV outbursts (CP17b)
appeared to have typical durations of 1-5~yr (though sometimes longer) and the spectroscopic characteristics of the two traditional outburst classes were 
sometimes mixed or swapped. This led to the tentative suggestion that a new ``MNor" outburst category might be needed to describe mixed outbursts (CP17b).
More recently \citet{guo21} showed that within the broad category of mixed outbursts there is a distinct class of long duration eruptions with EXor-like spectra, apparently 
with the accretion luminosity of FUors. In fact these long lasting but EXor-like outbursts were more numerous than FUors amongst the long duration events in that work.

In this work, we perform a thorough search of the 562~deg$^2$ VVV area for very high amplitude ($\Delta K_s \ge 4$) variable stars and transients. This threshold
was chosen because previous VVV-based searches had found very few eruptive YSOs with such high amplitudes. A complete search for all types of
variable sources is more difficult at lower amplitudes, if all false positives are to be removed, due to the larger number of candidates.
While our main interest was eruptive YSOs, unexpectedly we also discovered a population of aperiodic dipping giants in the Nuclear Disc of the 
Milky Way. Consequently, this paper focuses on these two topics. Supporting spectra of YSOs and giants are found in a companion paper (Guo et al. submitted, hereafter GLK).
Separately, in Contreras Pe\~{n}a et al.(submitted) we present results of a VVV-based search for eruptive YSOs with $\Delta K_s$$>$2, within published catalogues of YSO 
candidates.

VVV observed much of the Galactic bulge and the adjacent Galactic disc from 2010--2015 and this study benefits from the ongoing VVVX extension in the same
area, though we have not included the new areas of the Galactic plane area monitored by VVVX since 2016. This new search benefits not only from
longer duration light curves and larger area than the 119 deg$^2$ search of CP17a but also from the new VVV Infrared Astrometric Catalogue version 2 
(VIRAC2) profile-fitting photometry database (\citealt{smith18}, Smith et al., in prep.) which is deeper and more reliable than the standard VISTA data pipeline in crowded fields. 
VIRAC2 light curves have reduced noise due to a carefully optimised relative photometric calibration and the absolute photometric calibration is also improved, 
particularly in crowded inner-bulge fields (Smith et al., in prep.). 

In $\S$2 and Appendix \ref{sec:searches} we describe the data and the variable star searches. $\S$\ref{overview} summarises our initial classifications of the
variable sources and their distribution on the sky. In $\S$\ref{ysos} we examine the YSOs, with emphasis on the timescale of the rise to maximum light and the form
of the light curve in FUor-like events, and in section $\S$\ref{sec:dipgiants} we discuss the aperiodic dipping giants.
The dipping YSOs are presented fully in Appendix \ref{dipysos}. Similarly, a few noteworthy findings among other types of variable star (LPVs, novae and unusual sources) 
are summarised in $\S$\ref{others} and discussed more fully in appendices \ref{sec:s163}, \ref{sec:CVs} and \ref{unusual}. 
Microlensing events are discussed only briefly in Appendix \ref{micro}. Our conclusions are given in $\S$\ref{conc}.

\section{VVV(X) observations and databases}

\subsection{Observations}

The VVV and VVVX surveys employed the VISTA telescope \citep{sutherland15} at Cerro Paranal observatory and the VIRCAM camera \citep{dalton06}, which 
houses 16 HgCdTe 2048$^2$ Raytheon infrared array detectors. From 2010--2015, VVV surveyed a region of the Galactic bulge of at Galactic coordinates 
$|\ell|<10^{\circ}$, $-10^{\circ}<b<5^{\circ}$ and the adjacent southern Galactic disc at $295^{\circ}<|\ell|<350^{\circ}$, $|b|<2^{\circ}$. The region is divided into filled
1.5$\times$1.1~deg$^2$ tiles, each of which comprises six unfilled pawprint stack images. The tiling pattern is such that most parts of the tile are observed by at least two
pawprint stack images within a few minutes, with the two or more images of each star falling on different arrays or widely separated parts of the same array.   
VVV typically has between 50 and 80 epochs of observation for each tile in $K_s$, with additional observations in the $Z$, $Y$, $J$ and $H$ filters at the beginning and the end of the 
survey. The $K_s$ observing blocks in the VISTA observing queue each contained a group of four adjacent, slightly overlapping tiles that could be observed in about an hour.
From 2016-2022, VVVX extended the survey to a larger area, whilst continuing to take a few more epochs of $K_s$ data in the original VVV area, 
supplemented by some deeper  $J$ and $H$ observations in the inner bulge. Ultimately, most sources in the VVV area have light curves comprising 
$\sim$200 $K_s$ pawprint stack-based measurements over the 2010-2019 interval.

The VISTA $K_s$ pawprint stack images have a typical spatial resolution of 0.75\arcsec~ but data from images with much poorer seeing are retained in the VIRAC2 light curves. The sensitivity limit is
$K_s\approx 17.5$ (Vega system) in relatively uncrowded fields and the saturation limit is in the range $K_s$ = 11--12, 
with some variation depending on the seeing, the array detector, the brightness of the background and the transparency. Brighter sources are often included in the 
database, albeit with relatively few epochs of data, taken in poor seeing conditions.

\subsection{VIRAC2}

Our searches for highly variable stars and transients mainly used two preliminary versions of the VIRAC2 database, VIRAC2-$\alpha$ and VIRAC2-$\beta$,
which contain time series profile fitting photometry taken over 2010--2018 and 2010--2019, respectively.  The light curves presented herein are almost all from VIRAC2-$\beta$, 
supplemented by additional saturation-corrected photometry 
and photometry on stacked images in cases where this was needed for bright or very faint sources lying outside the single-epoch dynamic range of the 
survey. VIRAC2-$\alpha$ light curves are presented only for a few poorly sampled transients that were were not included in VIRAC2-$\beta$
because they were detected in fewer than ten pawprint stack images. The \textsc{dophot} software \citep{schechter93, alonso12} was employed for the photometry on each 
pawprint stack image.  Photometric and astrometric calibration were performed as part of the production of VIRAC2 (Smith et al., in prep.), see the summary in 
Appendix \ref{sec:searches}.

\subsection{VVV 4th Data Release}
\label{DR4}

An earlier search employed the SQL database of VVV tile-based aperture photometry available in the 4th VVV Data Release (VVV DR4), hosted at the VISTA Science Archive (\citealt{cross12}, \url{vsa.roe.ac.uk}), 
supplemented by further analysis using the VVV pawprint stack-based aperture photometry for each candidate variable source (see Appendix \ref{sec:searches}).
VVV DR4 comprises aperture photometry taken in the 2010--2013 time period, processed with version 1.3 of the {\sc casutools} VISTA data reduction 
pipeline \citep{gonzal18}. The 20 sources having $\Delta K_s >4$~mag in the VVV DR4 light curves were all recovered by the later VIRAC2-based searches. However, the VVV DR4 search was extended to lower amplitudes ($\Delta K_s >3$~mag), yielding a somewhat larger number of discoveries (see $\S$\ref{sec:DR4results}). 

\subsection{Search methods}

Details of how the above databases were searched for bona fide large amplitude variable stars and transients are given in Appendix \ref{sec:searches}.

\section{Overview of Results}
\label{overview}

In Table \ref{tab:table1a} we present the list of 222 bona fide variable stars and transients with $\Delta K_s > 4$~mag discovered by our VIRAC2-based searches. Light curves for 
all these sources are provided in catalogue and PDF forms in the Supplementary Online Information. Of these, 162 (73 per cent) 
are VVV discoveries, $\sim$14~per cent of which were previously reported in VVV-based searches for novae \citep{saito15, saito16, montenegro15, gutierrez16}, highly variable stars of all types 
\citep[CP17a,][]{medina18} and periodic variable stars \citep{guo22}\footnote{In addition, 127 of the 222 variable sources appear in the machine learning-classified 
``VIVACE" list of 1.4 million periodic VVV variable star candidates \citep{molnar22}, classified as LPVs in almost all cases. However, $72$ per cent of the 127 (i.e. 92 sources) are not in fact 
periodic variable stars. This suggests that the VIVACE classification procedures had some difficulty with aperiodic variable sources, which were not represented in the training 
sets used in that work. The authors of that work noted that some degree of contamination by aperiodic YSOs was expected in their catalogue; it now appears that other types
of aperiodic variable source such as the novae and microlensing events reported herein can also appear as contaminants, though YSOs are likely to be the most numerous 
type.}$^,$\footnote{For completeness, we note that 133 of the 222 variable sources are included in a list of 45 million VVV variable star candidates published by \citep{ferreira-lopes20}, 
based on the aperture photometry in VVV DR4 (2010 to 2013 time series). Bona fide variable stars are not clearly distinguished from the far more numerous false positives in that database,
nor given astrophysical classifications, but it provides a basis for searches, like VIRAC2.}. Table \ref{tab:table1a} gives a 
running number for each source and any other
name by which the source is known, the VIRAC2 J2000 equatorial coordinates of each source, a 
measurement of the amplitude of variability ($\Delta K_s$), the median $K_s$ magnitude, an initial classification of the type of astrophysical source, a description of the light curve, 
a comment, and the time
baseline in years over which the source was detected in the VVV/VVVX surveys.
In Table \ref{tab:table1a}, $\Delta K_s$ is
computed using flux measurements that are averaged within 1 day time bins\footnote{For source 185, $\Delta K_s$ was computed with the unbinned light curve due to large intra-night
variability of this brief transient.} to reduce measurement error, after rejecting any measurements with discrepant VIRAC2 astrometry
or a poor \textsc{dophot} fit to a stellar profile (see Appendix \ref{sec:searches})
and removing any measurements that differ by more than 3$\sigma$ from the mean in each bin. A small number of sources have $\Delta K_s$ just under 4~mag as measured in the
binned light curves (see Appendix B for details) since the sparse sampling and finite sensitivity of VVV would tend to cause the true amplitude of most sources to be slightly 
under-estimated. Lower limits on $\Delta K_s$ are given for sources that dropped out in part 
of the survey; these are based on upper limits on source flux in images constructed by stacking images in time groups, grouped by the calendar year of 
observation.

\begin{landscape}
\begin{subtables}
  \begin{table}
	\caption{Table of 222 variable sources with $\Delta K_s \ge 4$ mag}
	\begin{tabular}{lcccccccccc} \hline
No. & VVV designation & Other name & RA & Dec         & $l$ & $b$ & $\Delta K_s$ & Median $K_s$ & Initial              & Time      \\
      &                             &                     &  (deg) & (deg)   & (deg) & (deg)   &          &                         & classification & baseline (yr)  \\ \hline
                                                                                                                                                                                                 
  1    & VVV J114309.48-622113.2   &         & 175.78949 & -62.35366  &  295.10323  &  -0.52428   & 4.0  & 15.29  &  YSO    &     9.38  \\                                                                                                      
  2    & VVV J114613.98-634714.2 & OH 259.8-1.8 & 176.55824 & -63.78728 & 295.80639  & -1.82127   & 4.92 & 11.63   & LPV  &    9.38 \\                                                                                         
  3    & VVV J120447.85-622402.1      &       & 181.19936 & -62.40058  &  297.55928   &  -0.01635  &  4.09 & 14.31   &  YSO    &   9.26  \\                                                                                                                                    
  4    & VVV J122054.04-623821.9       &      & 185.22516 & -62.63943 & 299.43147 &  0.02459   &   6.91 & 15.42  &   YSO    &    9.52  \\                                                                        
  5    & VVV J123557.72-615111.9       &      & 188.99049 & -61.85329 &  301.10762  &  0.96363  &  4.4  & 12.66   &   YSO   &    9.37  \\
  6    & VVV J125457.41-610239.0       &      & 193.73920 & -61.04417 &  303.35803 & 1.82467  &  4.31 & 16.09    &   YSO   &    9.47  \\                                                                     
  7    & VVV J130946.62-625648.6        &     & 197.44425 & -62.94684 & 305.01540  & -0.14930  & $>$4.16 & 16.67  &  Sparse  &  1.12  \\
  8    & VVV J131222.41-640905.4 & VVV NV007 &   198.09336 & -64.15150 & 305.21194 & -1.37247  & $>$9.37 & 14.80  &  CV  & 1.06 \\                                                                                              
  9    & VVV J132049.76-623750.6 & Nova Cen 2005 & 200.20735 & -62.63073 & 306.30261 & 0.04851  & 4.56 & 14.21  &  CV  & 9.43  \\
  10  & VVV J132550.28-620446.7        &     & 201.45949 & -62.07963 & 306.94718 &  0.52364  &  4.02 & 15.38 &   YSO   &  9.40   \\
      ...\\
    		\hline
  	\end{tabular}
	\\
  	\begin{tabular}{ll} \hline
  Light curve & Comment \\
  description &                \\ \hline
  (1) Eruptive YSO, 6 year slow rise from $K_s$$\approx$16 to $K_s$$\approx$12. & Spectroscopically confirmed in GLK\\
  (2) LPV.   &   An OH/IR star, also noted in \citet{medina18}\\
  (3) Eruptive YSO light curve, rose from 2010-2011 then faded slowly & \\
  (4) Eruptive YSO, very high amplitude, relatively fast decline. & Spectroscopically confirmed in GLK. Spatially resolved circumstellar nebula. \\
  (5) Likely YSO extinction event. Faded progressively after 2015.  & Located in HII region IRAS 12331-6134 \& cluster [DBS2003] 78.\\
  (6) Eruptive YSO. 2MASS Ks=14.13 & Spectroscopically confirmed to be a YSO in GLK\\ 
  (7) Transient event in 2012, detected as faint source in 2013 &  Nature unknown.\\  
  (8) Peak of event was poorly sampled, later fading trend is clear. & Classical nova candidate, \citet{gutierrez16}\\
  (9) Event followed an 8~yr monontonic fade      & A known re-brightening event, \citet{delgado19}\\
  (10) Eruptive YSO with a slow rise. 2MASS Ks=15.05 & Spectroscopically confirmed in GLK\\
      ...\\
              \hline
              \label{tab:table1a}
        \end{tabular}\\
         Note: Only the first 10 rows of the table are shown here, split into two sections. The full table is available in the online supplementary information. J2000 \\
         coordinates are calculated at epoch 2015.0 for all sources in VIRAC2$-\beta$ but at epoch 2014.0 for sources 60, 103, 125 and 185, which were taken from\\
         VIRAC2$-\alpha$ due to not being properly recovered in VIRAC2$-\beta$.
  \end{table}

  \bigskip
  \begin{table}
  \vspace{8mm}
	\caption{Two additional eruptive variable YSOs with $3 <\Delta K_s < 4$ mag.}
	\begin{tabular}{lccccccccc} \hline
No. & VVV designation &  RA & Dec         & l & b & $\Delta K_s$ & Median $K_s$ & Initial              & Time      \\
      &                             &      (deg) & (deg)   & (deg) & (deg)   &          &                         & classification & baseline (yr)  \\ \hline
                                                                                                                                                                                              
  1   & VVV J163637.94-474444.1 & 249.15809 &  -47.74558  & 336.94797 & -0.312007 & 3.68  & 15.26 &  YSO  & 9.49    \\                                                                                                      
  2   & VVV J164011.76-484653.4 & 250.04901 & -48.78151 & 336.57292  & -1.44525 & 3.20  & 15.96  & YSO  & 9.49    \\   
   \hline
         \label{tab:table1b}
        \end{tabular}
  \end{table}
\end{subtables}

\end{landscape}

\noindent  
The initial classification of each source is based on an assessment of all available data, beginning with the light curve morphology, near infrared and mid infrared colours 
(the latter from the {\it Spitzer} Galactic Legacy Infrared Mid-plane Surveys, GLIMPSE I-II , \citealt{benjamin03, churchwell09}, and the Wide Field Infrared Survey Explorer, 
WISE, \citealt{wright10}). We then considered any past studies (assisted by SIMBAD, \citealt{wenger00}, Vizier and the International Variable Star Index, VSX), follow up spectroscopy (for many of 
the YSO candidates, see GLK), source location and the parallax and proper motion from {\it Gaia} Data Release 3 (DR3, \citealt{gaiadr3}) and VIRAC2. Source location can help to identify 
YSOs, if the star is projected in a star formation region, but this method is less useful on Galactic disc sight lines within 10$^{\circ}$ of the Galactic centre, where indicators of star 
formation are ubiquitous in SIMBAD and in the WISE mid-infrared images. To test for location in a star forming region, we employed the method as described in CP17a and \citet{lucas17}, 
but we treat the results only as one indicator amongst several when assigning initial classifications.

A period search was performed with elements of a software pipeline developed to search the entire VIRAC2-$\beta$ database for 
periodic sources (Miller et al., in prep). For sources in the present work, the pipelien uses three period search methods after first removing any overall linear trend: Phase Dispersion Minimisation 
\citep[][specifically the \textsc{PDM2} code]{stellingwerf78}, Lomb Scargle \citep{scargle82}, Conditional Entropy \citep{graham13} and selects the one that best fits the data. 
For this work, periods were confirmed by visual inspection of the light curves and independently confirmed by a novel 
machine learning procedure that computes a false alarm probability (Miller et al., submitted) after being trained to distinguish between aperiodic and periodic light curves.
Periods of LPVs are given in Appendix \ref{sec:s163} and the one other periodic source is discussed in Appendix \ref{dipysos}.

Our initial classifications break down as follows: 70 CVs, 40 YSOs, 38 microlensing events, 35 LPVs, 21 sources that we call ``dipping giants", 10 sources that we classify as ``unusual" 
and eight transients denoted ``sparse", where there are too few detections over time to attempt an astrophysical classification. We discuss each of these classes below in turn.
To note, there is relatively little useful optical time series data to aid classification of the new discoveries, from {\it Gaia} \citep{eyer23} or other public optical time domain surveys, 
owing to the red nature of the sample. Most of the 50 {\it Gaia} DR3 matches are novae (28 sources, previously published in all but one case), microlenses (11 sources, none of which
have multi-epoch photometry in Gaia DR3), or YSOs (7 sources, among which only source 78 has a Gaia DR3 light curve, presented in a single object study, Guo et al., submitted).
Additionally, source 192 had a Gaia Alert, Gaia~17bzk, though only the fading portion of this eruptive YSO event was sampled and the star is not in Gaia DR3.

\begin{figure*}
        \includegraphics[width=\textwidth]{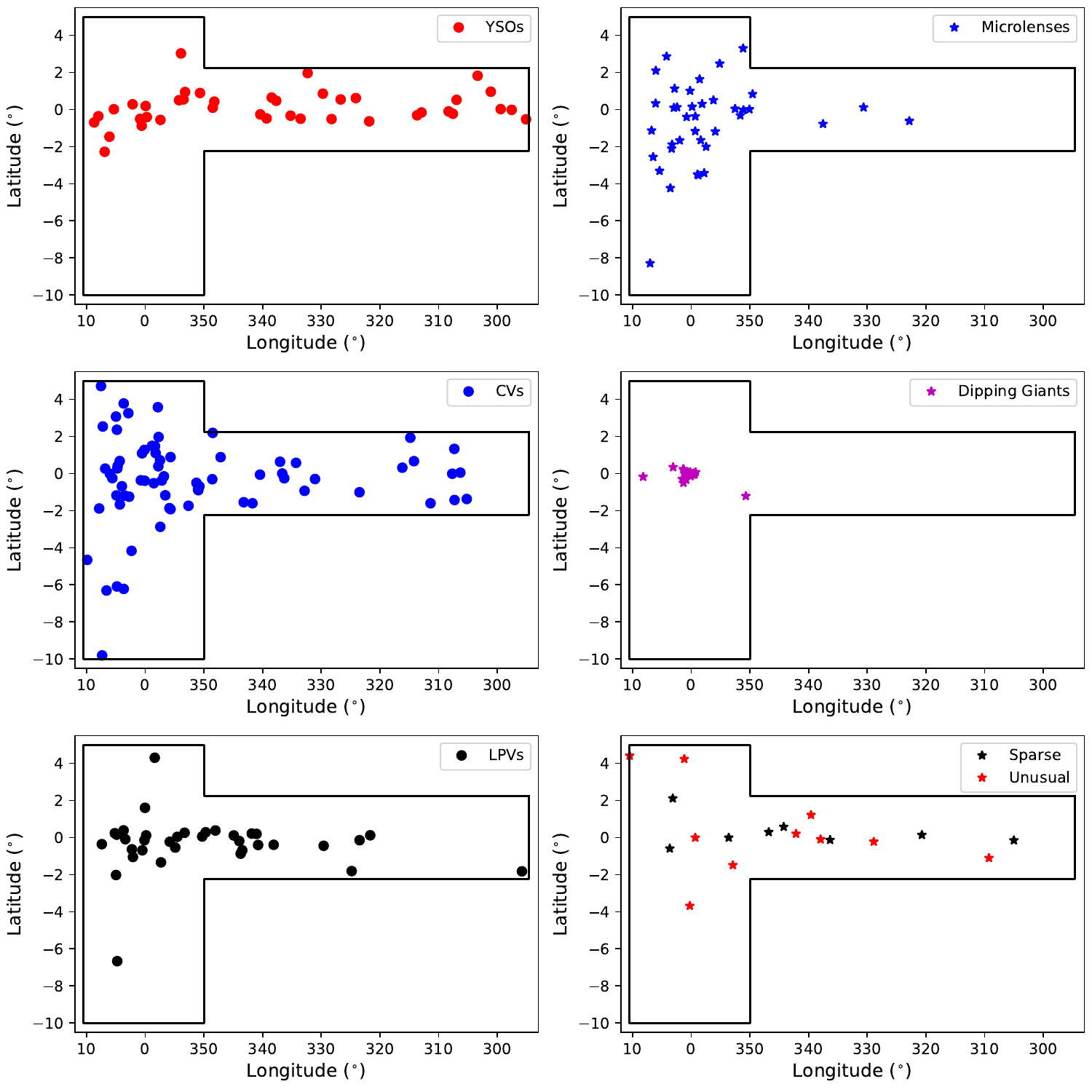}    
    \caption{Spatial distribution of variable sources in the $\Delta K_s > 4$ sample. The vertical scale is stretched for clarity. The "dipping giants" show a strong clustering 
    towards the Nuclear Disc of the Milky Way. VVV survey boundaries are indicated with black lines.}
    \label{fig:galdistrib}
\end{figure*}

Figure \ref{fig:galdistrib} shows the spatial distribution of all the sources in Galactic coordinates, with the VVV survey area indicated. The YSOs are fairly uniformly spread along the 
mid-plane of the inner Galactic disc. Most YSOs (35/40) lie at Galactic latitudes $|b|<1^{\circ}$, as expected for a young population with a small Galactic scale height.
By contrast, CVs are more concentrated in the VVV bulge region, especially the inner bulge. Microlenses are even more strongly concentrated in the bulge, consistent with 
past optical surveys such as OGLE \citep{wyrzykowski15} and prior work using VVV \citep{navarro18, navarro20a, navarro20b}.

The dipping giants stand out as a very tight cluster of 21 points near the Galactic centre, most of them projected in the Central Molecular Zone and the Nuclear Disc of the Milky Way
that lie between longitudes 358$^{\circ}$ and 2$^{\circ}$.  At first, there were three possible interpretations of these sources:
YSOs with long-lasting dips like AA Tau \citep{bouvier13, covey21}, giant stars undergoing unusually strong dimming by circumstellar dust, or eruptive YSOs (in cases where the VVV light curve morphology was ambiguous). 
However, spectroscopy of seven of these sources (see GLK) and use of photometry from earlier surveys (see $\S$\ref{sec:dipgiants})
helped to distinguish giant stars from YSOs and resolve the nature of ambiguous light curves: such sources are assumed to be undergoing a dip in VVV if they were bright in earlier surveys. In some cases there is supporting 
evidence of variable reddening from the VVV multi-filter observations. Moreover, the recognition that the Central Molecular Zone has a very high surface density of molecular cloud cores and clumps 
\citep[e.g.][]{parsons18} helped us to remove YSO status for some sources lacking spectroscopic data that are projected adjacent to one or more cloud cores. 

Given the spectroscopic evidence for evolved status of 
seven sources we opt for a giant star classification for dipping sources near the Galactic centre that presently lack spectra in all but one case (source 134) where YSO status is supported by a previous study, 
see $\S$\ref{sec:ysodipstub}.
In total we classify eight sources with long-lassting dips as YSOs but only source 134 is projected in the Nuclear Disc.
The dipping giants are discussed further in \S\ref{sec:dipgiants} and in GLK.

The LPVs are spread along the Galactic mid-plane but with a greater concentration towards the inner Galaxy than the YSOs. This is consistent the {\it Akari}-based study of \citet{ishihara11}, which was sensitive to 
LPVs at all Galactocentric radii and quantified the high concentration of O-rich LPVs in the inner Galaxy, where the more uniformly spread C-rich LPVs are greatly outnumbered.

Interestingly, the eight sparsely sampled transients (see lower right panel of Figure \ref{fig:galdistrib}) are spread along the mid-plane (7/8 lie at $|b|<1^{\circ}$) in a manner that 
appears different to the bulge-centric CV and microlens spatial distributions. This suggests that not all of these sources are poorly sampled examples of those 
two classes of variable sources. Finally, the 10 ``unusual" sources are a very diverse group (see Appendix \ref{unusual}) so 
their distribution has little meaning.

Table \ref{tab:table1b} lists two eruptive YSOs with amplitudes in the 3 to 4 mag range that were removed from Table \ref{tab:table1a} at a fairly late stage, after undertaking spectroscopic follow up (see GLK), when it was recognised that the amplitudes of the light curves were significantly below the 4 mag threshold after 1-day binning with 3$\sigma$ rejection of outliers. In the
case of VVV~J164011.76-484653.4, hereafter VVV~1640-4846, the YSO is embedded in a compact nebula so the VIRAC2-$\beta$ photometry were replaced with aperture photometry 
(aperture diameter 1.414\arcsec) using a carefully selected background annulus. 

\subsection{DR4 sources with $\Delta K_s>3$}
\label{sec:DR4results}

Our search of the public DR4 release for sources with $\Delta K_s>3$ mag (see $\S$\ref{DR4}) yielded 105 sources. These 105 sources are listed in Appendix C. They are denoted with the prefix 
``DR4\_v", consistent with \citet{guo21}.
Of the 105, the 20 with $\Delta K_s>4$ in the DR4 aperture photometry data were all recovered in the VIRAC2-based searches. A further 15 have $\Delta K_s>4$ in the deeper VIRAC2 
\textsc{dophot}-based data so 35 of the DR4 sources are also listed in Table \ref{tab:table1a}. We very briefly discuss the 70 DR4 sources with $K_s =$~3 to 4 mag in order to provide some information 
about the types of sources seen, since a general $\Delta K_s >3$~mag search has not yet been done with VIRAC2. Our initial classifications 
break down as follows: 23 LPVs, 22 YSOs, 1 CV (nova candidate VVV-NOV-012, \citealt{saito16}), 15 microlenses, five listed as unusual and four in the sparse category. 
We see that, as expected, there are larger proportions of YSOs and LPVs and a much smaller proportion of CVs with measured amplitudes in this range than in the VIRAC2 
$\Delta K_s>4$ sample. 

Spectroscopic follow up and analysis of 15 eruptive YSO candidates from this DR4 set was undertaken in 2019 (see \citealt{guo21}). Three eruptive YSO 
candidates in the set were noted earlier in CP17a and followed up in CP17b. 

\section{Young Stellar Objects}
\label{ysos}

\begin{figure*}
        \includegraphics[width=\textwidth]{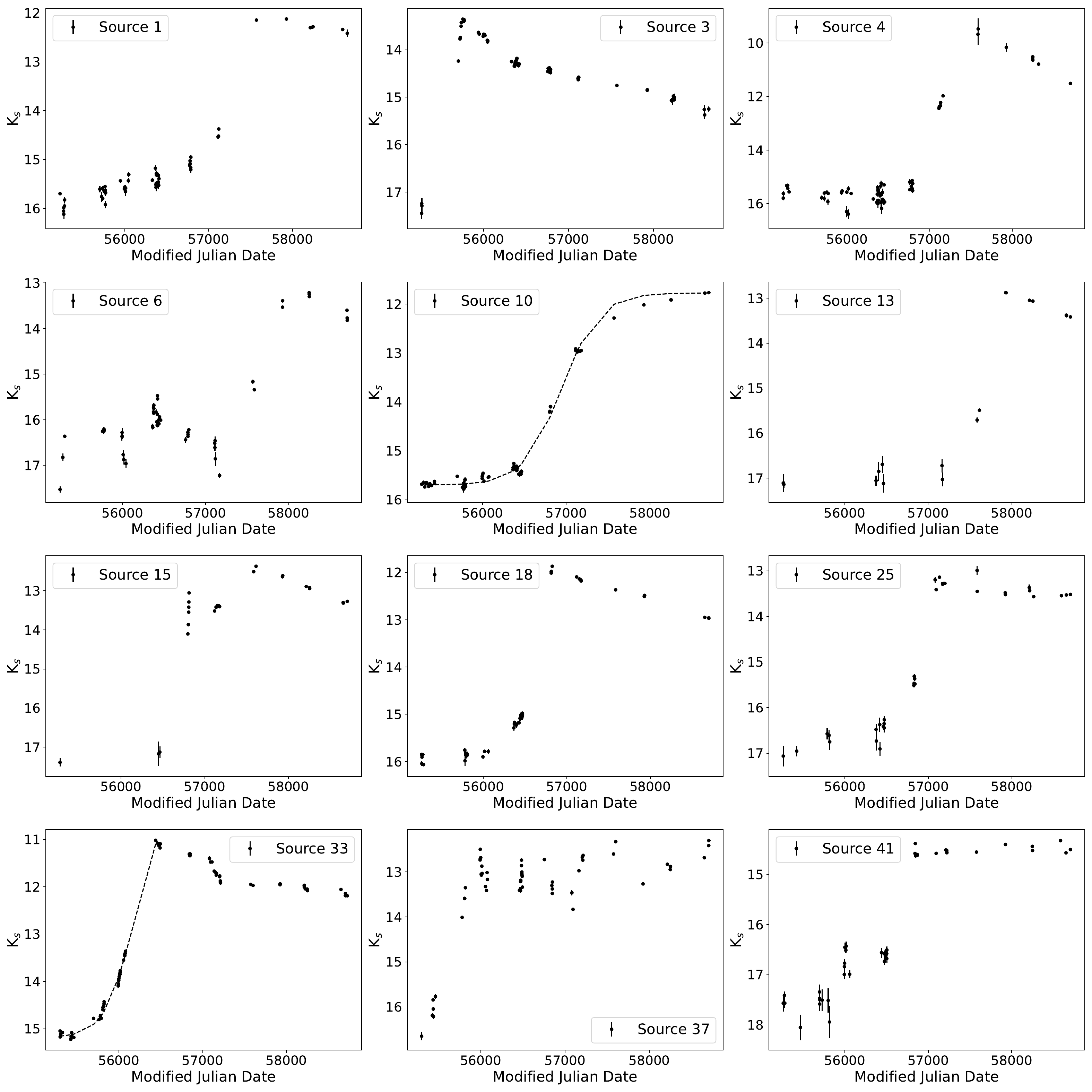}    
    \caption{Light curves of the ``classic" long duration outbursting YSOs. Examples fits to the rising portion of the light curves are shown for source 33 (using equations 1a and 1b) and for source 10
    (using equation 1a only), see main text.}
    \label{fig:erup12}
\end{figure*}

\renewcommand{\thefigure}{\arabic{figure} - continued}
\addtocounter{figure}{-1}
\begin{figure*}
        \includegraphics[width=\textwidth]{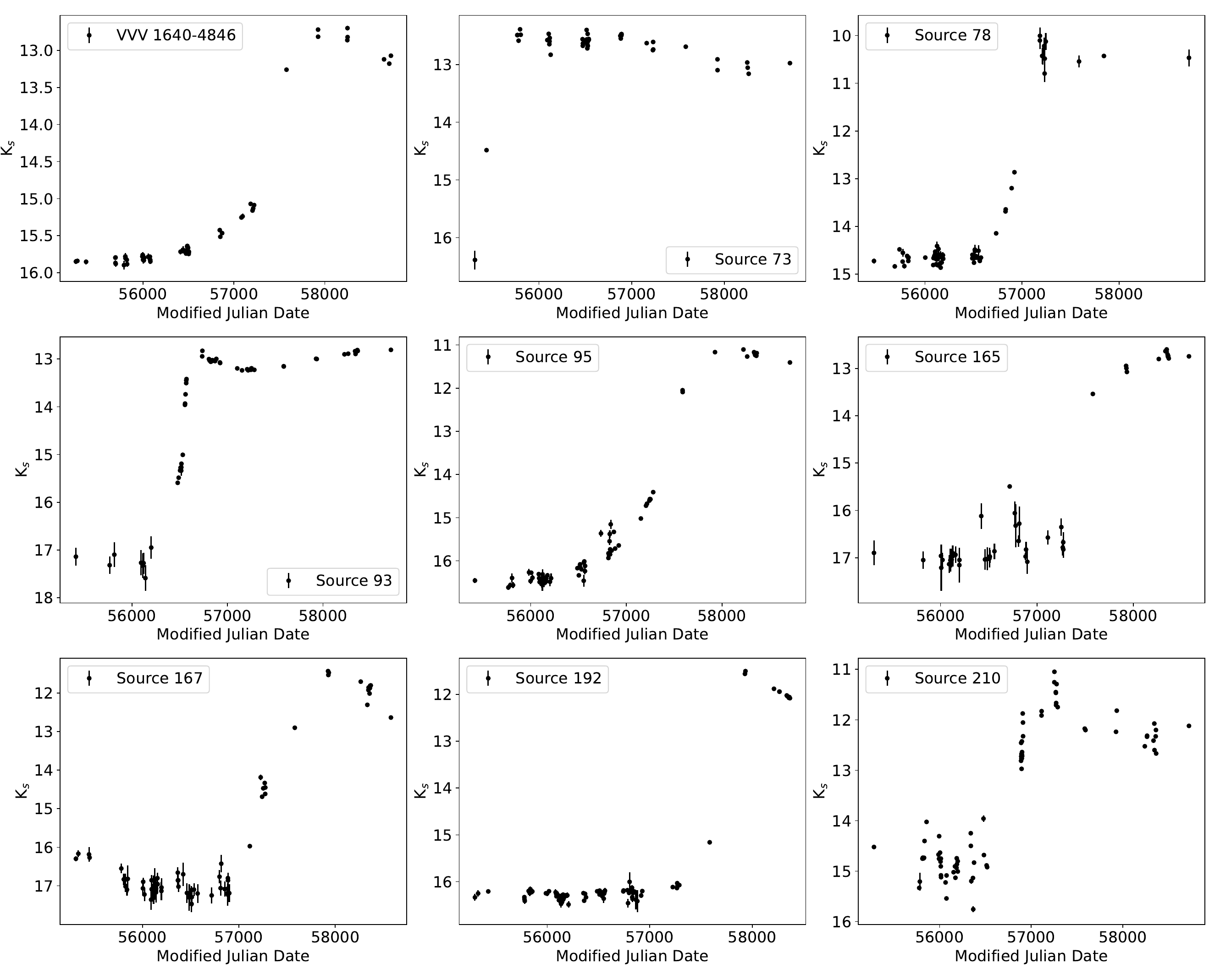}    
    \caption{Light curves of the ``classic" long duration outbursting YSOs.}
    \label{fig:erup9}
\end{figure*}
\renewcommand{\thefigure}{\arabic{figure}}

The 40 YSOs in Table \ref{tab:table1a}, comprise 32 eruptive YSOs and eight systems with dips that appear to be due to variable extinction. Our main focus in this work, and in GLK, is the eruptive YSOs.
The 32 eruptive systems have some variety in light curve morphology but 20 (63\%) can be broadly described as ``classic" eruptive YSO events, see Figure \ref{fig:erup12}. These have a reasonably smooth
rise from a faint state to a bright state that then lasts for at least several years, i.e. continuing after the end of the VVV survey, though some fade by $>1$~mag during that time\footnote{Source 165 
had a smaller burst two years before the main event and source 15 shows slightly more complicated behaviour after the initial rising stage but we include these in the "classic outburst" group despite these 
differences.}. Nine eruptive systems have more irregular and diverse light curves, see $\S$ \ref{sec:irregular}.

A further three eruptive systems may well have been of the classic type: these are cases where the outburst reached maximum in 2010, or earlier, so there was little or no sampling
of the rise by VVV and the high amplitude is due to fading of the outburst by several mag within the 2010 to 2019 time period. Such a rapid rate of decay is faster than the three historical sources
often called the ``classical FUors" \citep[FU Ori, V1057 Cyg and V1515 Cyg, see e.g.][]{hartmann96} but within the range of the 20 classic VVV events where the rise was observed, as shown by recent photometry 
in GLK. These three systems are source 19 \citep[= WISEA~J142238.82-611553.7,][noted for its $\sim$8~mag outburst]{lucas20b}, source 47 (= VVVv717, which has a FUor spectrum, see CP17b) 
and source 84, which lacks spectroscopic data.

The two slightly lower amplitude eruptive variables listed in Table \ref{tab:table1b}, comprise one classic eruptive system, VVV~1640-4846, and one irregular system, VVV J163637.94-474444.1 (hereafter
VVV~1636-4744). The latter is one of the few irregular systems having spectroscopic follow up (GLK) and our analysis of classic eruptions includes several other spectroscopically
confirmed lower amplitude events so these two are discussed along with the rest.

\subsection{Classic outbursts}

\begin{figure*}
        \includegraphics[width=0.49\textwidth]{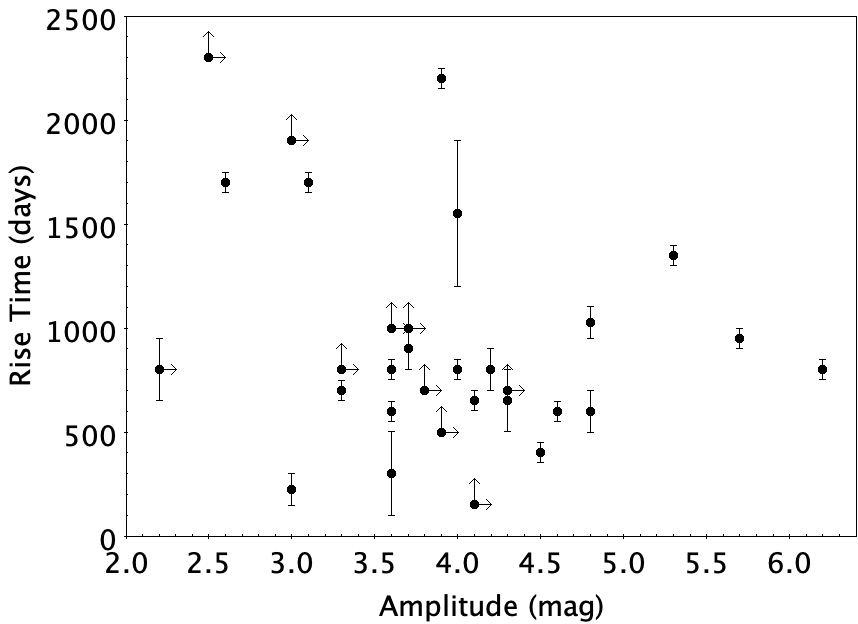}    
        \includegraphics[width=0.49\textwidth]{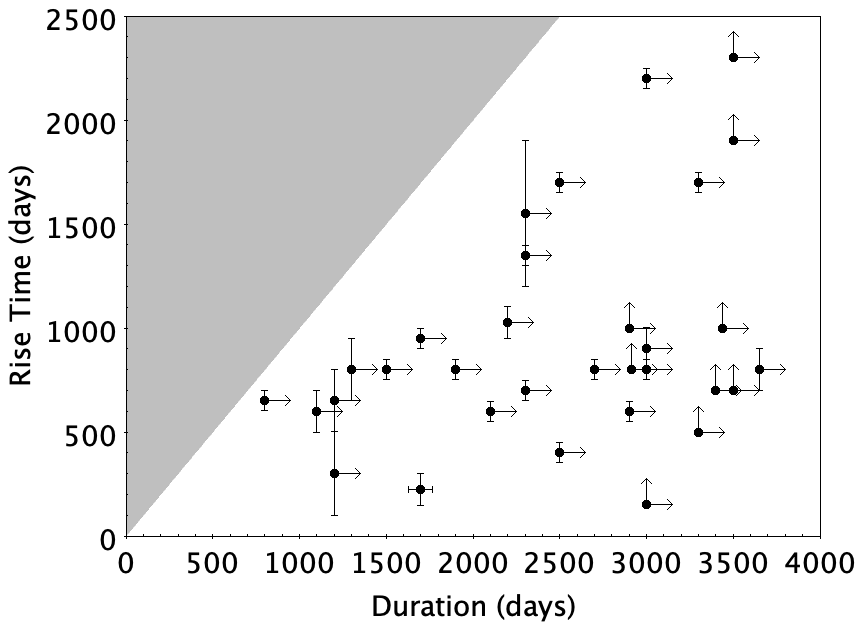} 
    \caption{Rise time plotted against amplitude (left panel) and outburst duration (right panel) for 32 classic events drawn from this work and earlier VVV discoveries. 
    Typical rise times are 600 to 1000 days, though there are faster rises and slower rises. Durations are usually lower limits since the events continued beyond 2019. The shaded region is empty because the 
    duration includes the rise time.}
    \label{fig:rise}
\end{figure*}

\begin{figure}
        \includegraphics[width=0.49\textwidth]{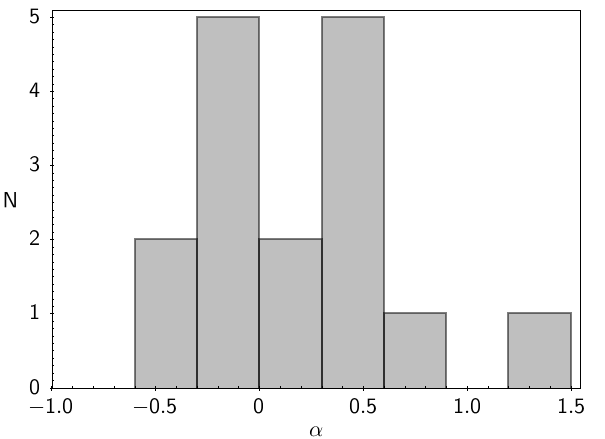}    
    \caption{Distribution of spectral indices for 16 classic eruptive variables.}
    \label{fig:alpha}
\end{figure}

A key finding of this work is that high amplitude outbursts with a long duration typically have a slow rise over $\sim$2 yr or more, contrasting with the classical FUor systems FU Ori and 
V1057 Cyg where the rise time was only about 6 months \citep[FU Ori, see ][]{kenyon00} or 13 months \citep[V1057 Cyg ][]{herbig77}. By contrast, the third classical FUor, V1515 Cyg, had a uniquely 
slow rise time of at least 15 years, perhaps longer \citep[see][]{herbig77}, considerably slower than any in this work or any other published FUor.
In Figure \ref{fig:rise} (left panel) we plot the rise time of the outbursts against amplitude for the 20 classic outbursts in Table \ref{tab:table1a}, and a further 12 spectroscopically confirmed VVV discoveries with 
classic morphology and amplitudes in the 2 to 4 mag range, drawn from Table \ref{tab:table1b}, Appendix C and CP17b, see \citet{guo21, guo20} and CP17b for spectra and VIRAC2 light curves of the Appendix C  
sources and earlier discoveries. Here we measure the rise time from the beginning of the outburst until the bright plateau or turnover is reached, neglecting any short timescale/low amplitude scatter in the faint state or the 
bright state. Due to the sparse VVV sampling there is some uncertainty in the rise times but the error bars encompass the plausible range of values based on visual inspection. The numerical data are given in 
Table \ref{tab:rises}, along with a fitted rise timescale, $T_{fit}$, see below. 

We see that the typical rise time is 600 to 1000~d, though in some cases it is $\sim$2000~d or more and in a smaller number of cases it is 
only $\sim$200 to 400~d.

\begin{table*}
	\caption{Rise times and quiescent-state spectral indices of 32 classic outbursts}
	\label{tab:rises}
	\begin{tabular}{lcccccc} \hline
Source & Amplitude &  Duration & Rise & Rise Time & $T_{fit}$ & $\alpha$\\
      &   (mag)   &  (d)   &  Time (d)          &  Error  (d)     & (d) & \\ \hline
  1      &      3.9  &    $\ge$3000  &   2200  &    50    &  768 &  0.47 \\
  3   	&    $\ge$4.1  &  $\ge$3000  &   $\ge$150   &     &  & \\
  4  	  &	 6.2   &   $\ge$1900  &   800   &    50     & 812 & 0.04 \\    
  6  	  &	 4.0   &   $\ge$1500   &  800   &    50   & 608 & \\  
  10 	  &	 4.0   &   $\ge$2300   &  1550 &     350  &  909 & -0.58 \\  
  13 	  &	4.1   &   $\ge$800   &   650    &   50   &  501 &\\  
  15 	  &	4.8   &   $\ge$2200  &   1025 &     75    & & \\  
  18  	  &	 4.0   &   $\ge$2700  &   800   &    50    & 461 &  -0.35\\  
  25 	  &	  3.3   &   $\ge$2300  &   700   &    50    & 676 &  \\  
  33 	  &	  4.2  &    $\ge$3650  &   800   &    100  & 706 & -0.28 \\   
  37  	 & $\ge$4.3   & $\ge$3400  &   $\ge$700   &      & & 0.55 \\
  41	  & $\ge$3.7  &    $\ge$2900  &   $\ge$1000  &      &  & \\    
  73 	  & $\ge$3.9  &   $\ge$3300  &  $\ge$500   &      & & 0.39 \\
  78    &	4.6  &  $\ge$2100  &   600   &    50   & 425 & \\   
  93  	  &	4.5  &    $\ge$2500  &   400  &     50  &  163 &\\    
  95 	  &	5.3  &   $\ge$2300  &   1350  &    50  &  997 &\\  
  165    &	4.3   &  $\ge$1200  &   650    &   150  &  340 &\\ 
  167    &	 5.7   &   $\ge$1700  &   950   &    50  &  700 & 1.26\\  
  192   &	 4.8  &  $\ge$1100  &   600   &    100  &  449 &\\ 
  210   &	3.6   &  $\ge$1200  &   300   &    200  &   &\\  
   VVV 1640-4846 & 3.1  &    $\ge$2500  &   1700   &   50  & 796 &  \\
  Stim 5  &    $\ge$3.8  &   $\ge$3500  &  $\ge$700   &    & & -0.27 \\    
  DR4\_v10  &  $\ge$3.6  &   $\ge$3437  &  $\ge$1000  &   &   & \\  
  DR4\_v15  &   3.6  &   $\ge$3000  &   800   &    50  & 969 & -0.01 \\  
  DR4\_v20  &  $\ge$3.3  &  $\ge$2913  &   $\ge$800   &       &   & -0.12 \\
  DR4\_v34  &   3.6  &  $\ge$2904 &    600  & 50       & & -0.23 \\ 
  V270      & 3.7   &  $\ge$3000  &   900  & 100    &  753 & 0.26  \\
  V322      & 3.0   &  1700  &   225    &   7    & &  \\
  V721       & $\ge$3.0  &   $\ge$3500  &  $\ge$1900  &   &    &  0.4 \\
  V631       & 2.6  &  $\ge$3300  &   1700   &   50    & &\\ 
  V800       &  $\ge$2.2  &   $\ge$1300  &   800   &   150   & 748 & 0.68 \\
  Stim 1     &  $\ge$2.5  &  $\ge$3500  &  $\ge$2300  &      &   & 0.31 \\ \hline
\end{tabular}\\
\raggedright Notes: the $K_s$ amplitudes in column 2 exclude short timescale variability so they are a little less than the $\Delta K_s$ values in Table~\ref{tab:table1a}.\\
The ``Rise Time" is based on visual inspection, measuring the total time from the start of an outburst until photometric maximum, discounting short-timescale
scatter. $T_{fit}$ is a fitted rise time that encompasses the bulk of the rise in brightness but excludes the asymptotic tail that occurs at the beginning, and
sometimes at the end also. $T_{fit}$ is only given if a useful fit was possible\par
\end{table*}

We inspected the literature on long duration outbursts, for which a helpful list was compiled in \citet{connelley18}. While most past eruptions have little photometry during the rise to maximum,
useful (optical) data exist for V2493 Cyg (=HBC722),  V2775 Ori (=CTF93 216-2), V582 Aur and V960 Mon, in addition to the three classical FUors mentioned above. The rise times from quiescence to 
the bright state are: 2 months (V2493 Cyg), $<18$ months (V2775 Ori), 1~yr (V582 Aur) and $\le8$~months (V960 Mon), based on reports by \citet{semkov12}, \citet{caratti11}, \citet{semkov13} and 
\citet{hackstein15}, respectively. These four rise times, like those of FU Ori and V1057 Cyg, are somewhat faster than is typical of the VVV sample. 
These six previously reported events are relatively bright at optical wavelengths, excepting only V2775 Ori which \citet{connelley18} argue suffers high reddening simply due to location at the back of the L1641 cloud rather than an earlier evolutionary status. 

It is possible that the slower rise times seen in the near infrared reflect a difference between optical and infrared measurements, if the accretion burst is detected at an earlier stage in the infrared, at lower disc temperatures,
and only later becomes obvious in the optical as the temperature rises further. For example, this occurs naturally in disc models where FUor outbursts are due to sudden triggering of magneto-rotational instability in the dead 
zone and the outburst then propagates inward \citep{cleaver23}.
However, for the more recent FUors Gaia~17bpi and Gaia~18dvy  \citep{hillenbrand18, szegedi-elek20}, inspection of the {\it Gaia} Alerts data  shows that these had total optical rise 
times of $\sim$900--1200~d (Gaia~17bpi) and almost 700~d (Gaia 18dvy), in the same range as the near infrared data for the VVV sample.

In view of these two recent {\it Gaia} discoveries, we cannot be certain that the slow infrared rise times seen in the VVV sample are different to those of optically bright, optically measured events. The
slightly faster rise times of the six historical outbursts mentioned above may simply reflect small number statistics rather than a real distinction between optical and infrared measurements, or a difference 
between optically bright events and embedded systems. 

In this context it is perhaps significant that there appears to have been a slower rise in the mid-infrared than the optical in Gaia~17bpi and Gaia~18dvy \citep{hillenbrand18, szegedi-elek20}, interpreted as evidence for 
inward propagation of the disc instability. At present, we tend to the view that either (i) there is a genuine difference in optical and infrared rise timescales in individual eruptive YSOs due to the different temperature 
sensitivities of the wavebands,
or (ii) the faster optical timescales in most historical outbursts are a fluke of small number statistics. Whilst the embedded nature of the VVV systems might also be significant, it is unclear whether these sources are truly
at an earlier evolutionary stage than optically bright FUors. Historical FUors are usually associated with large amounts of circumstellar matter, seen in the far infrared and mm wavebands \citep{green13, feher17}
so their less embedded nature may simply be due to the lower foreground interstellar extinction towards nearby star forming regions or better clearing of circumstellar material along the individual lines of sight
by the stellar wind.

The spectral index, $\alpha$, measured at a time prior to the outburst for 15 systems is given in Table \ref{tab:rises} and Figure \ref{fig:alpha}. This was measured using either 3--22~$\mu$m WISE photometry from 2010 (preferred if available and pre-outburst) or 3--24~$\mu$m {\it Spitzer} photometry otherwise.\footnote{Bright and highly structured nebulosity is typically not well handled by photometric pipelines so it was necessary to visually inspect the images from WISE, GLIMPSE~I-II, MIPSGAL \citep{rieke04, carey09} and VVV in order to verify catalogued detections, many of which have warning flags, and remove any sources where the data were compromised by blending.} 
(VVV $K_s$ photometry were excluded in order to reduce the potentially significant effect of foreground extinction in the Galactic plane, see CP17b). We also include the index for a sixteenth system, source 37, which appears to have been in the early stages of the rise at the time of the measurement in 2010. The mean spectral index is $\alpha = 0.16$, or 0.13 if source 37 is excluded, and (14/16) systems
fall in the Class I category ($\alpha >0.3$) or Flat Spectrum category (-0.3 < $\alpha <0.3$), see \citet{greene94}.  Most historical FUors lack a measurement of the spectral index in the quiescent state, preventing a
direct comparison between optically bright and embedded systems. These data are therefore a useful indicator of the evolutionary stage of the progenitors of classic outbursts, for cases where the system is not too deeply 
veiled to be detected in the near infrared. The systems lacking a spectral index due to non-detection in the [24] and $W4$ passbands were in most cases veiled from view by strong nebulosity gradients or blending, so their spectral energy distributions (SEDs) cannot be assumed to be bluer.

The morphology of the $K_s$ light curves in the rising phase of classic outbursts can be described as a rise that begins slowly, accelerates to a faster, approximately constant rate of increase and then changes slope abruptly as the bright state is reached. There is some variation here: a minority of light curves show a gradual decline in gradient as the rise approaches the 
bright state, rather than an abrupt change. An example of this "S-shaped curve of growth" (to borrow a term from economics) is found in source 10, which rises asymptotically towards
maximum light over several years (making it hard to define when the bright state is reached). Other examples are source 165 and perhaps sources 6, 95 and VVV~1640-4846, though the sampling is too sparse for certainty in these last three cases. Gaia~18dvy has a sufficiently well-sampled {\it Gaia} light curve to inform this topic: the outburst has the characteristic slow start over $\sim$400~d, followed by a fast linear rise to within $\sim$0.5 mag of peak brightness over $\sim$200~d, but the final stage of the rise then proceeds at a more gradual rate over $\sim100$~d. Given the sparse VVV sampling, a similar brightening profile to Gaia~18dvy is certainly possible in many cases. Finally, we note that source 167 lacks the slow start and shows an 
approximately linear rise from quiescence to the bright state.
 
 In Figure \ref{fig:rise} (right panel) we plot the rise time against total duration, though almost all of these outbursts were ongoing at the end of the VIRAC2 time series so the durations 
are usually lower limits. We see that the outbursts typically have a total duration of at least several years, though the decay rates vary considerably between systems (discussed further in GLK). 
In terms of duration then, these high amplitude outbursts appear to match the expectation for FUor events. In GLK we confirm spectroscopically that most of the highest amplitude classic events are 
indeed FUors, the remainder being eruptive YSOs of other types. Only sources 3 and 41 presently lack spectra.
The companion paper by Contreras Pe\~{n}a et al. (submitted) includes a small subset of the high amplitude sample presented here, likewise noting their long outburst durations in comparison 
with lower amplitude events, which have a spread of durations (as seen in \citealt{guo21} also). 
 
For 18 sources in Table \ref{tab:rises} where the full rise was adequately sampled from quiescence, we fitted the source magnitude, $m(t)$, during the rising portion of the outbursts with the following formalism:

\begin{subequations}
\label{eq:e1a}
\begin{align}
& m(t) = m_q - \frac{s}{1 + e^{-(t-t_0)/\tau}}     \hspace{2.2cm}      t < t_0 \\
& m(t) = m_q - s(0.5 + 0.25(t-t_0)/\tau)    \hspace{0.7cm}         t_0 \le t \le t_0 + 2\tau \label{eq:e1b}
\end{align}
\end{subequations}

\noindent where $m_q$ is the quiescent magnitude, $s$ is the amplitude of the outburst, $t_0$ is the time at which the source has brightened by $0.5s$ and $\tau$ is the
e-folding timescale.  With this form, the bright state is reached at $t=t_0 + 2\tau$, at which point, $m(t) = m_q - s$. To aid the fitting of $\tau$ and $t_0$, we fixed $m_q$ and
$s$ and discarded data taken after the end of the rise and data taken more than 1--2 yr before the rise began (which otherwise had a tendency to dominate the fit due to denser VVV 
sampling in 2012-2013 than later years). Levenberg Marquardt fitting was performed with the Python \texttt{lmfit} package and the \texttt{piecewise} routine in \texttt{Numpy}.
Sources that were better described by an S-shaped curve of growth (sources 10, 165, 6 and 95) were fitted using equation 1a for all times, up to the end of the VIRAC2-$\beta$
time series. Equation 1a is simply a hyperbolic tangent, scaled by the outburst amplitude and shifted appropriately. Examples of these fits are shown in Figure \ref{fig:erup12}.

\begin{figure}
        \includegraphics[width=0.49\textwidth]{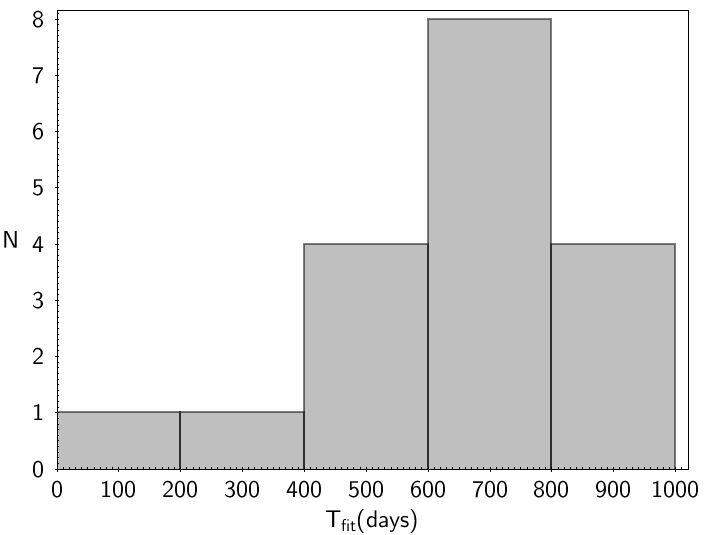}    
    \caption{The distribution of the fitted rise timescale for the classic outbursts. The quantity $T_{fit}=4\tau$ encompasses most of the rise (see main text) so it gives a similar but
    slightly shorter timescale than the full rise timescales plotted in Figure \ref{fig:rise}, which were estimated from visual inspection.}
    \label{fig:risehist}
\end{figure}

\begin{figure*}
        \includegraphics[width=\textwidth]{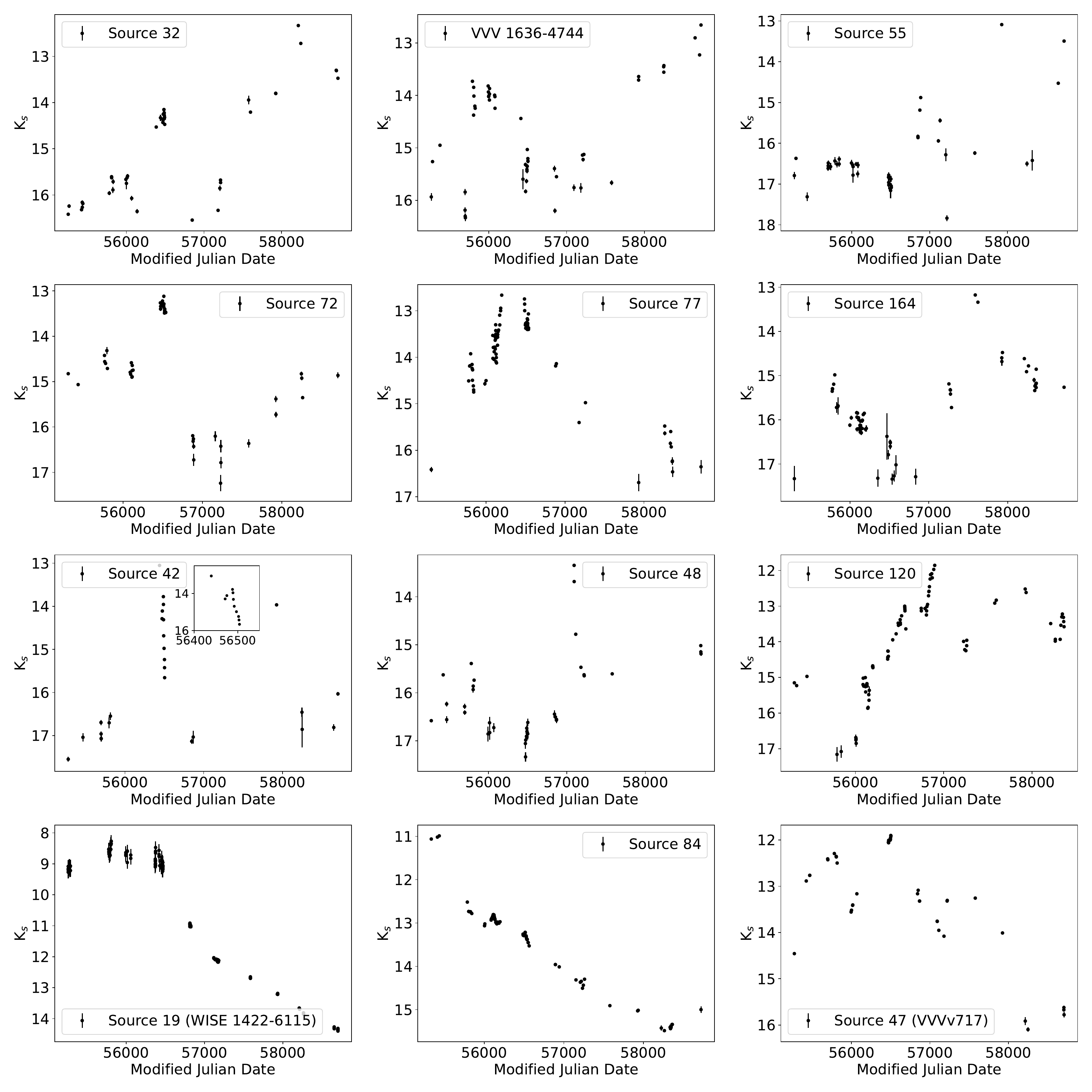}    
    \caption{Light curves of the nine irregular outbursting YSOs (upper three rows) and three others (bottom row) that mainly show a fading trend in VVV.
    An inset is included for source 42, to better illustrate the rapid variation of the first outburst of that source.}
    \vspace{-4mm}
    \label{fig:other12}
\end{figure*}

Figure \ref{fig:risehist} shows the distribution of rise times resulting from these fits, plotted as a histogram of $T_{fit}=4\tau$. Using the form in equations 1a and 1b, the time interval
$T_{fit}$, from $t_0 - 2\tau < t < t_0 + 2\tau$, encompasses most of the rise in flux, specifically $0.88s$. When using equation 1a only, for an S-shaped rise, this time interval 
corresponds to $0.76s$, whereas for a linear rise, $T_{fit}$ corresponds to the full rise, $s$. We see that the timescale represented by $T_{fit}$ is typically between 400 and 1000 
days, with the peak at 600 to 800 days. The distribution and this typical timescale are broadly consistent with the 600-1000 day typical rise time found by visual inspection, given that (i) 
the full rise does take longer than $T_{fit}$ and (ii) three of the slowest rises plotted in Figure \ref{fig:rise} are lower limits and therefore not fitted. Rise times longer than 1000 
days are more significantly under-represented by $T_{fit}$ but since this parameter encompasses most of the change in brightness, aside from the slowest-changing portions of the
S-shape, it may be of use to theorists. GLK use piecewise linear fits to describe the full light curves and come to similar conclusions about the rise timescale.

The faint pre-outburst magnitudes shown in Figure \ref{fig:erup12} allow us to infer the relevant range of stellar masses. The median $K_s = 16.5$ and source distances given in 
GLK for classic outbursts are usually between 2 and 8~kpc according to two separate methods (see Tables~1 and 2 of that work). Absolute magnitudes then lie mostly in the range $2 < M_{K_{s}} < 5$. 
Comparison with YSOs with well measured SEDs, similar spectral indices ($-0.3 < \alpha<0.6$) and absolute magnitudes in the ``Cores to Disks" sample of \citet{evans09} suggests that extinction-corrected 
bolometric luminosities are probably in the range $0.2<L_{\rm bol}/L_{\odot}<20$. This corresponds to low mass YSOs: if the quiescent luminosity is dominated by the star rather than accretion, the range of 
masses is approximately $0.2 < M/M_{\odot} < 3$, based on the 0.5 to 1~Myr isochrones of \citet{baraffe15} and \citet{felden16}. If accretion dominates then the masses would be lower.

The effect of FUor outbursts on source SEDs has been modelled in detail by \citet{macfarlane19}, \citet{rodriguez22}, \citet{hillenbrand22a} and \citet{liu22}. The first of these studies performed a full 
hydrodynamic simulation and radiative transfer modelling, focussing on class 0 YSOs, while the latter three studies explore a larger parameter space and focus on optical to mid-infrared observations. 
These works show that the outburst amplitude behaviour depends on many factors, such as the properties of the central star, the temperature structure of the disk, and the envelope density structure. 
\citet{hillenbrand22} find a dependence of the amplitude of the outburst with wavelength, with optical amplitudes being typically larger than infrared ones, in agreement with observations \citep{hillenbrand18,
szegedi-elek20}. \citet{liu22} model the full range of stellar masses mentioned above, and quantify the changes in $W1$ and $W1-W2$, finding a smaller difference between the infrared and optical amplitudes.
The authors attribute this to the assumption of a dipole magnetic field, which leads to a fainter quiescent disk and larger mid-infrared amplitudes. Nevertheless the amplitudes in their models depend on the maximum 
accretion rate reached during the outburst and the mass of the central star. From Figure \ref{fig:erup12}, only sources 1, 6, 10, 18 and 37 have well-sampled WISE/unTimely light curves \citep{meisner23} and these 
typically have $\Delta W1\ge 3$~mag (not shown). Inspection of Figure 16 in \citet{liu22} then suggests at least a 1.5-2 dex increase in accretion rate.

Unfortunately, the bolometric luminosity in outburst is not very well constrained for the VVV sources by the available VVV and WISE photometry at $\lambda \le 4.6~\mu$m and the somewhat uncertain
individual distances. Moreover, most of these sources have pre-outburst $W1-W2$ colours that are considerably redder than predicted and change (become bluer) by a larger amount than predicted, a behaviour also seen 
in the embedded FUor SPICY~97855 \citep{cp23}. \citet{liu22} note that the $W1-W2$ colours of FUors are sometimes considerably redder than predicted, which they attribute to various effects of the circumstellar envelope 
that were not included in the model. High and variable extinction alone cannot resolve these issues in the VVV sources, given the measured $H-K_s$ colours (see GLK). Some recently discovered embedded FUors 
actually become redder when brighter during outburst \citep[listed in][]{cp23}. In view of these difficulties, we do not attempt to quantify the accretion rate in outburst or analyse the systems fully at this stage.

\subsection{Irregular Eruptive Behaviour}
\label{sec:irregular}

In Figure \ref{fig:other12} we show the light curves of nine eruptive variable YSOs that display irregular outbursts and three (in the bottom row) that show fading behaviour within VVV because the
start of the outburst was missed. Eight of the nine irregular outbursts in Figure \ref{fig:other12} are drawn from Table \ref{tab:table1a} and one slightly lower amplitude system, VVV~1636-4744, is drawn from 
Table \ref{tab:table1b}.
A tenth irregular outburst, in source 148, is not shown here. That system displays strong periodic oscillations atop a rising trend, somewhat resembling
the recent outburst of LkH$\alpha$ 225 South, also known as V1318 Cyg South \citep{hillenbrand22, magakian19}. Source 148 is discussed further in GLK, where a
spectrum confirming its nature as an eruptive YSO is presented.

The light curves of the six sources plotted in the upper two rows of Figure \ref{fig:other12} (sources 32, VVV~1636-4744, 55, 72, 77 and 164) have some visual resemblance to each other. There are signs of two or more photometric outbursts, or at least two times when each source is relatively bright (though this is unclear in source 77) and significant variability on short timescales ($<100$~d). None of these
six sources have a detection in 2MASS PSC, indicating that they were relatively faint at the time of that survey. This supports the view that the variability is mainly due to accretion driven luminosity increases
from a faint state rather than a reduction in luminosity due to variable extinction. There is no useful data on the near infrared colour changes of these sources (all of which have red 1 to 5$~\mu$m SEDs): in most cases VVV detected them only in the $H$ and $K_s$ filters, with only one $H$ band detection and non-detections at other times when they were fainter.
Next we give some detail on individual sources, most of which are not discussed in GLK, before analysing the timescales of variability collectively.

\subsubsection{Individual sources}

{\it VVV~1636-4744 and source 72}. We consider the eruptive status of these two stars to be doubtful. In VVV~1636-4744 this is because the 2MASS images shows signs of an uncatalogued, heavily blended detection at $K_s \approx 14$--$14.5$, slightly brighter than the mean brightness level for this source, based on visual comparison with adjacent sources. The source has a spectrum in GLK, observed in an intermediate state of brightness with a modest signal to noise ratio. It shows fairly strong Br$\gamma$ emission and a red continuum, consistent with a veiled YSO undergoing magnetically controlled accretion, but no CO emission. Source 72 simply has an ambiguous VVV $K_s$ light curve morphology. There is no spectrum available and little additional information in other bandpasses.

{\it Sources 32, 55, 77 and 164}. 
Source 32 has an ``outflow dominated" spectrum (see GLK)\footnote{Outflow dominated spectra have been seen in several VVV YSOs with high amplitude variability \citep{guo21}. While these lack the HI recombination lines and the CO emission/absorption associated with classical EXors and FUors, many are thought to be cases of accretion-driven eruptive variability. GLK discuss further the relation between outflow dominated spectra and light curve morphology.}, with exceptionally strong H$_2$ emission lines. Sources 55, 77 and 164 lack spectra but the WISE unTimely and NEOWISE data \citep{meisner23, mainzer14} are useful for 
the latter two sources and for source 32, see Appendix \ref{irregysos}.
Sources 32 and 77 become bluer when they are brighter, with magnitude changes approximately in the ratio $\Delta W2/\Delta K_s = 0.25$ and 0.52, respectively (with some uncertainty since the measurements are not quite contemporaneous). Source 164 becomes redder when it is brighter, with $\Delta W2/\Delta K_s \approx 1.7$. According to \citet{wang19} the expected ratio in the case of variable extinction is $\Delta W2/\Delta K_s = 0.33$, 
based on measurements of Cepheids with high extinction \citep{chen18}. Only source 32 has colour variations close to this ratio and the exceptionally strong H$_2$ emission indicates the presence of a strong 
molecular outflow or disc wind, implying a high time-averaged accretion rate. Source  55 is blended with brighter neighbours in the $\sim$$6\arcsec$ WISE beam (as are VVV~1636-4744 and source 72) but in this case the 
light curve has a convincing eruptive morphology. Hence the available data tend to support an accretion driven origin for the variability of these four systems.

{\it Sources 42 and 48}. These two stars display only relatively brief outbursts. In VVV the decline timescales were $\sim$100~d but the rise and the photometric maximum were not sampled. For source 
42, WISE provides data for the second outburst sampled by VVV, at MJD=57925 (see Appendix \ref{irregysos}) indicating a duration of 1~yr. There is evidence for earlier outbursts of source 42: the 2MASS survey measured 
$K_s=13.52$ in 1999 and the {\it Spitzer} GLIMPSE survey in 2003 measured $[4.5] = 9.58$ at 4.5~$\mu$m, similar to the WISE/unTimely measurement $W2=9.48$ at 4.6~$\mu$m during the outburst in 2017. 
During the decline of the first outburst sampled by VVV there was a brief $\sim$0.5 mag re-brightening (see inset in Figure \ref{fig:other12}).
In source 48, four outbursts are seen in the WISE unTimely data, each observed at a single unTimely epoch (two with $\sim$2 mag amplitude in $W$2 and two with $\sim$1 mag amplitude).
This indicates that the durations were of order 6 months. Source 55 may also have outburst durations of only 6 to 12 months but the sparse sampling makes this uncertain. Among previous VVV discoveries, only 
VVVv118 shows somewhat similar behaviour \citep[CP17b, ][]{guo20} though in that case the duration of the outbursts was only $\sim$$50$~d.

{\it Source 120 (=DR4\_v67)}. This star shows rapid variability, in addition to larger inter-year variability, resembling that seen in the ``multiple timescale variables" (MTVs) reported in \citet{guo20, guo21}. Spectra
shown in \citet{guo21} and GSK are outflow dominated, with unusually strong H$_2$ emission. Comparison of the VVV and WISE light curves, shown in Appendix \ref{irregysos}, shows bluer-when-brighter behaviour with
ratios $\Delta W2/\Delta K_s = 0.48$ and $\Delta W1/\Delta K_s = 0.75$, somewhat larger than would be expected for variable extinction by interstellar dust.

\begin{figure}
        \includegraphics[width=0.5\textwidth]{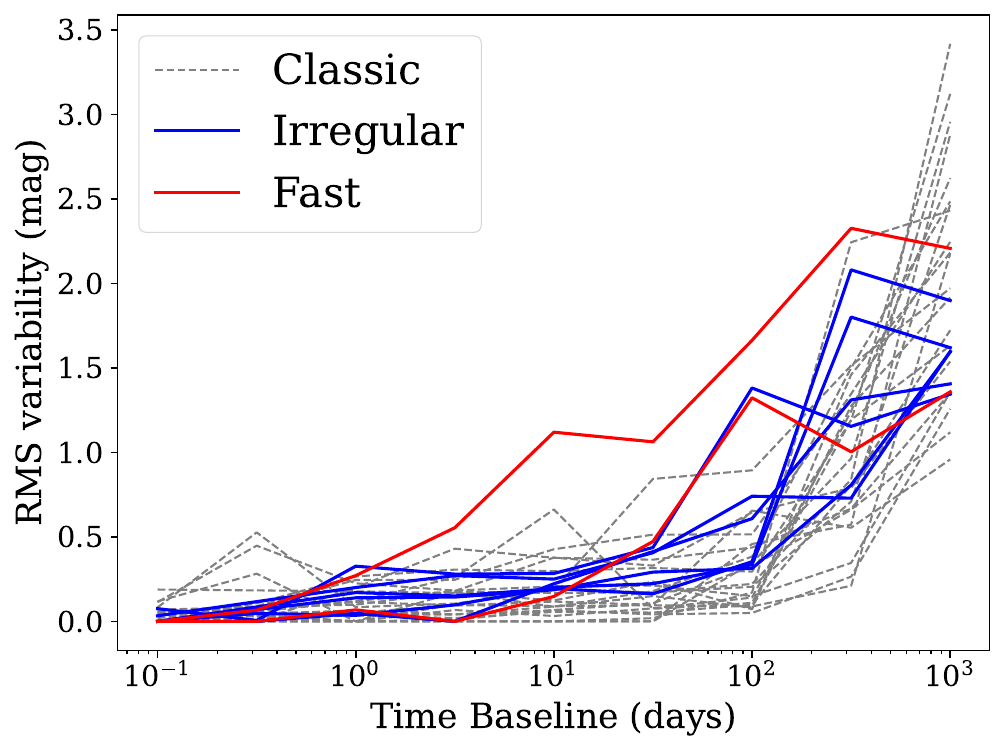}    
        \includegraphics[width=0.5\textwidth]{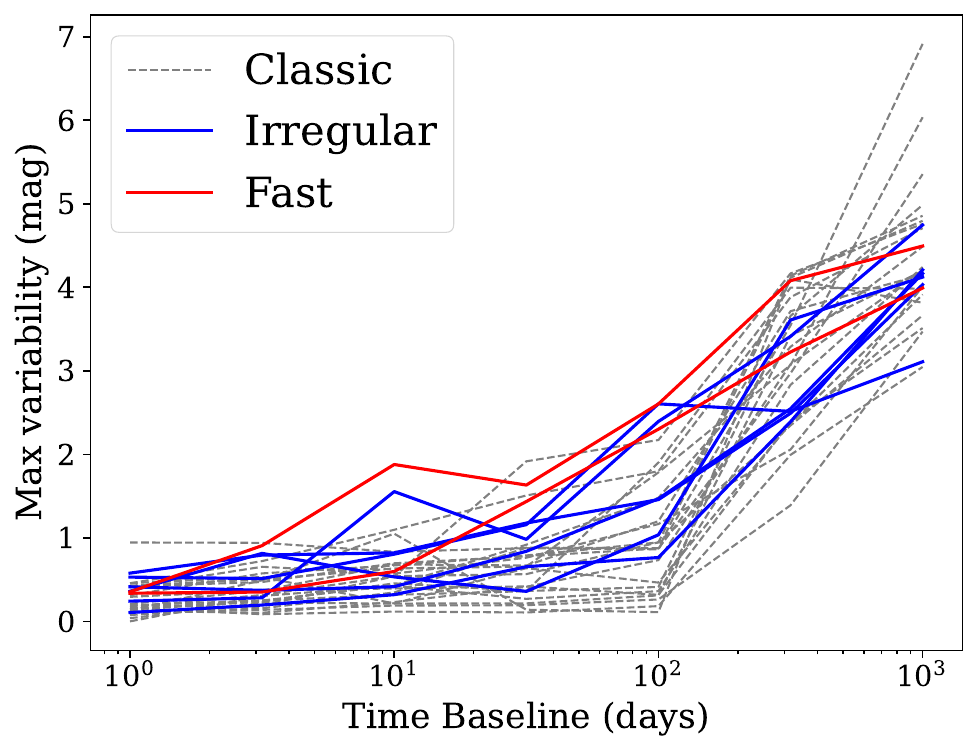}
        \vspace{-4mm}
    \caption{Variability of eruptive YSOs as a function of time baseline. The 21 classic outbursts from plotted in Figure 3 (grey dashed lines) are compared with the 6 sources with ``irregular" 
    light curves (blue lines) in the upper two rows of Figure 7 and two ``fast" sources (sources 42 and 48, red lines) also shown in Figure 7. {\it (Upper panel)}: r.m.s. variability of the 
    unbinned light curves from 0.1~d to 1000~d. {\it (Lower panel)}: maximum variability of the 1-day-binned light curves, over time baselines from 1~d to 1000~d. The irregular 
    and fast eruptive systems tend to have more variability than classic outbursts on timescales up to 100~d.}
    \label{fig:timescales}
\end{figure}

\subsubsection{Timescales}

Figure \ref{fig:timescales} illustrates the variability of the eruptive YSOs as a function of time baseline, comparing 21 classic outbursts (grey dashed lines) with the six irregular eruptive systems in the upper
two panels of Figure \ref{fig:other12} (solid blue lines) and sources 42 and 48 (solid red lines). These plots are constructed by computing the r.m.s. variability (upper panel) or maximum variability (lower panel)
over all pairs of $K_s$ points, $m_i$, $m_j$, in each light curve and binning the results in 0.5 dex bins in the logarithm of time baseline. 

The r.m.s. variability was computed using the unbinned pawprint light curves and the uncertainty in each value of $(m_i - m_j)^2$ was subtracted from the mean square variability in order to debias the result. 
The r.m.s. variability of source magnitudes is the square of the structure function as defined in \citet{lakeland22}, though some authors have preferred to use flux differences, see e.g. the YSO optical 
variability studies of \citet{sergison20}, \citet{venuti21}. An approach based on magnitudes, i.e. flux ratios, is more suitable than flux differences for large amplitude variability, as discussed in \citet{lakeland22}. The maximum 
variability vs. time baseline (also called the ``accumulation function'', see \citealt{guo20}) is a useful alternative for the sparsely sampled VVV light curves. This was computed using the 1-day binned light curves plotted 
in this work, in order to reduce the effect of individual 2--3$\sigma$ outliers.

We see that the six irregular systems and the two fast eruptive systems tend to have more variability than classic outbursts on timescales up to 100~d. The irregular outbursts generally show maximum
brightness changes of 1 mag or more on a 100~d timescale (Figure \ref{fig:timescales}, lower panel). The variability of the classic outbursts rises sharply on baselines from 100--1000~d. Source 42 displays the 
largest r.m.s. variability on 2~d to 300~d time baselines, though this is partly due to the good sampling of the first outburst seen by VVV. Source 93 stands 
out slightly from the other classic outbursts over 30~d to 100~d time baselines due to the unusually fast rise time of the event (see Figure \ref{fig:erup12}).

YSOs having irregular high amplitude accretion-driven behaviour clearly require further study and spectroscopic follow up, as do the MTVs reported in \citet{guo20} and \citet{guo21}. The connection
to classical EXors \citep{giannini22} may then become clearer. There appear to be two main types of these irregular light curve:  those with repeating outbursts that last 1~yr or less (sources 42, 48, VVVv118 and perhaps 
source 55) and those those with both intra-year and inter-year variability.

\subsection{YSO Extinction Events}
\label{sec:ysodipstub}

In eight YSO candidates there is good evidence that the variability is due to changing extinction. Further details and light curves are shown in Appendix \ref{dipysos}. We highlight
one case of special interest: source 207 appears to resemble KH~15D \citep[=V582 Aur,][]{herbst02, chiang04}, a system wherein the two bright components of a binary YSO are 
extinguished by a circumbinary ring for much of their orbits but each becomes visible in alternation at certain phases.

\begin{figure}
        \includegraphics[width=0.48\textwidth]{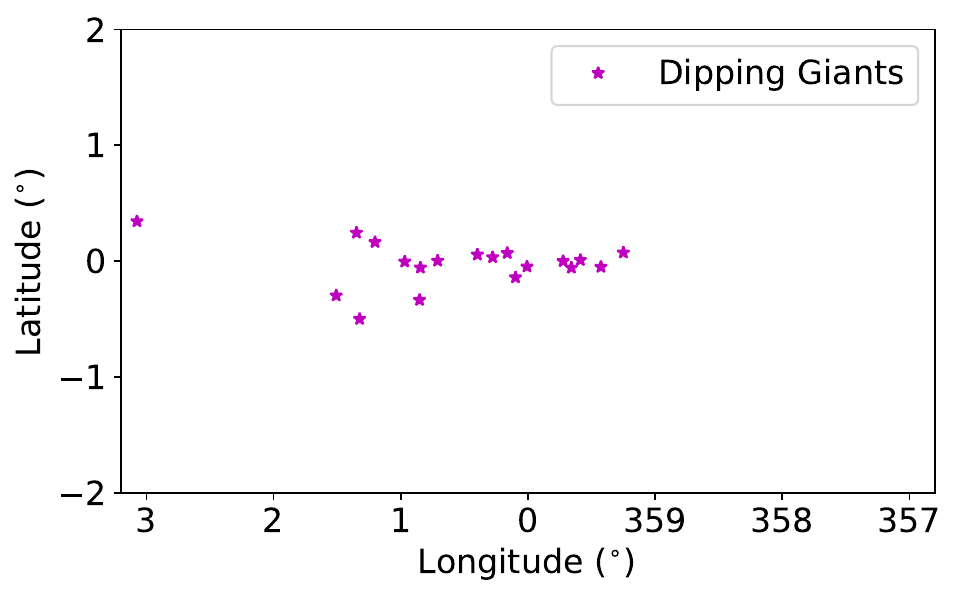}    
    \caption{Locations of the 18 dipping giants in the Nuclear Disc. A 19th case is plotted at $l = 3.1^{\circ}$ and two others lie a little further from the Galactic centre, at ($l$,$b$)=(350.7$^{\circ}$,-1.2$^{\circ}$) 
    and (8.2$^{\circ}$,-0.2$^{\circ}$), respectively.}
    \label{fig:dipgiants}
\end{figure}

\section{Dipping Giants in the Nuclear Disc}
\label{sec:dipgiants}

\begin{figure*}
        \includegraphics[width=\textwidth]{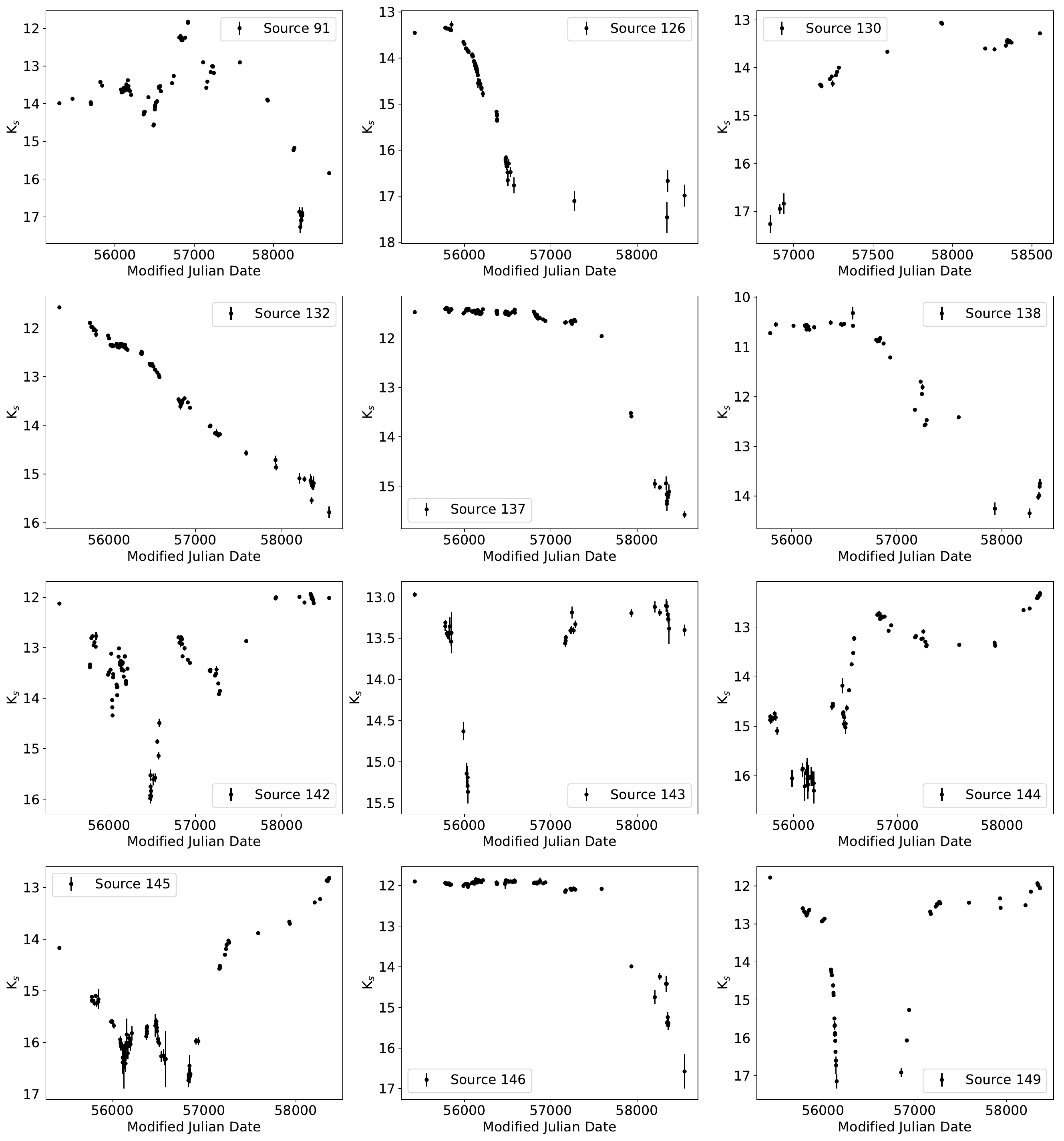}
    \caption{Light curves of the dipping giants}
    \label{fig:dipgiant12lc}
\end{figure*}

As discussed in $\S$\ref{overview}, we classify 21 sources as dipping giants. Of these, 18/21 are projected in the Nuclear Disc of the Milky Way, see Figure \ref{fig:dipgiants}. This structure is roughly coincident with 
the Central Molecular Zone, with radius 220~pc and scale height $\sim$45~pc, corresponding to Galactic longitudes $358.45^{\circ} < l < 1.55^{\circ}$, and scale height $\sim$$0.3^{\circ}$ \citep{launhardt02, fritz21}.
The $K_s$ light curves of the 21 sources are shown in Figure \ref{fig:dipgiant12lc}. They are very diverse but all can be described as a single, apparently aperiodic flux dip with a duration of at least a year. In fact most of the dips
either started before or ended after the 2010 to 2019 VIRAC2 time baseline. Only 6/21 dips can be said to be fully contained within the time series. 

\renewcommand{\thefigure}{\arabic{figure} - continued}
\addtocounter{figure}{-1}
\begin{figure*}
        \includegraphics[width=\textwidth]{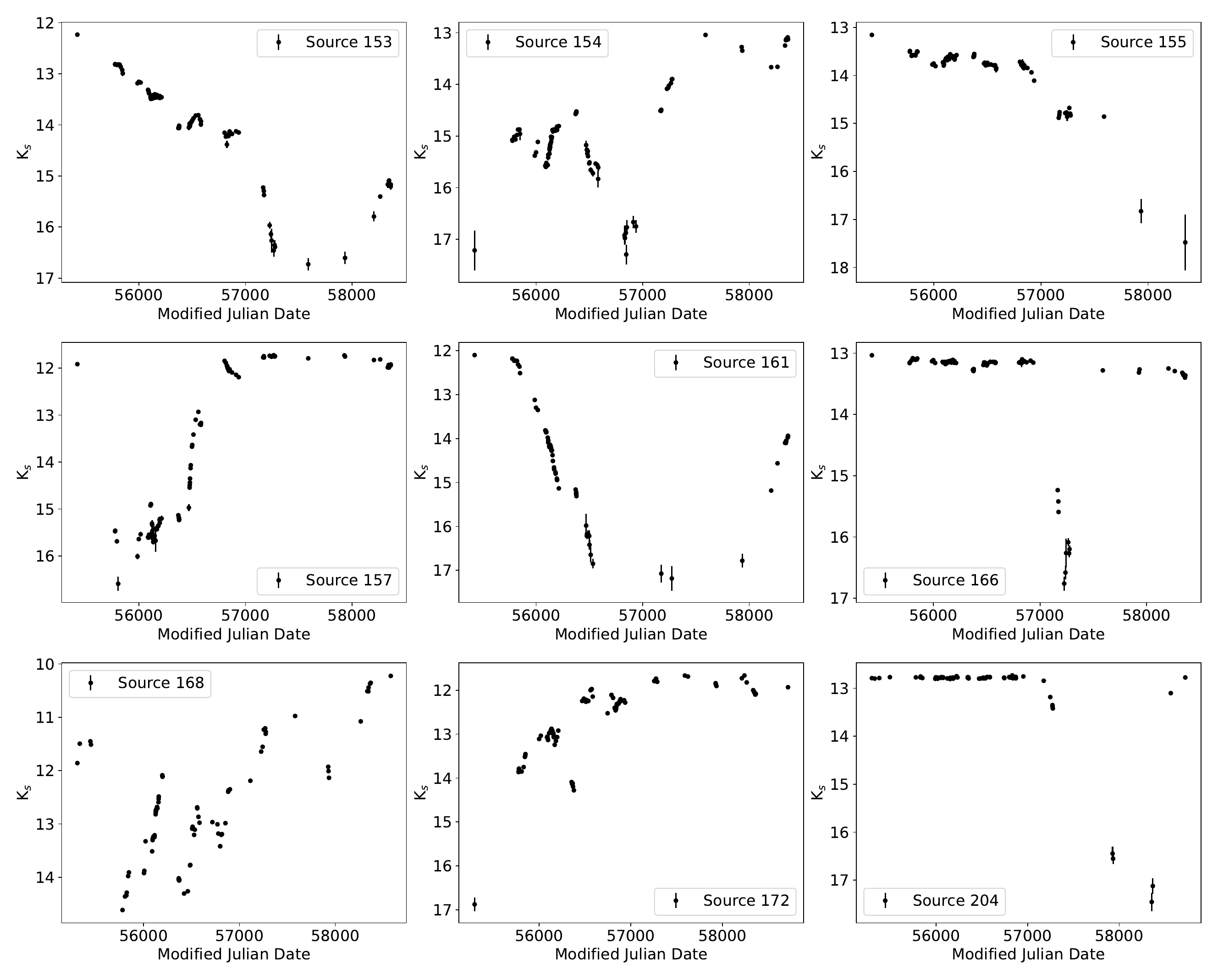}    
    \caption{Light curves of the dipping giants.}
    \label{fig:dipgiant9lc}
\end{figure*}
\renewcommand{\thefigure}{\arabic{figure}}

A cross match to 2MASS PSC and the United Kingdom Infrared Deep Sky Survey Galactic Plane Survey \citep[UKIDSS][]{lawrence07, lucas08} finds that 18/21 sources were
detected in a bright state by one of the two surveys some years prior to the VVV survey (i.e. close to or brighter than the brightest VVV $K_s$ measurements). This helps to verify that we are indeed observing dipping 
events rather than photometric outbursts. N.B. non-detection in 2MASS PSC is sometimes due to source confusion in this region, which also increases scatter in the photometry of detected sources. 2MASS PSC
typically reports magnitudes slightly brighter than either UKIDSS or the brightest VVV measurement for the same source in this set, most likely due to blending. Only 16/21 sources lie in the UKIDSS footprint but 
14/16 were in a bright state in 2006 or 2007.

\begin{figure}
        \includegraphics[width=0.48\textwidth]{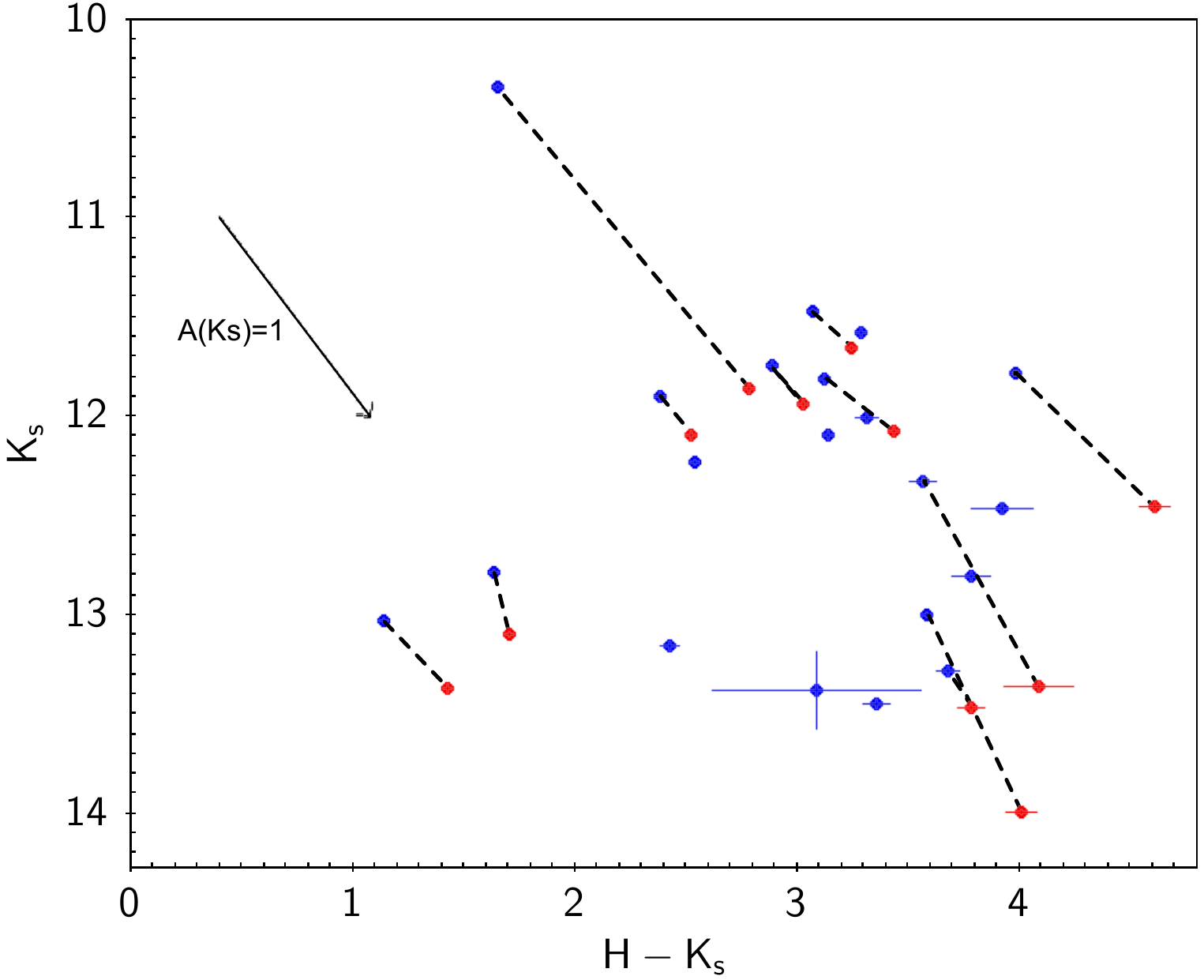}    
    \caption{$K_s$ vs. $H-K_s$ colour-magnitude diagram for 20 of the dipping giants. Blue points correspond to the brightest available epoch of multi-filter VVV data. Red points show the 
    faintest available VVV epoch, plotted for 11 sources. Elements of a pair are connected by dashed lines. The changes are roughly parallel to the extinction vector.}
    \label{fig:dipgiant_HK}
\end{figure}

GLK present near infrared spectra of seven of the dipping giants (sources 130, 144, 145, 149, 154, 168, 172), six of which were taken during the bright state after the dip had ended. All of these are 
late-type giant spectra suffering high infrared extinction, such that high quality data were obtained only at $\lambda>2$$~\mu$m. The initial analysis in that work finds effective temperatures from 
$3100$~K$ < T_{eff} < 4100$~K and suggests that these are O-rich stars: there was no sign of C-rich features such as the 1.77~$\mu$m C$_2$ absorption band, with the caveat that the spectra 
are of poorer quality in the $H$ bandpass.
The high foreground extinction towards the 21 sources is further evidenced by their very red ($H-K_s$) colours. In Figure \ref{fig:dipgiant_HK} blue points are plotted for 20 sources, corresponding to data taken near 
maximum brightness within the VIRAC2 time series. The 11 red points correspond to data taken at the faintest available multi-filter detection in each light curve, the remaining sources lacking a second detection in $H$.
A reddening vector is also plotted, based on the \citet{wang19} extinction law, showing that the colour-magnitude changes are approximately consistent with changes in extinction due to sub-micron-sized dust grains.

Using the \citet{wang19} extinction law, the extinction towards these 21 sources outside the dips is 
typically 4 to 5 mag in $K_s$, based on observed colours $3 < H$-$K_s < 4$ (see Figure \ref{fig:dipgiant_HK}) and intrinsic $H$-$K_s$ $\approx$ 0.2 to 0.4 for post-main sequence stars at these temperatures, according to the
PARSEC-COLIBRI isochrones \citep{chen15, marigo17}. This is supported by fitting the dereddened spectra to comparison giant stars from the Xshooter Spectral Library \citep{verro22}, showing no obvious contribution of emission
by circumstellar dust to the red continuum (to be discussed further in a future work).  Observed magnitudes are $K_s \approx$~11 to 13 mag outside the dips. After correcting for extinction and adopting an 8~kpc distance 
for the Nuclear Disc, the distance modulus is 14.5 and the absolute magnitudes are in the approximate range $-8<M_{K_s} <-6$. From inspection of the PARSEC-COLIBRI isochrones, the range of $M_{K_s}$ and $T_{eff}$ is 
consistent with an AGB interpretation.

\begin{figure*}
        \includegraphics[width=0.48\textwidth]{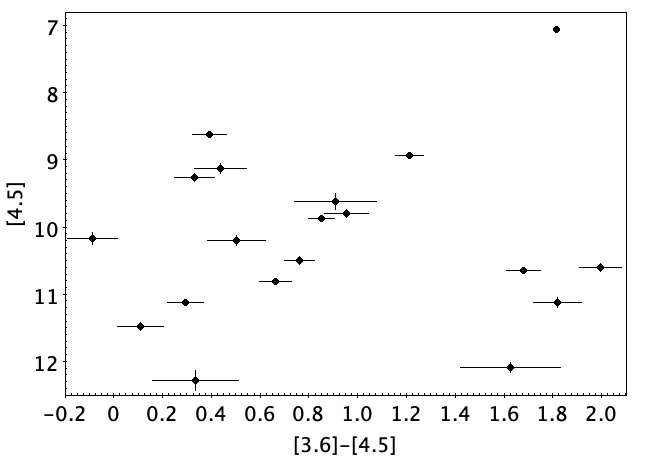}    
        \includegraphics[width=0.48\textwidth]{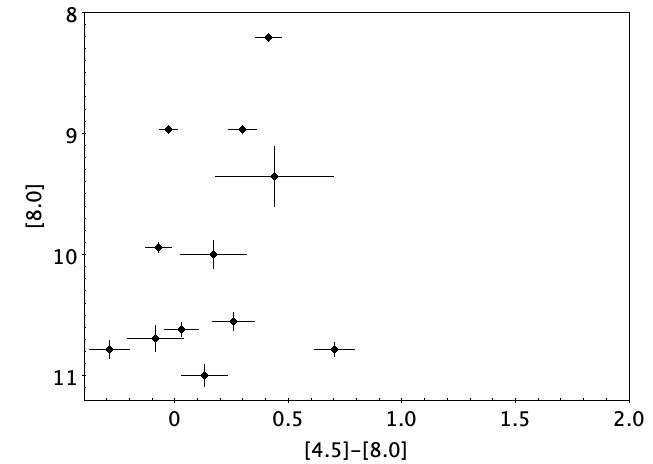}  
    \caption{{\it Spitzer}/IRAC colour-magnitude diagrams for the dipping giants. ({\it left panel}): [3.6]$-$[4.5] vs. [4.5] plot for 19 sources. {(\it right panel}): [4.5]$-$[8.0] vs. [8.0] plot for 12 sources, with the
    abscissa scaled to emphasise the relatively blue nature of the population in this colour, which is almost independent of extinction. Many sources have [3.6]-[4.5]~$<1$, also indicating little emission by hot dust at the 
    time of the GLIMPSE-II survey in 2005 to 2006, but hot dust is required to explain some sources with [3.6]$-$[4.5]~$>1$.}
    \label{fig:dipgiant_Spitcols}
\end{figure*}

Mid-IR colours from {\it Spitzer}/GLIMPSE-II and GLIMPSE-3D, where available, are plotted in Figure \ref{fig:dipgiant_Spitcols}. These were measured prior to the dips observed by VVV. The very high foreground extinction 
prevents us from distinguishing O-rich and C-rich AGB stars via the colours but we see that only a minority (6/19) of the sources in the left panel have [3.6]~$-$~[4.5]$>1$. The very red colours of this smaller subset suggest
the presence of hot dust close to the photosphere. However, no source has [4.5]~$-$~[8.0]$>1$ (the threshold for extremely dusty AGB stars and other intrinsically red sources listed in \citealt{robitaille08}). 
None are detected in the {\it Spitzer} MIPSGAL survey or the far-infrared {\it Herschel} Hi-Gal survey \citep{molinari10, elia17}, and there is only one detection at 22~$\mu$m in WISE (source 91, $W$4=1.2).
This indicates that there is no large mass of cooler dust in almost all cases. There is little other useful data from the WISE satellite and the associated unTimely light curves, owing to severe source confusion. 

The VIRAC2 proper motions in Galactic coordinates have a cluster around $\mu_l = -7$~mas/yr, $\mu_b= -3$~mas/yr (not shown). The $V$ and $W$ components of Galactic velocity can be closely approximated by 
adopting the Galactic centre sky coordinates for these sources so that the velocity components could be determined simply from the proper motions, assuming $d=8$~kpc. 
The $V$ component (parallel to Galactic rotation in the solar neighbourhood) is then usually between 200 and 300 km/s, suggesting locations 
on the near side of the Nuclear Disc and consistent with typical Nuclear Disc stellar motions measured by VIRAC2 \citep{sormani22} for the Nuclear Disc stars in the spectroscopic sample of \citet{fritz21}. 
However, some sources have highly discrepant velocities. On investigation, this appears to be due to a rare type of astrometric
error that affects only highly variable sources that are in very crowded fields, or otherwise blended with adjacent stars. A comparison with Hubble Space Telescope astrometry has confirmed that the uncertainties on VIRAC2 
proper motion and parallax measurements are usually correct even in crowded fields \citep{luna23}, partly because the error calculations include a comparison with Gaia DR3 rather than internal errors alone
(Smith et al., in prep). However, the effect of large changes in source flux is to change the number of stars in a blend that are detected by the \textsc{dophot} software, which can cause a systematic shift in 
source coordinates in sources like dipping giants that are faint for a significant fraction of the light curve.

Variability due to changing extinction by circumstellar matter is often seen in pulsating dusty Mira variables. In these stars, mass loss is attributed to radiative pressure on dust grains that form in the outer layers of the stellar
atmosphere (and above the photosphere) and drag the gas along with them. Pulsations and convection are thought to assist the process by providing an initial upward motion and causing shocks to form, creating high density 
regions that enhance dust formation \citep{hofner18}. This process is inherently three-dimensional, a fact confirmed by recent simulations \citep{freytag23} and several high resolution observations at infrared and radio 
wavelengths \citep[see e.g.][]{adam19, matthews23}. Consequently, ejection of ``dust puffs" can be directional, providing an explanation for the deep aperiodic dips occasionally seen in Mira variables that lack
the very thick dust shell and the corresponding very red colours seen in OH/IR stars (the reddest Mira variables). E.g. \citet{whitelock06} note that such dips are seen in C-rich Miras, appearing observationally similar to the 
R Cor Bor events seen in H-deficient C stars, where random fading events of up to 7 mag in the visible ($\sim$0.7 mag in $K_s$) are attributed to ejection of puffs of C-rich dust at random times and in random directions 
\citep{clayton96,clayton12}.
They went on to note that dust puffs had not been seen in solitary O-rich Miras but occur commonly in O-rich symbiotic Miras \citep{whitelock87}, thought to be caused by interaction between the binary components. 
While the spectra presented in GLK do not show the emission lines that would be associated with accretion on to a compact companion, interaction with a main sequence companion is certainly a 
possibility to consider. 

In Appendix \ref{sec:s163} we highlight two Mira variables with exceptionally deep dips, also projected in the Nuclear Disc. Deep dips in infrared light curves due to circumstellar matter in non-pulsating giant stars have been 
seen much more rarely hitherto. Outside the nuclear disc, perhaps the best examples are found in \citet{whitelock06}, where the carbon-rich star IRAS~09164-5349 was discussed. That system showed a slow fading trend of 
at least 2~mag in $K$ after previously showing only very low amplitude semi-regular variations. Two other carbon-rich giants, IRAS~10136-5743 and IRAS~16406-1406, showed dips of 1 to 1.7 mag in $K$ and no obvious
pulsations, though the light curves were sparsely sampled. In a single 11.5$\times$11.5 arcmin field close to the Galactic centre, aperiodic dips with $\sim$1~mag depth in $K_s$ were reported in several apparently 
non-pulsating giant stars in the VVV-based study of \citet{molina19}, along with two cases of $\sim$2.5~mag dips (the stars NV232 and NV261). One of the stars with a $\sim$1~mag dip, NV062, was spectroscopically 
classified as M4-5 III type (though the 2.0-2.4~$\mu$m spectrum did not unambiguously distinguish O-rich and C-rich giants).
The prior study of the Galactic centre region by \citet{matsunaga09} listed a large number of variable giant stars for which a period could not be determined, though these were not discussed. In fact, two of the
21 dipping giants in our sample, sources 137 and 146, are listed as variable stars with no known period in Table 5 of \citet{matsunaga09}. These are presently listed in SIMBAD as [MKN2009]~37 and  
[MKN2009]~1313, respectively, arguably misclassified there as LPVs. 

Interestingly, [MKN2009]~37 shows a 1.3 mag rising trend from 2002 to 2008 in the published light curve, at which time it reached the flat plateau at $K_s=11.5$ seen in the VIRAC2 light curve from 2010 to 2014, 
prior to the deep dip that followed (see source 137 in Figure \ref{fig:dipgiant12lc}). In this star then, we have evidence for an earlier dip, where 
the minimum occurred at least 17 years prior to the minimum of the dip seen by VVV. By contrast, [MKN2009]~1313 showed only small variations in $K_s$ in the 2002 to 2008 interval.

The key new feature of the present sample is the clustering within the Nuclear Disc. The metallicity distribution in this region is largely in the super-solar range $0<$~[Fe/H]~$<1$ \citep{fritz21}, exceeding
even the nuclear bulge of the Milky Way \citep{queiroz21} which has a broader distribution. In view of this difference from other parts of the Milky Way, it seems plausible that high metallicity is aiding the mass loss e.g. 
by strongly enhancing
production of refractory dust species. Increased dust opacity in the photosphere might also lead to more energetic convection. The analysis of the seven dipping giant spectra in GLK, based on CO and NaI
indices, appears to favour high metallicity. Additional observations, searches for less extreme dipping events in the VIRAC2 database, 
and 3D modelling of mass loss in metal-rich atmospheres will be needed to better understand this population and distinguish the various possibilities. 

An explanation related to high metallicity would indicate that these stars are O-rich, since dust production in O-rich stars is metallicity dependent, whereas dust production in C-rich stars is due to nuclear burning 
and dredge-ups from the interior, rather than the initial metallicity of the star \citep[see e.g.][]{bladh19}.

To finish this topic, we should mention the possibility of periodic dips due to eclipses by either (i) a circumsecondary disc around a very low mass companion on a wide orbit, or (ii) matter spread around the orbit of 
such a companion. The high amplitude samples presented herein (Table \ref{tab:table1a} and Appendix C) include three late type giant stars displaying a deep dip attributed to a circumsecondary disc: VVV-WIT-08 (=source 199 
in Table 1a), VVV-WIT-10 (=DR4\_v16 in Appendix C) and VVV-WIT-11 (=source 101 in Table \ref{tab:table1a}), see \citet{smith21}. Two of those three sources show a single symmetric dip, whilst the third (VVV-WIT-11) was undetected 
for a year during the dip and is regarded as a candidate of that type. None of the three are located in the Nuclear Disc. The light curves of the 21 dipping giants discussed above differ from those three sources in that the dips 
are not symmetric and in most cases there is significant variability outside the deep dip. Consequently, an eclipse by a circumsecondary disc can be ruled out. We also note that periodic dips (a ``long secondary period") are seen 
in up to a third of LPVs \citep{wood99, soszynski13, pawlak23}, these having a longer timescale than the pulsation period. These events are due to either option (i) or option (ii) above \citep{wood99, soszynski21}, 
the latter option being the favoured explanation 
for asymmetric and longer lasting dips. Option (ii) is a possible explanation for 21 dipping giant systems. However, the range of periods seen in Miras with a long secondary period is usually only 200~d to 1600~d 
\citep[see Figure 2 of][]{soszynski21}, whereas the 21 dipping giant systems would have to repeat on a longer timescale, if they are periodic. A few Mira variables are known with very long secondary periods, see
Appendix \ref{sec:s163} so this explanation cannot be ruled out entirely. However, long secondary periods are found in Miras all over the Milky Way so it seems more likely that the explanation is related to the high metallicity 
of the Nuclear Disc, with no obvious requirement to invoke a companion object as yet.

\section{Other Variable Sources}
\label{others}

\subsection{Long Period Variables}
\label{LPVstub}
The 35 sources classified as LPVs are dusty Mira variables with extremely high amplitudes, $\sim$5 mag from peak to trough in $K_s$ in two cases. Separately, 
sources 163 and 147 are very interesting in the context of the Dipping Giants discussed above, since both show a very deep, apparently aperiodic dip and both are projected in the Nuclear Disc 
Other LPVs with dips that are slightly less deep are also of interest in this context. See Appendix \ref{sec:s163} for the light curves and further discussion.

\subsection{Novae and other transients}
\label{novastub}
In Table \ref{tab:table1a}, 70 sources are classified as CVs. Of these, 45 have been previously classified as classical novae, recurrent novae or 
nova candidates and one other, source 219 (aka ASASSN-17fm) is a dwarf nova. Individual references and identifications are given in Table \ref{tab:table1a}. 
In Appendix \ref{sec:CVs} we briefly discuss four of the 24 new discoveries with unusual light curve features, such as the oscillations and cusps occasionally
seen at optical wavelengths \citep{strope10} but to our knowledge not previously seen in the infrared.

\subsection{Unusual sources}
\label{unusualstub}
Of the 222 sources in Table \ref{tab:table1a}, there are 10 that have fairly well sampled light curves but do not fit into the categories discussed hitherto. 
Two of these were known as unusual sources independently of the VVV survey (V4334~Sgr (= Sakurai's star) and the X-ray source MAXI~J1348-630). The nature of two 
others (VVV-WIT-08 and VVV-WIT-10) is thought to be at least partly understood \citep{smith21} in terms of an eclipse of a giant star by a circumsecondary disc. The nature of 
the remaining six is unclear at present. Each of the 10 is discussed briefly in Appendix \ref{unusual}.

\section{Summary and Conclusions}
\label{conc}

A thorough search of the VVV/VIRAC2 survey database covering 562~deg$^2$ of the Galactic bulge and adjacent Galactic disc has yielded a sample of 222 variable
stars and transients with $\Delta K_s \ge 4$ mag, most of which are new discoveries. Novae and nova candidates make up the largest proportion but YSOs
are slightly more numerous in the disc fields at $295^{\circ} < l < 350^{\circ}$ (see Figure \ref{fig:galdistrib}). Whilst detection of eruptive YSOs was the principal
motivation for this work, this was the first panoramic infrared search for variable sources of all types to cover the most obscured regions of the Milky Way, so it is
not entirely surprising that it resulted in detection of an unexpected new population: the aperiodic late-type dipping giant stars in the Nuclear Disc. 
This discovery comes soon after the recent detection of the population of post-Red Giant Branch stars \citep{kamath16, oudmaijer22}, thought to be 
products of post-common envelope evolution.

Our conclusions regarding YSOs are as follows.
\begin{itemize}
\item Of the 40 sources classified as YSOs, 32 are eruptive variables presumed to be undergoing an episodic accretion event. We describe most of these as ``classic"
outbursts with a single long-lasting event, the majority of which are spectroscopically confirmed to have FUor-type or EXor-type spectra in the companion paper by 
GLK. 

\item The rise time of these long-lasting accretion-driven outbursts as measured in $K_s$ is
most often between 600 and 1000~d, notably slower than the 6 to 12 month timescale reported in the optical for the small number of historical FUor events where the 
timescale is known. The typical outburst duration is at least 5~yr, rather longer than many lower amplitude events (see the companion paper by Contreras Pe\~{n}a et al. submitted for a
sample with a broad range of amplitudes.) There is considerable diversity in the light curves of the classic outbursts, in terms of rise time, rate of decline post-maximum,
intra-year variability and overall morphology. The spread of rise times might be explained by the onset of the accretion disc instability occurring at different radial locations, or
different disc viscosity parameters \citep{liu22}. However, the overall diversity may indicate that more than one disc instability mechanism can trigger long-lasting outbursts.

\item The form of the classic outburst light curves during the rise is usually described by a slow start that then accelerates to a rapid linear rise in brightness that stops rather
quickly when the bright state is reached. A minority of classic outbursts have a more symmetric ``S-shaped" rise, with a slow start and a slow finish.

\item A small proportion of the eruptive YSOs have irregular light curves, showing strong variability on both intra-year and inter year timescales and typically at least two
clear maxima within the 9.5~yr VIRAC2 time series. These somewhat resemble the Multiple Timescale Variables detected in \citet{guo20, guo21}. Most of these sources 
presently lack spectroscopic follow up.

\item Periodic outbursts are not seen at these high amplitudes, despite making up a significant proportion of outbursts up to 4~mag in amplitude \citep{guo22}. However,
one YSO, source 148, shows periodic 2~mag outbursts atop a 3~mag classic outburst light curve (discussed further in GLK).

\item Eight YSOs with deep, long-duration extinction events (dippers) are seen wherein the source is dimmer and usually redder for a year or more, attributable to inner disc 
structure \citep[e.g.][]{bouvier13, rice15}.

\item Notable amongst the dipping YSOs is source 207, a candidate KH~15D-like periodic system in the Lagoon Nebula Cluster (NGC~6530) that showed two almost equal peaks in 
flux within each 59~d period, attributed to alternating extinction of the two components of a binary YSO by a circumbinary disc \citep{chiang04, arulanantham16}.
\end{itemize}

The aperiodic dipping giants will be an interesting topic for future study. The main things to note concerning these are as follows.
\begin{itemize}
\item These sources are strongly clustered in the Nuclear Disc, evidenced by the projection of 18/21 dipping giant candidates in that region and their high infrared extinction even outside 
the dipping event, Of the 21 candidates, 7 are spectroscopically confirmed as late type giants in GLK. This informed the classification of the 21 candidates, indicating
that the remaining 14 that lack spectra at present are unlikely to be YSOs. 

\item The light curves of the dips are diverse but they are generally asymmetric (hence not caused by a circumsecondary disc) and last for at least a year, more often a few years or longer. 
Their ($H-K_s$) colours approximately follow the interstellar reddening law, indicating that the variable extinction is due to small dust grains. Some variability is almost always seen outside the 
main dip. Only a minority of the dipping giants have red mid-infrared colours in the {\it Spitzer}/GLIMPSE-II survey, indicating that there is no pre-existing large mass of warm dust in most cases.
This tends to suggest a scenario of dust puffs along the line of sight.

\item In view of their location and the uniquely high metallicity (within the Milky Way) of many stars in the Nuclear Disc, we suggest that the aperiodic dips are caused in some way
by super-solar metallicity, e.g. by an increase the production of highly refractory dust species. The initial spectroscopic analysis presented in GLK appears to support this, and
the spectra, effective temperatures and luminosities appear to be consistent with an AGB population that is probably O-rich. The radiatively driven mass loss in AGB stars proceeds via 
radiative pressure on dust but is usually linked to pulsation \citep{hofner18}. However, it depends on luminosity, chemical composition and the 3-D process of dust production \citep{dellagli15}, 
which is influenced by convection and shocks \citep{freytag23}. Further observational and theoretical work are clearly needed to understand this population, including a search for lower amplitude 
examples of the phenomenon.
\end{itemize}

Other discoveries to note in the sample are as follows.
\begin{itemize}
\item The set of 35 LPVs includes several dusty Mira variables with deep, long-lasting dips or fading trends. All three of the LPVs that are projected in the Nuclear Disc display this behaviour,
two of these having the deepest dips in the set (3.5 to 4 mag in $K_s$) which far exceed the amplitude of the pulsations in these stars.

\item The set of 70 novae and nova candidates includes 24 new candidates. Some of the light curves display unusual features within the overall decline, such as a cusp or oscillations, to 
our knowledge previously reported only in the optical waveband \citep{strope10}.

\item We identify 10 unusual variable sources, several of which are not yet understood and warrant further investigation.
\end{itemize}

\section*{Acknowledgements}

We thank the referee for their work to improve this paper.
We also thank Mike Kuhn for pointing us to the recent discovery of post-RGB stars and Chiaki Kobayashi for pointing out that a dependence of dust production on metallicity would indicate that the dipping 
giant stars are O-rich rather than C-rich. We gratefully acknowledge data from the ESO Public Survey programs 179.B-2002 and 198.B-2004 taken with the VISTA telescope, and products from the Cambridge 
Astronomical Survey Unit (CASU) and the Wide Field Astronomy Unit at the Royal Observatory, Edinburgh. This research has made use of the NASA/IPAC 
Infrared Science Archive, which is funded by the National Aeronautics and Space Administration and operated by the California Institute of Technology.
We also acknowledge use of NASA's Astrophysics Data System Bibliographic Services and the SIMBAD database operated at CDS, Strasbourg, France. 
We also made use of the VizieR catalogue access tool, CDS, Strasbourg, France (DOI : 10.26093/cds/vizier). The original description of the VizieR service was published in 
2000, A\&AS 143, 23. PWL and ZG acknowledge support by STFC Consolidated Grant ST/R00905/1 and CM was supported by an STFC studentship funded by grant ST/S505419/1.
DM gratefully acknowledges support from the ANID BASAL projects ACE210002 and FB210003, from Fondecyt Project No. 1220724, and from CNPq Brasil Project 350104/2022-0.,  the BASAL 
Center for Astrophysics and Associated Technologies (CATA) through grant AFB170002, and the Ministry for the Economy, Development and Tourism, Programa Iniciativa Cientifica Milenio grant 
IC120009, awarded to the Millennium Institute of Astrophysics (MAS). Support for MC is provided by ANIDÕs Millennium Science Initiative through grant ICN12\_009, awarded to the Millennium Institute of 
Astrophysics (MAS); by ANID/FONDECYT Regular grant 1231637; and by ANID's Basal grant FB210003. RKS acknowledges support from CNPq/Brazil through projects 308298/2022-5, 350104/2022-0
and and 421034/2023-8. ACG has been supported by grants PRIN-MUR 2022 20228JPA3A ``The path to star and planet formation in the JWST era (PATH)" and INAF-GoG 2022 ``NIR-dark Accretion 
Outbursts in Massive Young stellar objects (NAOMY)".\\

\noindent{\bf Data Availability}\\
The VVV and VVVx images are publicly available in the VISTA Science Archive (VSA) (\url{vsa.roe.ac.uk}) and the ESO archive \url{http://archive.eso.org/cms.html}. VSA facilitates generation of
cut-out images around a specified coordinate so we have not included images in this work. The final version of the VIRAC2 database, representing a minor improvement on VIRAC2-$\beta$, is currently 
being prepared for publication and upload to ESO archive, expected to occur within the next few months. UKIDSS data are available at the WFCAM Science Archive \url{wsa.roe.ac.uk}.
The WISE, {\it Spitzer} and 2MASS datasets are publicly available, see the Acknowledgements above.



\bibliographystyle{mnras}
\bibliography{v195} 



\appendix

\section{Searches}
\label{sec:searches}

\subsection{VIRAC2$-\alpha$ and VIRAC2-$\beta$ differences}

As noted in the main text VIRAC2$-\alpha$ light curves have data from 2010--2018, whereas VIRAC2-$\beta$ light curves have data from 2010--2019.
The absolute photometric calibration of VIRAC2-$\beta$ is anchored to the 2MASS Catalogue of Point Sources (PSC) \citep{skrutskie06} in a region of the lower bulge having low extinction and negligible 
source confusion, thereby circumventing absolute calibration issues in crowded star fields of the inner bulge \citep{hajdu20}. This calibration was propagated across the VVV survey area via the field overlaps. VIRAC2-$\alpha$ was calibrated using the VVV \textsc{dophot}-based catalogue of \citet{alonso18}, which provides ZYJHKs photometry averaged over two epochs, ultimately calibrated to 2MASS PSC
on a per-field basis. The astrometric calibration of both versions of VIRAC2 is anchored to the absolute reference frame of the {\it Gaia} second data release (DR2).

VIRAC2-$\beta$ benefits from a more sophisticated detection-matching scheme for each source than VIRAC2-$\alpha$, allowing for proper motion and parallactic motion (Smith et al., in prep). This led to a modest but useful increase in the number of matched detections for a substantial minority of sources, within the 2010--2018 time interval. However, VIRAC2-$\beta$ contains a small fraction of cases where two adjacent stars were incorrectly matched as a single detection with high proper motion and/or large parallax. These occasional mismatches required some additional effort to remove false positive detections whilst making the searches as complete as possible.

\subsection{VIRAC2 searches}
\label{sec:searchesA1}

\begin{figure*}
        \includegraphics[width=0.49\textwidth]{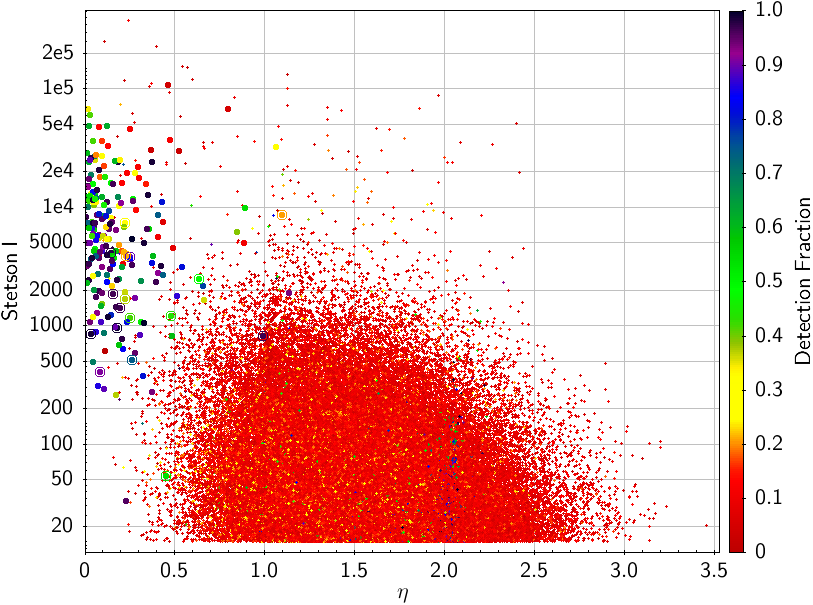}
        \includegraphics[width=0.49\textwidth]{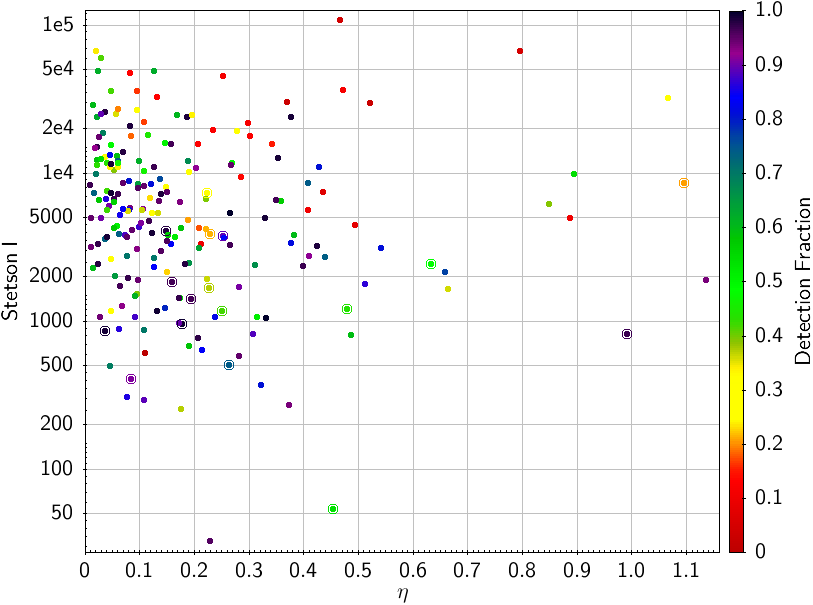}       
    \caption{Stetson $I$ vs. von Neumann $\eta$ plots for candidate variable sources (small red $+$ symbols) and bona fide variables (filled circles). The points are colour coded by 
    a parameter that represents (approximately) the fraction of images of the field in which the source was detected. The plotted data distribution drives most of the   
    logic of our search methods. The left panel includes numerous false positive candidates, whereas the right panel shows more clearly the distribution of the bona fide variables
    in our sample VIRAC2 sample of 222 $\Delta K>4$~mag sources.
    Data are taken from VIRAC2-$\alpha$ except for 17 sources discovered using VIRAC2-$\beta$ which are highlighted with a ring around the filled circle. These 17 tend to have slightly 
    lower Stetson $I$ values, see Appendix A.}
    \label{fig:search}
\end{figure*}

The VIRAC2 databases contain a number of light curve parameters that can help to distinguish bona fide variable stars and transients from false positives. The most
productive searches were based on a combination of the well-known Stetson $I$ index \citep{welch93} and the slightly less well-known von Neumann $\eta$ parameter 
\citep{neumann41}. The Stetson $I$ index, when applied to times series measurements in a single filter, has a large value if pairs of magnitude measurements taken very 
close together in time differ from the light curve mean in the same direction, by amounts that are large in comparison to the measurement error. Defining $m_{a,i}$, $m_{b,i}$ 
and $\sigma_{a,i}$, $\sigma_{b,i}$ as the magnitude values and uncertainties in the $i$th pair of measurements in the light curve, and $n$ as the number of pairs, the Stetson 
$I$ index is computed as:

\begin{equation}
       I = \sqrt{ \frac{1}{n(n-1)}} \Sigma_{i=1}^{n}\delta m_{a,i} \delta m_{b,i}
	\label{eq:stetson}
\end{equation}
where $\delta m_{a,i} =  \frac{m_{a,i} - \bar{m}}{\sigma_{a,i}}$, $\delta m_{b,i} =  \frac{m_{b,i} - \bar{m}}{\sigma_{b,i}}$ and $\bar{m}$ is the mean magnitude for the source. In VIRAC2, pairs of measurements are required to be separated by $<1$ hour in order to contribute to the sum, though the interval is more typically a few minutes for consecutive pawprint stack images within a six point tile. If there were more than two measurements taken within an hour, pairs of measurements were formed by first selecting the two that were closest together in time, then forming any additional pairs by successively selecting the two remaining measurements that are closest in time. Most locations within the VVV survey area have two or more pawprint stack observations taken within an hour during the course of a six point tile observation (or during observation of an adjacent, slightly overlapping tile). However, a few highly variable sources found in our searches have relatively low Stetson $I$ values (see Figure \ref{fig:search}) because the number of suitable pairs of observations in the light curves is small, e.g. due to location at the edges of the group of four tiles that made up the observing block.

The von Neumann $\eta$ parameter is conceptually somewhat similar to the Stetson index in that it has an outlying (small) value if the consecutive values in a light curve (whether close together in time or not) typically differ by less than the variance of the whole light curve. The $\eta$ parameter is defined as:

\begin{equation}
       \eta = \frac{1}{(N-1)\sigma_{lc}^2}\Sigma_{i=1}^{N-1}(m_{i+1}-m_i)^2
	\label{eq:eta}
\end{equation}
where $m_i$ are the individual magnitudes, $N$ is the number of measurements in the light curve and $\sigma_{lc}$ is the standard deviation of all the magnitudes in the light curve. 
Unlike the Stetson index, the $\eta$ parameter is robust against the problem of sources that have only a single pawprint stack image at each observing epoch so it helps
to pick out bona fide variable stars in the lower left quadrant of Figure \ref{fig:search}. However, the Stetson index performs better than the $\eta$ parameter for detection of poorly 
sampled transient events, e.g. some microlensing events, which tend to lie toward the upper right of Figure \ref{fig:search} (right panel).\footnote{The parameter 1/$\eta$ 
is sometimes used so that larger values of the parameter correspond to sources that are more likely to be variable. In this case, a plot of $\eta$ more clearly illustrates the value of 
the threshold at $\eta=0.5$ that we used in several of our searches.} 

A simplistic search of the VIRAC2 databases for highly variable sources using only the value of $\Delta K_s$ inevitably finds numerous false-positives where $\Delta K_s$ is inflated 
by bad data at one or more epochs. The databases also include numerous spurious sources located in 
the wings of the image profiles of very bright stars, though VIRAC2-$\beta$ mitigates this somewhat by including only sources having detections in at least 10 pawprint stacks. (This
problem is partly due to the choice of \textsc{dophot} parameters in VIRAC2 because it was decided to opt for completeness in the database rather than exclude sources within the seeing halo of saturated stars). These spurious sources (including those relating to diffraction spikes and ghost images) are usually very obvious on inspection of the images. For example, ghost images of bright stars where the image falls just off the edge of an array typically appear as a streak or "jet" (see section 13.4 of \citealt{sutherland15}), though occasionally they resemble a point source. We also noticed that bright Long Period Variables (LPVs, which are often saturated in VVV), were a common source of false positives because \textsc{dophot} detected time-variable spurious sources in the image halo, the diffraction spikes or ghost images. The light curves of these false positives have the appearance of a very noisy sinusoid, the detections associated with each datum often having large values of the \textsc{dophot} $Chi$ parameter (indicating that the image profile was not well fit by a point source) and large values of the VIRAC2 $ast\_res\_chisq$ parameter, which quantifies the significance of the residual to the VIRAC2 five parameter astrometric fit as an astrometric $\chi^2$.

A selection of $\sim$122 000 VIRAC2-$\alpha$ false positive candidates having Stetson $I$>15, $\Delta K_s>4$~mag and detections spanning at least a three year time baseline 
is plotted in the left panel of Figure \ref{fig:search}. The colour coding illustrates the fact that most false positive sources are detected in only a small fraction of the $K_s$ 
pawprint stack images 
covering their location. It is also obvious that bona fide variable sources are concentrated at the upper left of the panels in Figure \ref{fig:search}, typically with Stetson $I$>500 and $\eta$<0.5. Since it is impractical to visually inspect time series images of very large numbers of sources, the Stetson $I$, $\eta$ and detection fraction parameters, supplemented by a few others, were combined in various ways to slice the large pool of candidates in all promising regions of the parameter space and yield an essentially complete list of all bona fide variable stars and transient sources in VIRAC2 with $\Delta K_s>4$~mag. 

The precise details of our numerous searches of VIRAC2-$\alpha$ and then VIRAC2-$\beta$ are given in the following sub-sections. Any VVV sources that were missed are either very poorly sampled or have variability that fell outside the dynamic range of the \textsc{dophot}-based photometry in the database. A large proportion of VVV transients are saturated during part of their light curves, requiring additional processing for this work, so a search using saturation-corrected photometry would probably yield some additional discoveries. E.g., the bright transient VVV-WIT-05, discovered by \citet{saito16b} at $K_s \sim 4$, is not included in VIRAC2-$\beta$.

Causes of false positives that were encountered during our searches were, in decreasing order of frequency:
\begin{enumerate}
\item bright stars, including bright LPVs (see above)
\item stellar blends
\item asteroid $+$ star blends
\item real lower amplitude variable stars having exaggerated amplitude due to one or two bad data points
\item a bad image showing gross electronic defects or loss of guiding
\item blends involving a high proper motion star
\item defective regions at the edges of the arrays
\item small image defects affecting only a few pixels. 
\end{enumerate}
We attribute the last category to cosmic rays or bad pixels. Regarding category (iv) we retained sources in Table \ref{tab:table1a} if their amplitude in the one-day binned light curves remained close to or larger than 4~mag,
computed after rejection of bad data (see Appendix \ref{sec:lcs} and 3$\sigma$ rejection of outliers during the binning. The logic here is that VVV is sparsely sampled and will therefore typically not measure
the full amplitude of a variable source, so retaining borderline cases helps to give a more complete view. Here ``close to" means $\Delta Ks > 3.6$~mag in the case of transients and stars that dropped below the detection
threshold when faint whilst in other cases we required that the amplitude can exceed 4~mag within the 1$\sigma$ errors of the photometry.

The asteroid$+$star blends (category iii) occur in pawprint stack images taken at a single epoch in a light curve, corresponding to the spatial coincidence of a fairly bright main belt asteroid and a fainter star in the Galactic bulge, typically within a few degrees of the intersection of the ecliptic and the Galactic equator. These blends were often not detectable via inspection of the images because the spatial match was within $0.1\arcsec$. They were instead identified using the \textsc{Minor Planet Checker} software available
at the Minor Planet Centre of the International Astronomical Union.

Fortunately, the precision astrometry of VIRAC2 greatly facilitates identification of false positive candidates in all the above categories because any type of image defect, blend or spurious detection due to a bright star is often associated with large values of the $ast\_res\_chisq$ parameter, corresponding to astrometric shifts large in comparison to the 5--30 milli-arcsecond precision of the individual detections. The VIRAC2 data release paper (Smith et al., in prep) gives examples 
of astrometric elimination of false positives by colour coding each datum in VIRAC2 light curves according to the $ast\_res\_chisq$ value.

\subsection{VIRAC2-$\alpha$ searches}

As described in Appendix \ref{sec:searchesA1}, the most important search parameters were the Stetson $I$ and von Neumann $\eta$ parameters, 
in combination with a parameter approximating to the fraction of $K_s$ images of a given sky coordinate in which a source is detected. In VIRAC2-$\alpha$
the latter parameter is called {\it pp2frac}. To explain this term, we first need to define a ``pawprint set" as a set of VVV$+$VVVX pawprint stack images 
that have the same telescope pointing coordinates (within 30\arcsec). Then we recall that each location in a VVV tile is normally covered by at 
least two pawprint stack images so the great majority of sources have photometry in two or more pawprint sets. For each VIRAC2-$\alpha$ source,
the pawprint sets having detections were ranked according to the number of detections. (Commonly there are detections in only two pawprint sets, with
a similar count in each). {\it pp2frac} is defined as the fraction of images in the second-ranked pawprint set for each source where there is a detection.
For most sources this rather convoluted parameter is similar to the detection fraction in the full sequence of pawprint stack images covering the
location on sky. The reader may be relieved to know that VIRAC2-$\beta$ enabled us to use just such a simple detection fraction in later searches.

Another VIRAC2-$\alpha$ parameter employed in our searches is {\it epoch\_baseline}, the time in years between the first detection and the last detection 
of each source. This parameter was helpful in searches for transients having low values of {\it pp2frac}. 

Below we list the parameters employed in of the VIRAC2-$\alpha$ 
searches in turn. All searches required $\Delta K_s>4$ mag in addition to the other cuts listed. Except where otherwise stated, the initial choice of parameter
values was based on inspection of the parameter distributions with the aim of removing the bulk of false-positive candidates, some of which are illustrated
in Figure 1 (left panel).

\begin{enumerate}
\item Search 1a. $pp2frac > 0.2$ AND $I>15$ AND $n>2$ AND median $K_s >11.25$ AND ($I>1000$ OR $\eta<0.5$). To remind, $n$ is the number of pairs of 
contemporaneous pawprint stack images that have detections and therefore contribute to the calculation of Stetson $I$. The cut on median $K_s$ magnitude was 
designed to limit the number of false positive candidates caused by saturation. This was the most successful search, yielding 248 candidates of which 175 were 
confirmed to be real by visual inspection of the images. Sources found in this search tend to have fairly well-sampled light curves. They lie towards the upper half 
or the left side of Figure 1 (right panel) and are colour-coded in any hue except red. Eight of the false positive candidates were real variable sources found to have
amplitudes $\Delta K_s<4$ mag after removal of a few outlying data points or, in one unusual case, the use of a separate photometric procedure (see Appendix B).

Inspection of the parameter distribution of bona fide variable sources found by this search found that the subset with $\eta<0.5$ AND ($I>1000$ OR $pp2frac>0.35$)
included 167/175 bona fide variable sources and only eight false positives. This refined set of parameters that describe the richest region of parameter space
helped to define Searches 4a and 5a, see below, where the cuts on $pp2frac > 0.2$ and median $K_s >11.25$ were removed.

The threshold median $K_s$ value of 11.25 was estimated from results of the prior VVV DR4 search (see Appendix B) using aperture photometry data. Later 
investigation of the parameter distribution in VIRAC2-$\alpha$ showed that a stricter threshold at median $K_s = 12.0$ would have more reliably removed false 
positive candidates caused by saturation.

\item Search 2a. $pp2frac<0.2$ AND $epoch\_baseline<1$~yr AND $I>15$ AND $n>2$ AND median $K_s>11.25$ AND ($I>1000$ OR $\eta<0.5$). This search was 
one of a few that attempted to detect transient or poorly sampled sources, which are often more difficult to pick out from the larger number of false positive candidates.
Here the $pp2frac$ parameter is relaxed, potentially adding a very large number of false positives detected at only a few epochs, but the $epoch\_baseline<1$~yr cut 
served to limit the number to a manageable level. This search for relatively brief transients yielded 44 candidates, seven of which were found to be real.

\item Search 3a. $pp2frac<0.2$ AND $epoch\_baseline=1$ to 3~yr AND $I>15$ AND $n>2$ AND median $K_s>11.25$ AND ($I>500$ OR $\eta<0.5$). This search for
transients detected over a longer time baseline yielded 235 candidates, 11 of which were found to be real variable sources. Of these 11 sources, 10 have $\eta<0.5$.

\item Search 4a. $pp2frac<0.2$ AND $epoch\_baseline>3$~yr AND $I>1000$ AND $\eta<0.5$ AND median $K_s>11.25$ AND $n>2$. Most real variables detected over
a time baseline longer than 3~yr have  $pp2frac > 0.2$ and would be found in Search 1 so this search aimed to catch any unusual cases, aided by the requirement for 
both a high $I$ value and a low $\eta$ value that was informed by the rich region of parameter space identified in Search 1. This search yielded 41 candidates (after 
removing 2 duplicates in the database with almost identical coordinates) and seven of them were found to be real. Some of these are transients or eruptive YSOs that 
erupted during the later stages of the VVV/VVVX surveys, when the number of observations each year was smaller than in earlier years.

\item Search 5a. $I>15$ AND $n>2$ AND $\eta<0.5$ AND ($I>1000$ OR $pp2frac>0.35$) AND median $K_s<11.25$. This search reversed the mean magnitude cut 
used in Searches 1 to 4 with the aim of finding any real variable sources that are likely to be saturated in part of their light curves by exploring only the rich region
of ($I, \eta, pp2frac$) parameter space identified in Search 1a. This effort was motivated by the discovery of the extremely high amplitude eruptive variable YSO 
WISEA~J142238.82-611553.7 \citep{lucas20b} using time series data from the Wide-field Infrared Survey Explorer 
($WISE$) and {\it Spitzer} satellites. We noted that the source had been missed by Search 1a because it had median $K_s = 11.24$.
Search 5 yielded 22 candidates and three of them were found to be real, one of which is WISEA J142238.82-611553.7.

\item Search 6a. $n \le2 $ AND $Kepochs>4$ AND $\eta<0.2$ OR ($\eta<0.3$ AND $Kepochs>10$). Here $Kepochs$ is the number of pawprint stack images 
having a detection in $K_s$. This search was designed to recover any well sampled variable sources that lack a useful Stetson $I$ index value due to a location at the 
outer boundary of the VVV survey area or on the edge of one of the groups of four tiles that make up a typical VVV/VVVX observing block. (Some variable sources at 
the edge of a group of tiles were recovered in earlier searches employing the Stetson index because adjacent tile groups were sometimes observed consecutively, 
depending on the vagaries of the VISTA observing queue.) The above selection yielded 73 candidates, of which three were confirmed as real.
\end{enumerate}

\subsection{VIRAC2-$\beta$ searches}

Our searches of VIRAC2-$\beta$ were mostly a repeat of the VIRAC2-$\alpha$ searches for sources with $\Delta K_s>4$ mag, with some necessary adjustments. 
These searches aimed to benefit from the additional year of data and the more sophisticated source matching of the newer database. A feature of VIRAC2-$\beta$ is that 
some light curves contain values with very bright or very faint magnitudes, well outside the range of meaningful VVV \textsc{dophot} measurements, even for saturated stars. 
These sources typically have $\eta>1.99$. All the searches listed below therefore included cuts against sources where the brightest magnitude was $K_s<9$ or the faintest 
magnitude was $K_s>85$ or $\eta>1.99$, in order to reduce the number of candidates requiring visual inspection. Cuts against sources with a very large absolute value of 
parallax ($|\varpi| > 500$~mas) or proper motion ($\mu > 500$~mas yr$^{-1}$) were implemented in all searches; a check for bona fide very high proper motion stars not 
already detected in VIRAC1 had already shown that there are none in the database. An additional limitation of VIRAC2-$\beta$ is
that it contains a large number of false positive candidates, caused by mis-matches, in a region of 8 contiguous VVV bulge tiles ($\sim$12~deg$^2$) at $1.6^{\circ}<l<7.5^{\circ}$,
$-3.6^{\circ}<b<-1.5^{\circ}$ where the survey has a much higher sampling cadence than elsewhere. The combination of high cadence and high source density in this region is 
presumed to be responsible. It was necessary to exclude this area from all the VIRAC2-$\beta$ searches and rely on VIRAC2-$\alpha$.

VIRAC2-$\beta$ enables calculation of the fraction of $K_s$ images of a given sky coordinate in which a source is detected via the ratio of the {\it ks\_n\_detections} and
{\it ks\_n\_observations} database parameters. We denote this ratio as $f_{det}$ and it is used in VIRAC2-$\beta$ searches in place of the $pp2frac$ parameter.

\begin{enumerate}
\item Search 1b. $f_{det}>0.2$ AND $I>7.5$ AND $n>2$ AND median $K_s >11.25$ AND ($I>100$ OR $\eta<0.5$). Here an important change from Search 1a was the
reduction of the thresholds in Stetson $I$ from 15 and 1000 to 7.5 and 100, respectively. This was required due to the typically lower values of this index in VIRAC2-$\beta$ 
than VIRAC2-$\alpha$, owing to the larger photometric errors in the newer database that include the calibration uncertainty. Stetson $I$ is usually a factor of $\sim$2
lower in VIRAC2-$\beta$ than VIRAC2-$\alpha$ but in some cases the difference is larger. This search yielded 613 candidates not selected in earlier searches,
of which 17 were found to be real.

\item Search 2b. $f_{det}<0.2$ AND $epoch\_baseline<1$~yr AND $I>15$ AND $n>2$ AND median $K_s>11.25$ AND ($I>100$ OR $\eta<0.5$). This search for transient
sources yielded only one new candidate, which did not pass visual inspection. The small number of candidates is likely to be due to the requirement of a minimum of 10 
detections in papwrint stack images for a source to be included in VIRAC2-$\beta$.

\item Search 3b. $f_{det}<0.2$ AND $epoch\_baseline=1$ to 3~yr AND $I>7.5$ AND $n>2$ AND median $K_s>11.25$ AND ($I>50$ OR $\eta<0.5$). 
This selection yielded 64 new candidates, none of which were found to be real variable sources.

\item Search 4b. $f_{det}<0.2$ AND $epoch\_baseline>3$~yr AND $I>100$ AND $\eta<0.5$ AND median $K_s>11.25$ AND $n>2$. This search yielded 78 new candidates,  
none of which were found to be real variable sources.

\item Search 5b. $I>7.5$ AND $n>2$ AND $\eta<0.5$ AND ($I>100$ OR $f_{det}>0.35$) AND median $K_s<11.25$. This search yielded two new candidates, neither of which 
were found to be real variable sources.

\item Search 6b. $n \le2 $ AND $Ks\_n\_epochs>4$ AND $\eta<0.2$ OR ($\eta<0.3$ AND $Ks\_n\_epochs>10$). This search yielded 28 new candidates,  none of which were 
found to be real variable sources.
\end{enumerate}

Whilst most of the above VIRAC2-$\beta$ searches were fruitless, the 17 sources found in search 1b included a few sources that are of very rare or unique nature within the full 
VIRAC2 sample of 222 sources presented in this work so the effort can be deemed to be justified. These are discussed in Appendix \ref{unusual}. Examples include a born-again 
giant undergoing a late thermal pulse (source 175 = V4334 Sgr, aka Sakurai's star) and a black hole X-ray binary candidate (source 16 = MAXI~J1348-630). In addition, one of the 
``false positive" candidates from search 1b was also of astrophysical interest and helped to motivate further work, though its 3.5~mag amplitude fell below the $\Delta K_s=4$~mag 
threshold for inclusion in the list published here, after removal of a bad datum. This source, recovered separately by \citet{guo22} as VVV\_PB\_41, was an early example of a 
periodically outbursting YSO candidate and the amplitude is among the highest of many such objects presented in that work.

\subsection{VVV DR4 searches}

Candidate variable stars in VVV DR4 (comprising 2010 to 2013 VVV data) were selected in two stages. First, a search of public SQL database at the VSA at Edinburgh was done for all 
sources with an amplitude exceeding a 4 mag threshold (later reduced to 3 mag) with the additional criteria that in each light curve all points lie in the range $0 < K_s < 18$. These cuts 
removed a large number of spurious faint detections at the noise level of the survey, whilst also removing non-detections, reported as negative magnitudes in the VSA database. Non-detections 
appear to be numerous in the database so it was necessary to exclude them. A final criterion was that there were at least 5 good quality $K_s$ observations of the field, i.e. the parameter 
{\it ksnGoodObs $\ge 5$}. 

The first stage provided 460 candidates with $\Delta K_s>4$, and later 3343 candidates with $\Delta K_s>3$. Since VVV DR4 was based on the pipeline aperture photometry from the 
VISTA tiles, i.e. images created by coadding the six pawprint stack images taken for each field at each epoch, it was not possible to compute a Stetson index with the DR4 data. Therefore
a local database was created at Hertfordshire by co-author Smith, using only the pipeline aperture photometry performed on the pawprint stack images by the CASU team at Cambridge. The 
database contained matched photometry for every source over the 2010 to 2015 time period. With this database prepared, the second stage of the search was simply to compute the
Stetson $I$ index for each candidate, discard those with $I<15$ and then visually inspect the much more manageable  number of images remaining. The $I=15$ threshold was based
on inspection of the data distribution and the images for a  subset of candidates. Stetson $J$ performed similarly well, but Stetson $I$ produced a marginally cleaner selection for
this data set. For candidates observed to be bright on a single date and otherwise faint and non-variable, a final step was to check for asteroids in the same way as for VIRAC2 candidates.
This step removed several false positives of that type.

\section{Additional processing and digital format light curves}	
\label{sec:lcs}	

Light curves are provided for all sources in two tar sets. As noted in the main text, these are usually based on VIRAC2-$\beta$ data, with further processing,
except in three cases, sources 60, 125 and 185, where VIRAC2-$\alpha$ was used because a source was missing from VIRAC2-$\beta$. 
In addition, the final publication version of VIRAC2 became available in the final stages of this project so it was used for sources 103 and 156, which are blended sources for which
VIRAC2-$\alpha$ and VIRAC2-$\beta$ light curves were inadequate. In each case it was necessary to amalgamate measurements from light curves for two separate sources, 
aided by the per-epoch astrometry supplied in the final version of VIRAC2. Due to the substantial work it would entail for very little benefit, the publication version was not used for other
sources.

The first tar set contains the VIRCAM pawprint-based light curves in the $Ks$,  $H$,  $J$,  $Y$ and  $Z$ filters.
The second tar set contains the binned version of the light curves that are used in the figures in this work, using 1 day bins and giving inverse variance weighted mean magnitudes.
The binned version includes the most sensitive upper limit (see below), for sources that dropped out in one or more calendar years.
A PDF file of all the light curves is also included in the Supplementary Information, using the binned light curves except for source 185, which has strong intra-night variability. 
The plots in the PDF do not include the upper limits since these tend to compress the time axis to an undesirable extent.

Both tar sets benefit from additional processing as follows.
All VIRAC2 light curves were inspected for signs of saturation. Bright detections having values of the \textsc{dophot} stellar profile parameter $Chi>5$ or the VIRAC2 astrometric 
parameter $ast\_res\_chisq>30$ were automatically saturation corrected. An additional check was performed using the ratio of fluxes in concentric apertures of radius 1.0$\arcsec$ and 1.414$\arcsec$~
and detection with unusually low values of this ratio (compared to other detections in the same image) were also assumed to be saturated. Where saturation occurred, we replaced the datum 
with a saturation-corrected value computed with the \textsc{fitsio\_cat\_list} \textsc{FORTRAN} script available at \url{casu.ast.cam.ac.uk}, which estimates source magnitudes using the flux in 
ring-shaped apertures outside the saturated core of the image profile. Images were visually inspected to select suitable apertures for this and ensure that the fluxes were not contaminated by 
adjacent stars. All flux measurements having $Chi>5$ or $ast\_res\_chisq>30$ were set to zero in the light curves and not used further.

A few sources are spatially resolved, e.g. the eruptive YSO VVV~1640-4846 displays compact nebulosity whilst in outburst. Aperture photometry in a 1.414$\arcsec$~ diameter aperture 
was used in such cases since the VIRAC2 \textsc{dophot}-based light curves display large scatter.

Many sources such as transients, dipping giants and dipping YSOs faded below the single VIRCAM pawprint-based detection limit of VIRAC2 in some calendar years. In an attempt to
increase the dynamic range of this work, we stacked cut-out images for such sources in each calendar year where there was no detection or only a single pawprint detection (in the latter
case the aim is to provide confidence in the photometry, given that there are usually two or more pawprint images at each epoch.) Aperture photometry was then performed on the image
and any detections were added to the binned version of the light curve. In practice, this process led to only a small number of detections since (i) the annual stacks are not significantly 
deeper than single pawprint images in the more crowded VVV star fields (such as bulge fields that contain most of the novae and dipping giants) and (ii) the annual stacks are often
only 1 or 2 mag deeper than single pawprints even in uncrowded fields. For sources with non-detections in a calendar year the variability amplitudes given in the 8th column of Table~\ref{tab:table1a}, 
$\Delta K_s$, is given as a lower limit, e.g. $\Delta K_s>4.16$ for source 7. This is based on either (i) the faintest detection in the binned light curve, or, (ii) approximate upper limits computed
using the annual stacks by generating artificial stars with a Gaussian image profile at the source's location and performing forced aperture photometry there. Artificial sources were counted as 
detected if the measured source magnitude agreed with the injected magnitude to within 0.5~mag. 
		
\begin{landscape}
\section{Table of variable sources detected in VVV DR4}

  \begin{table}
  \label{tab:DK3}
	\caption{Table of 105 variable sources from VVV DR4 with amplitude $\Delta K_s \ge 3$ mag}
	\begin{tabular}{lcccccccc} \hline
ID	 & VVV designation & Other name & RA & Dec         & $\Delta K_s$ & Median $K_s$ & Initial              & Time      \\
      &                             &                     &  (deg) & (deg)   &                     &                          & classification & baseline (yr)  \\ \hline
DR4\_v1  &  VVV J122329.90-614253.5  &  IRAS 12207-6126             & 185.87459 &  -61.71487 &   3.81 &  14.46 &  LPV   &  8.46              \\
DR4\_v2   &  VVV J124951.81-620038.3   &                              		  & 192.46589   &-62.01063 &   3.08 &  16.49 & Unusual    & 9.47           \\ 
DR4\_v3   &   VVV J131222.41-640905.4  & VVV-NOV-007, source 8 in Table 1a  & 198.09336 &  -64.15150  &  4.73  &  14.86 & CV   & 1.06  \\       
DR4\_v4   &  VVV J132205.42-623504.0   &                              		   &200.52260  & -62.58445 &   3.95  &  15.09 &  Unusual  & 2.23            \\  
DR4\_v5   &  VVV J132926.26-622326.4   & [RMB2008] G307.3187+00.1570 &  202.35944 &  -62.39067  &  3.97  & 12.55 &  YSO & 9.4    \\ 
DR4\_v6   &   VVV J134213.89-620039.1   &                                	    &205.55789  & -62.01086 &  3.38  &  15.30  & YSO  &  9.39    \\                  
DR4\_v7    &  VVV J135753.24-622012.6   &[RMB2008] G310.5597-00.4531     &   209.47185 & -62.33682 &  3.70 & 16.46   & YSO &  9.28      \\ 
DR4\_v8    &  VVV J135911.81-610523.4   & [RMB2008] G311.0300+00.7121     &  209.79920  & -61.08984  & 3.15  & 15.82 & YSO  & 8.19          \\
DR4\_v9    &  VVV J142359.16-600942.4   &Source 20 in Table 1a             &  215.99648 &  -60.16179 &  5.27 & 12.43  & CV  &  0.27    \\ 
DR4\_v10   &  VVV J142513.97-602020.0   & [RMB2008] G314.2893+00.4481  & 216.30820  & -60.33888 &  3.68 & 13.74 &YSO   &  9.41    \\     
  ...\\
    		\hline
  	\end{tabular}
	\\
  	\begin{tabular}{ll} \hline
  Light curve description & Comment \\ \hline
(DR4\_v1)   &  No OH maser detected in \citet{telintelhekkert91}             \\
(DR4\_v2) Rapid rise in 2011. Rapid fade in 2012. Faint at Ks~16.5 in other years. & No mid-infrared excess in GLIMPSE\\
(DR4\_v3)     &                                                                                                              \\
(DR4\_v4) Slow rising transient seen from 2013-2015        &                             \\                                             
(DR4\_v5) Dipped and then brightened into outburst & Spectroscopically confirmed eruptive YSO in \citet{guo21}    \\
(DR4\_v6) Eruptive YSO light curve & WCI shows a yellow-green source in an SFR.\\
(DR4\_v7) Fader  & \citet{robitaille08} YSO candidate. WCI shows a yellow-green source in an SFR.  \\
(DR4\_v8) Irregular, a multiple timescale variable    &    \citet{robitaille08} YSO candidate. WCI shows a reddish green star in an SFR.   \\
(DR4\_v9) Unusual transient, see Table 1a     &            \\
(DR4\_v10) Eruptive YSO light curve & Spectroscopically confirmed eruptive YSO in \citet{guo21}\\
      ...\\
              \hline
        \end{tabular}\\
         Notes: Only the first 10 rows of the table are shown here, split into two sections. The full table is available in the online supplementary information. J2000 coordinates\\
         are calculated at epoch 2015.0 for all sources in VIRAC2$-\beta$.  Here $\Delta K_s$ is the amplitude measured in VIRAC2-$\beta$.\\
         ``[RMB2008]" denotes red sources from \citep{robitaille08}. \\
         ``WCI" is an abbreviation of ``WISE colour image".
\end{table}
\end{landscape}

\section{VVV and WISE light curves of irregular eruptive YSOs }
\label{irregysos}
\begin{figure*}
        \includegraphics[width=0.48\textwidth]{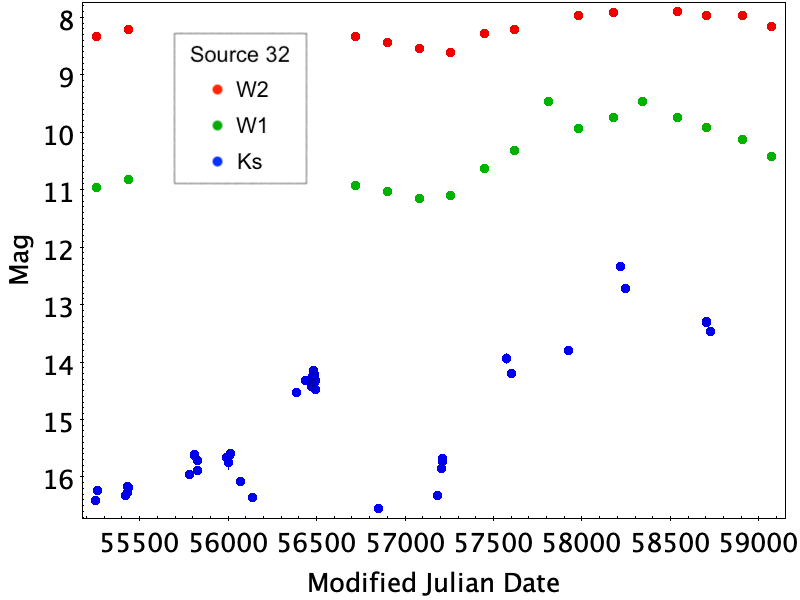}
        \includegraphics[width=0.48\textwidth]{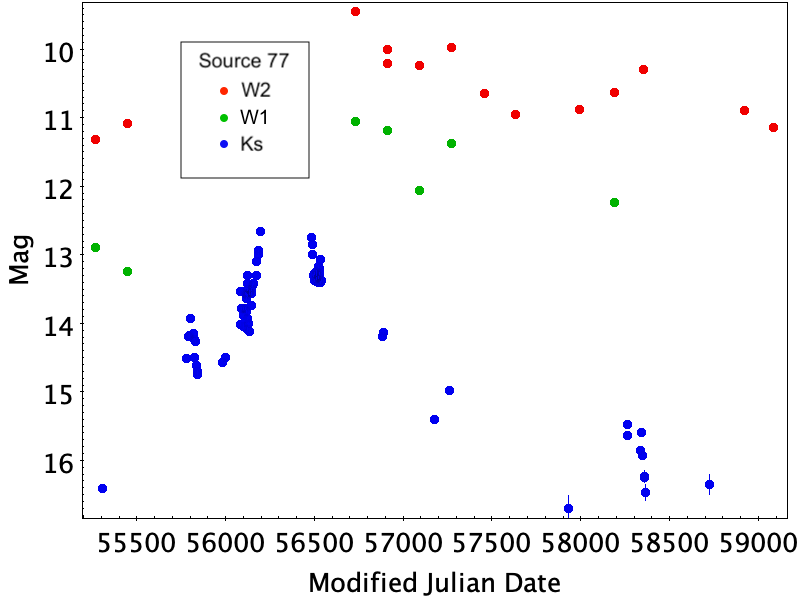}
        \includegraphics[width=0.48\textwidth]{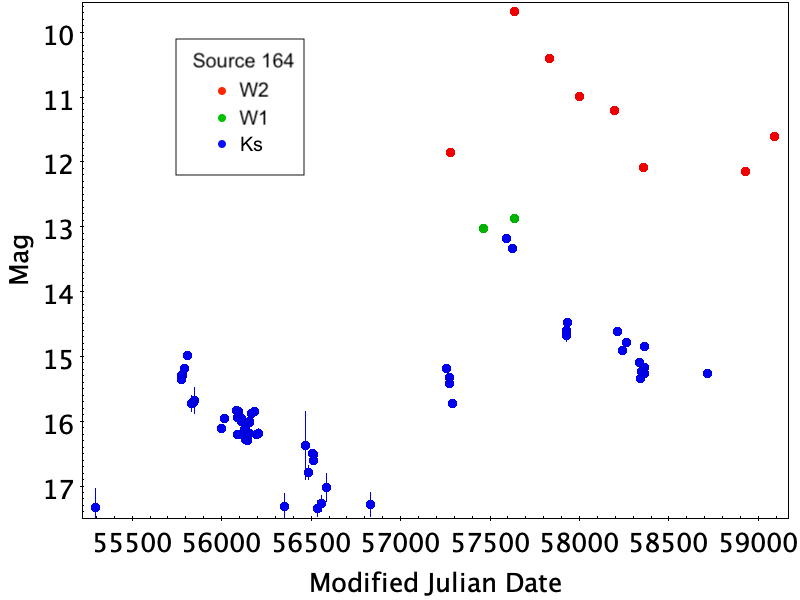}
        \includegraphics[width=0.48\textwidth]{s42_3bandf.png}
        \includegraphics[width=0.48\textwidth]{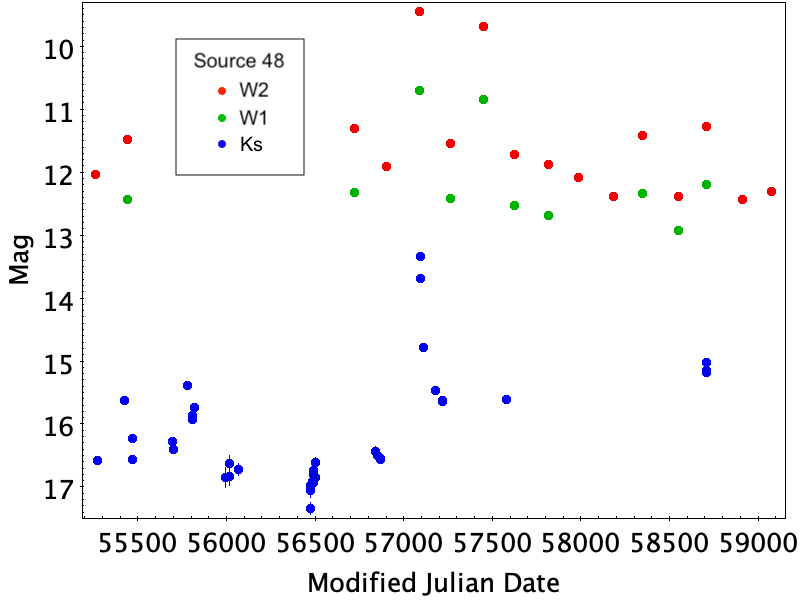}
        \includegraphics[width=0.48\textwidth]{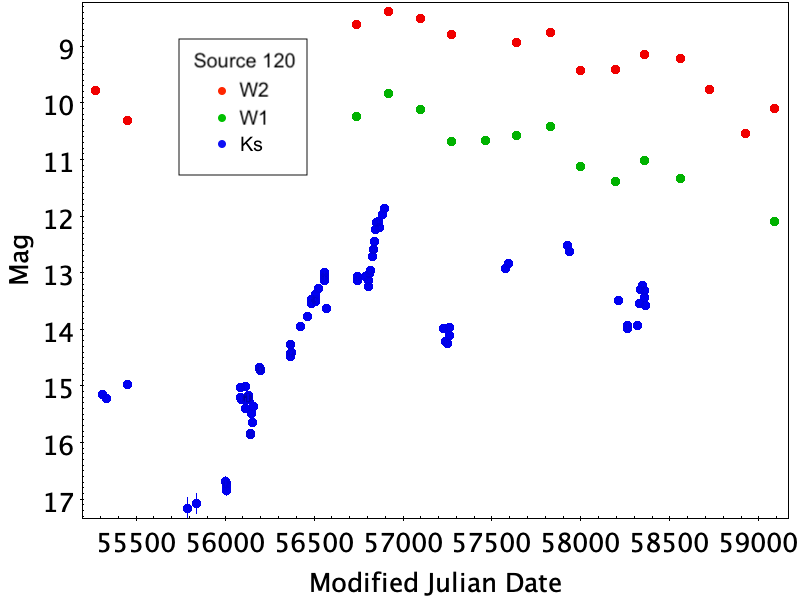}
    \caption{VVV and WISE light curves of six irregular eruptive YSOs.}
    \label{fig:irreg5}
\end{figure*}
Multi-filter light curves are shown in Fig. \ref{fig:irreg5}, for six irregular eruptive YSOs having useful data in the WISE/unTimely database.
Source 42 shows the $\sim$1~yr duration of the second outburst near MJD=58000 more clearly in the WISE $W2$ data than in VVV.
Source 48 shows four brief, single epoch outbursts of 1 to 2 mag amplitude in the WISE data, suggesting a duration of order $\sim$6 months in each case.

\section{VVV and WISE variability of Dipping YSOs}
\label{dipysos}
\begin{figure*}
        \includegraphics[width=\textwidth]{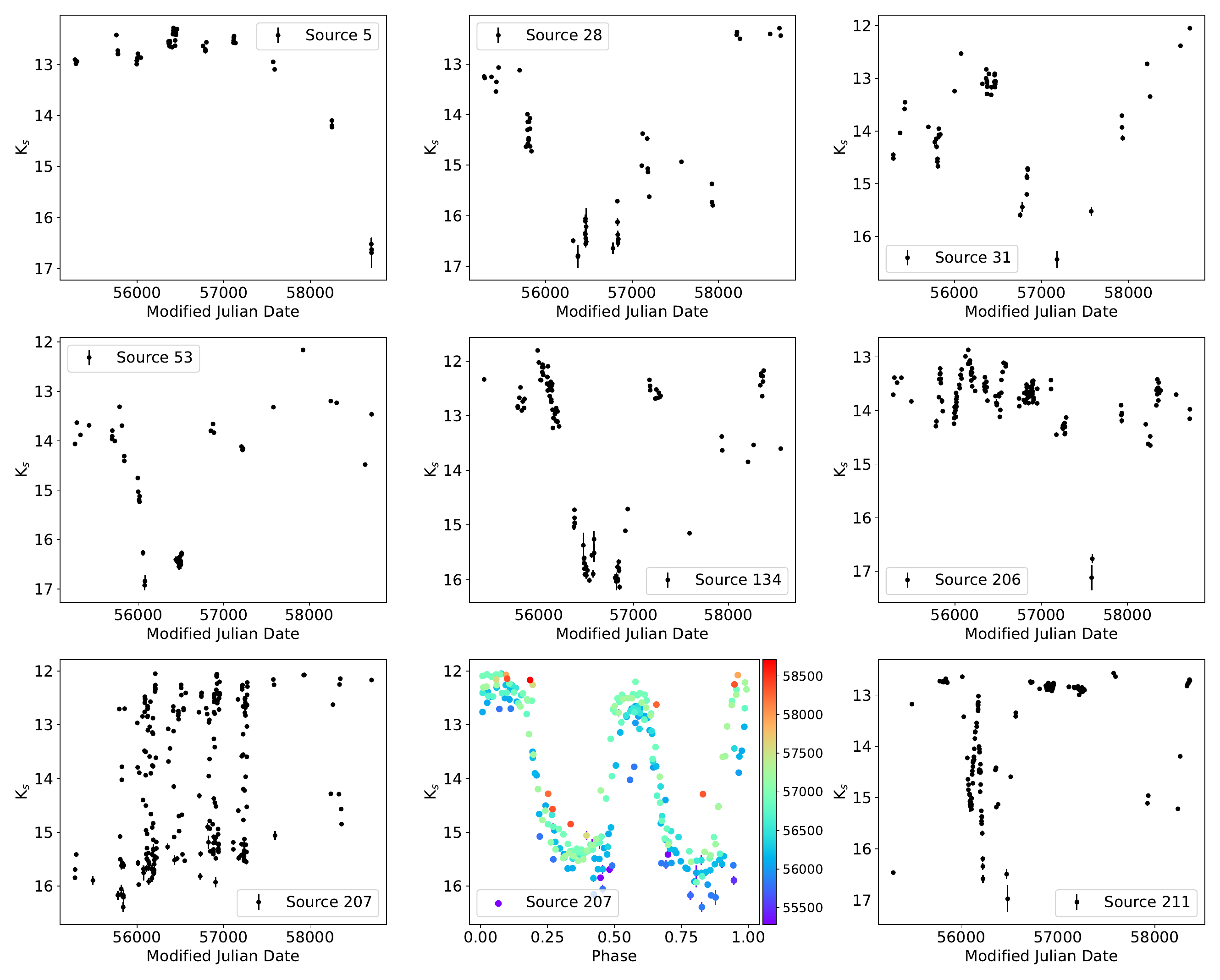}    
    \caption{Light curves of the eight dipping YSOs. The variation in these systems is thought to be due to variable extinction by circumstellar matter.
    Source 207 appears to be a KH15D-like system with periodic dips due to a circumbinary disc. The phase-folded light curve (lower middle panel) is colour-coded
    by Modified Julian Date.}
    \label{fig:dip8}
\end{figure*}

\begin{table}
	\caption{Supporting evidence for an extinction-driven dip}
	\label{tab:dippers}
	\begin{tabular}{lccccc} \hline
Source & Bright in & VVV colour & WISE colour \\
      & 2MASS  &  changes & changes  \\ \hline
 5   &  yes(I) & yes   & - \\
 28   &  yes &  yes  & yes\\
 31  &  yes &  -  & ? \\
 53  &  yes(I) &  yes  & yes?\\
 134  &  no & yes  & - \\
 206  &  yes(I) &  yes & - \\
 207  &  - &  no  & - \\
 211  &  yes &  yes  & -\\ \hline
\end{tabular}\\
Note: ``(I)" denotes an uncatalogued 2MASS detection in the images, ``no" indicates contrary evidence, ``-" indicates no useful data.\\
\end{table}

In eight YSO candidates it is likely that the variability is due to changing extinction. With one exception, the light curves plotted in Figure \ref{fig:dip8} show at least one long duration extinction event lasting a year or more. In one case the source remains in a faint state at the end of the VVV time series. In terms of their high infrared extinction, most of these eight systems can be classified as extreme examples of the aperiodic dipper phenomenon, a term used by 
\citet{morales-calderon11} and \citet{rice15} to distinguish aperiodic dippers from AA Tau-like dippers that show display periodic dips
(typically $P<15$~d) attributed to a magnetically induced warp in the accretion disc \citep{bouvier07} and sometimes long duration aperiodic dips also. Source 207 is different, being periodic. It is discussed in more detail below. 

\begin{figure}
        \includegraphics[width=0.49\textwidth]{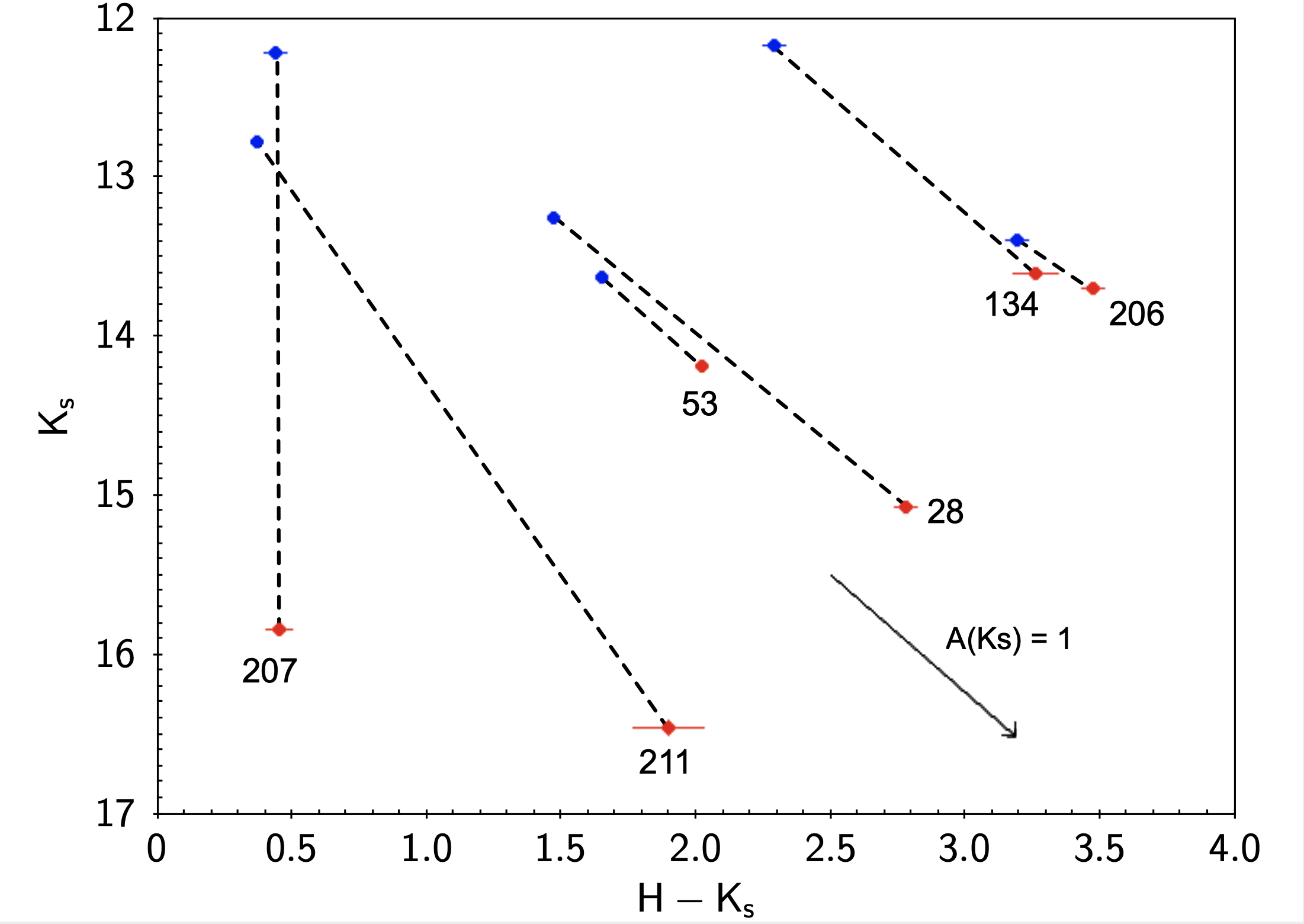}    
        \vspace{-4mm}
    \caption{$K_s$ vs. $H-K_s$ colour-magnitude diagram for six of the dipping YSOs, labelled with their source numbers. Blue and red points correspond to the brightest and 
    faintest available epochs of multi-filter data, respectively. Elements of a pair are connected by dashed lines. The changes are roughly parallel to the extinction vector in most cases. One source 
    is clearly different: source 207 (the candidate KH15D-like object in the Lagoon Nebular Cluster) shows negligible colour change, see text.}
    \label{fig:dipysoHK}
\end{figure}

In addition to the dips in the light curves, in most cases there is further evidence that the variability is due to extinction, based on inspection of data from 2MASS, the Deep Near Infrared Sky Survey \citep[DENIS,][]{epchtein99}, VVV, and to a lesser extent the WISE survey. The evidence is summarised in Table \ref{tab:dippers}, see also Figure \ref{fig:dipysoHK}. Seven of the nine dippers were detected in $K_s$ by 2MASS or DENIS between 1996 and 2000, 
all at a flux level similar to or slightly brighter than the brightest measurements in the VVV time series.\footnote{Sources 5, 53 and 206 are not in the 2MASS PSC due to blending or the presence of nebulosity in the field but they can be seen in the images. Source 207 is blended with a brighter star in 2MASS, so non-detection is to be expected.}. 
This supports the view that VVV observed dips below the typical brightness level rather than a gap between accretion-driven eruptions of the type seen in V346 Nor \citep{kospal20}. 

Most of the dipping YSOs also become redder when fainter in the $K_s$ vs ($H-K_s$) colour magnitude diagram in a manner that approximately follows
the \citet{wang19} reddening law for interstellar extinction, see Figure \ref{fig:dipysoHK}. (Some allowance should be made for departures from an idealised reddening behaviour due to changes in the accretion disc
over timescales of years, see e.g. \citet{covey21}.) Figure \ref{fig:dipysoHK} illustrates the colour changes for sources 28, 55, 136, 207, 208 and 212, the six sources having colour 
measurements on two different dates. They are plotted with blue points in the brighter state and red points in the fainter state, each pair being joined by a dashed line. 
The exception is source 207, the KH15D-like candidate, which shows no significant colour change despite a change in brightness of almost 4 magnitudes. 
Only source 207 and source 55 have detections in $J$, $H$ and $K_s$ that are contemporaneous (measured on the same night). Source 207 actually 
becomes bluer in the faint state ($J-H=0.71$, compared to $J-H=0.94$ in the bright state), which can perhaps be attributed to scattered light.

\begin{figure}
        \includegraphics[width=0.49\textwidth]{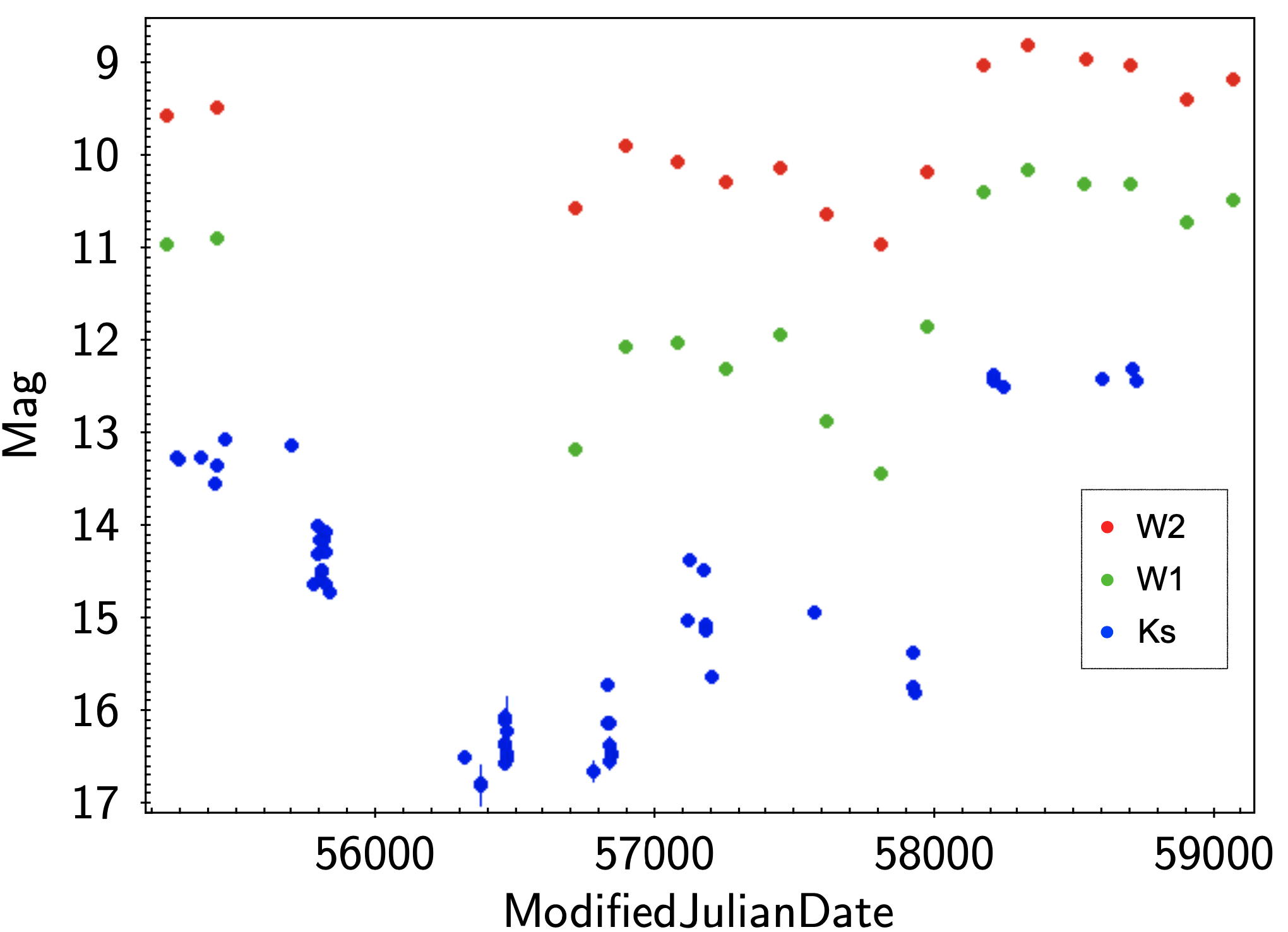}    
        \vspace{-4mm}
    \caption{VVV and WISE/unTimely light curves of the dipping YSO source 28 in $K_s$ (blue), W1 (green) and W2 (red).}
    \label{fig:s28}
\end{figure}

Finally, for one of the eight dippers, source 28,
the WISE/unTimely time series data show large changes in $W2$, $W1$ and VVV $K_s$ magnitudes that are consistent with variable reddening based on the \citet{wang19} extinction law, see Figure \ref{fig:s28}.
Supporting evidence from the unTimely or NEOWISE data is found in a further two systems but these cases are less clear, either because the flux changes sampled by WISE are small, or because the WISE and VVV
light curves are not contemporaneous and have subtly different trends over time (not shown).
In the other five systems the sources are not genuinely detected by WISE due to blending, or at least not detected when they are faint. Careful inspection of the WISE, GLIMPSE and VVV images, and any
astrometric shifts in WISE detections over time, was required to guard against being misled by blends in the WISE data.\\

\textbf{Source 207}. This is a periodically dipping system (period $P=59.35$d) that appears to resemble KH~15D \citep[=V582 Aur,][]{chiang04, herbst02}, a system wherein the two bright components of a binary YSO are extinguished by a circumbinary ring for much of their orbits but each becomes visible in alternation at certain phases. A similar type of behaviour has been seen at lower amplitude in WL4 and YLW16A in the $\rho$~Ophiuchi dark cloud \citep[][though these two are triple systems]{plavchan08, plavchan13} as well as CHS~7797, ONCvar 149 and ONCvar 1226 in the Orion Nebular Cluster \citep{rodriguez-ledesma12, rodriguez-ledesma13, rice15} and in the recent candidates Bernhard-1 and Bernhard-2 \citep{zhu22}. Source 207 is currently mis-classified in the SIMBAD database as a ``Cataclysmic Binary" (designated OGLE BLG-DN-652) due to being listed as a 
photometric dwarf nova candidate, within a list of 1091 candidates \citep{mroz15} drawn from the OGLE survey (most of which are indeed high quality dwarf nova candidates). In this case though, the light curve does not resemble that of a dwarf nova (see Figure \ref{fig:dip8}, bottom left and bottom centre panels) and the system is projected in the Lagoon Nebula Cluster (NGC~6530), with cluster membership confirmed by {\it Gaia} parallax and proper motion (and VIRAC2 proper motion) consistent with that of the cluster, at distance $d \approx 1.3$~kpc.\footnote{The {\it Gaia} DR3 parallax and proper motion of source 207 are, 
$\varpi = 0.71 \pm 0.13$~mas, $\mu_{\alpha cos \delta}= 1.95\pm0.09$~mas/yr, $\mu_{\delta}=-2.05\pm0.06$~mas/yr. The {\it Gaia} DR3 cluster parameters, are $\varpi = 0.76$~mas, $\mu_{\alpha cos \delta}= 1.28$~mas/yr, $\mu_{\delta}=-2.06$~mas/yr, with standard deviations of 0.66 mas/yr and 0.70 mas/yr for $\mu_{\alpha cos \delta}$ and $\mu_{\delta}$ and a systematic uncertainty on $\varpi$ of order 0.02 mas. 
The cluster parameters are based on median values from the set of cluster members in \citet{prisinzano19} having 5$\sigma$ parallaxes and reduced unit weight error, $ruwe<1.4$.} 

The two components of the binary have similar but not identical luminosities, see Figure \ref{fig:dip8}, such that the ``outburst frequency" given by \citet{mroz15} is double the true frequency of the periodic variability.
The system is also listed with half the true period in separate VIRAC2-$\beta$-based searches for periodic systems by \citet{molnar22} and \citet{guo22}. (The latter study noted it as a dipping YSO but analysis was 
deferred to this study because the system had already been classified as KH~15D-like in the initial stages of this work). The peak near phase 0.6 in the phase-folded plot is narrower as well as shorter, but the 
colour-coding by Modified Julian Date shows that it broadened over the course of the survey. The few red-coloured points, corresponding to the sparse sampling at the end of the time series, indicate that the 
phasing of the variability continued to change and the dips may have become shallower. In KH~15D changes of this sort are explained as precession of the warp in the circumbinary ring, see e.g. 
\citet{arulanantham16}, so this is consistent with our interpretation of the system.

Data from {\it Spitzer}/GLIMPSE-3D indicate an infrared excess but this is hard to quantify because 
source 207 is blended with a $K_s=15.26$ neighbour in the $\sim$$2\arcsec$ GLIMPSE-3D beam and the observations were taken at a time when source 207 was faint. The reported colours are $[3.6]-[4.5]=0.48$, 
$[4.5]-[8.0]=1.52$. (In the WISE/unTimely data, source 207 is detected only when in the bright state, usually in $W2$ only, owing to blending with an additional $K_s=14.50$ neighbour in the larger WISE beam). In VVV, source 207 is not very red, with ($J-H \approx 0.8$), ($H-K_s \approx 0.45$) and its variability is almost wavelength independent in the five contemporaneous epochs of $JHK_s$ photometry, suggesting 
either grey extinction by large dust grains or a contribution by scattered light in the faint state, as is seen in KH~15D. The system will be analysed in more detail in a future publication, with the aid of follow up spectroscopy.

\section{Long Period Variables}
\label{sec:s163}

\begin{figure}
        \includegraphics[width=0.48\textwidth]{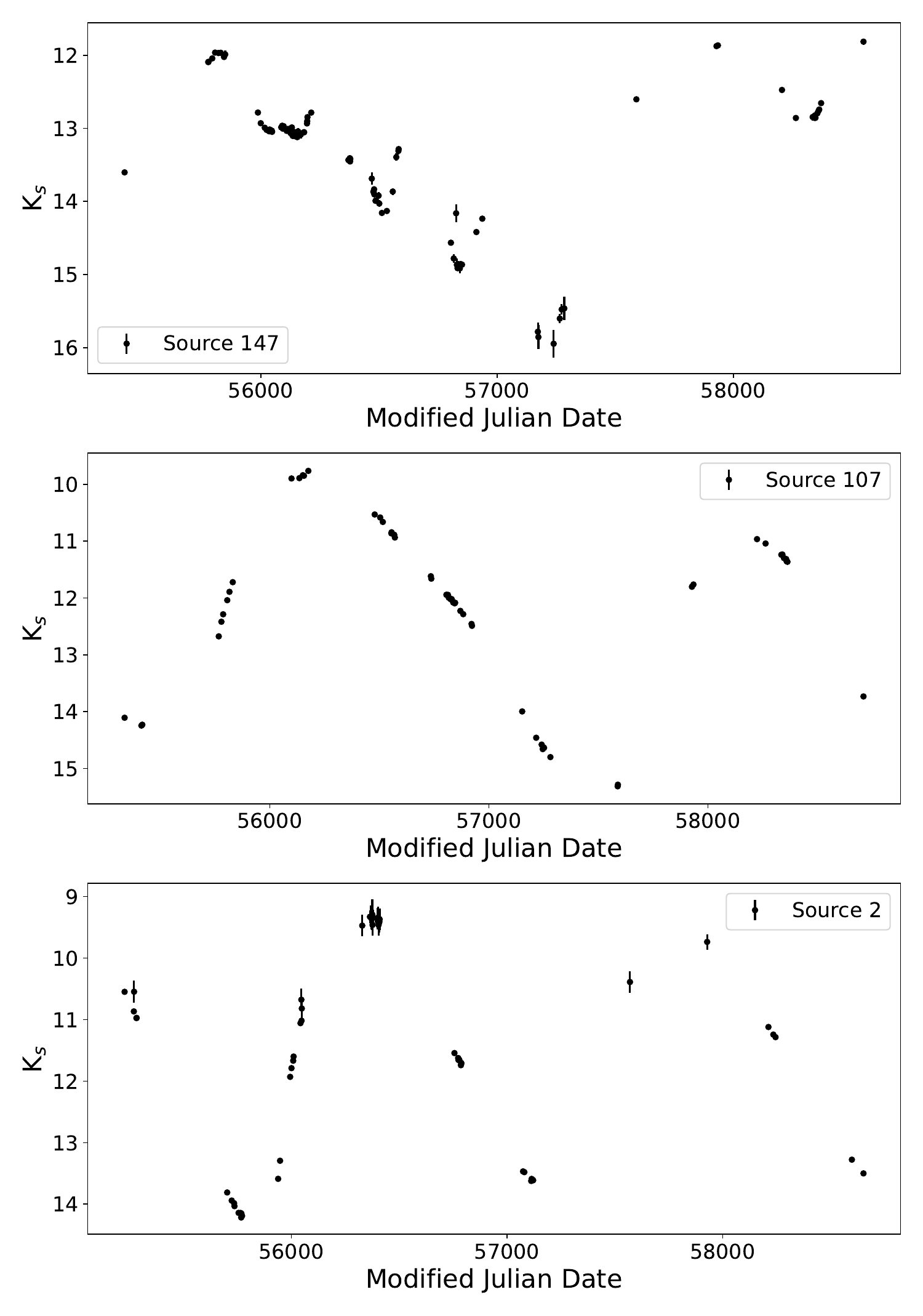}    
        \vspace{-4mm}
    \caption{VVV light curves of three notable LPVs. Source 147 shows a $\sim$3.5 mag fading trend over 4 years before recovering. Source 107 and source 2 have the highest amplitudes of
    pulsation in the LPV sample, $\sim$5 mag in $K_s$.}
    \label{fig:3LPVs}
\end{figure}

The 35 sources classified as LPVs are dusty Mira variables with unusually high amplitudes. These represent the upper end of the amplitude distribution in such sources and consequently given the
period vs. amplitude relation \citep[see e.g.][]{nikzat22} they also tend to have unusually long periods. The periods are given in Table \ref{tab:LPVs}. Given the inner Galaxy locations (see Figure \ref{fig:galdistrib}), 
we expect that most of the 35 sources are O-rich. Some (10/35) are known OH/IR stars and most (24/35) have previously been identified as LPVs, either in VVV-based variable star searches 
(e.g CP17a), or earlier work (see references in Table \ref{tab:table1a}). 

A notable example, shown in Figure \ref{fig:3LPVs}), is source~107 (=OH 355.815-0.226 = IRAS 17328-3234) which has a peak to trough amplitude of $\sim$5 mag in $K_s$ ($\Delta K_s \approx 5.6$~mag 
in total, due to a slow fading trend) and a period $P \approx 2000$~d. Source 2 (=OH~259.8-1.8 = IRAS~11438-6330) also has a $\sim$5 mag amplitude but a shorter period and no clear long term trend. 
These are the highest $K_s$ amplitudes of pulsation that we are aware of in Mira variables. OH/IR stars often have large amplitude pulsations and they are associated with high mass loss rates 
\citep[see e.g.][]{gaylard89, hofner18}. Source 2 is listed as d001-79 in a previous VVV-based study of variable stars \citep{medina18} but the full amplitude was not apparent at that time.

\begin{figure}
        \includegraphics[width=0.48\textwidth]{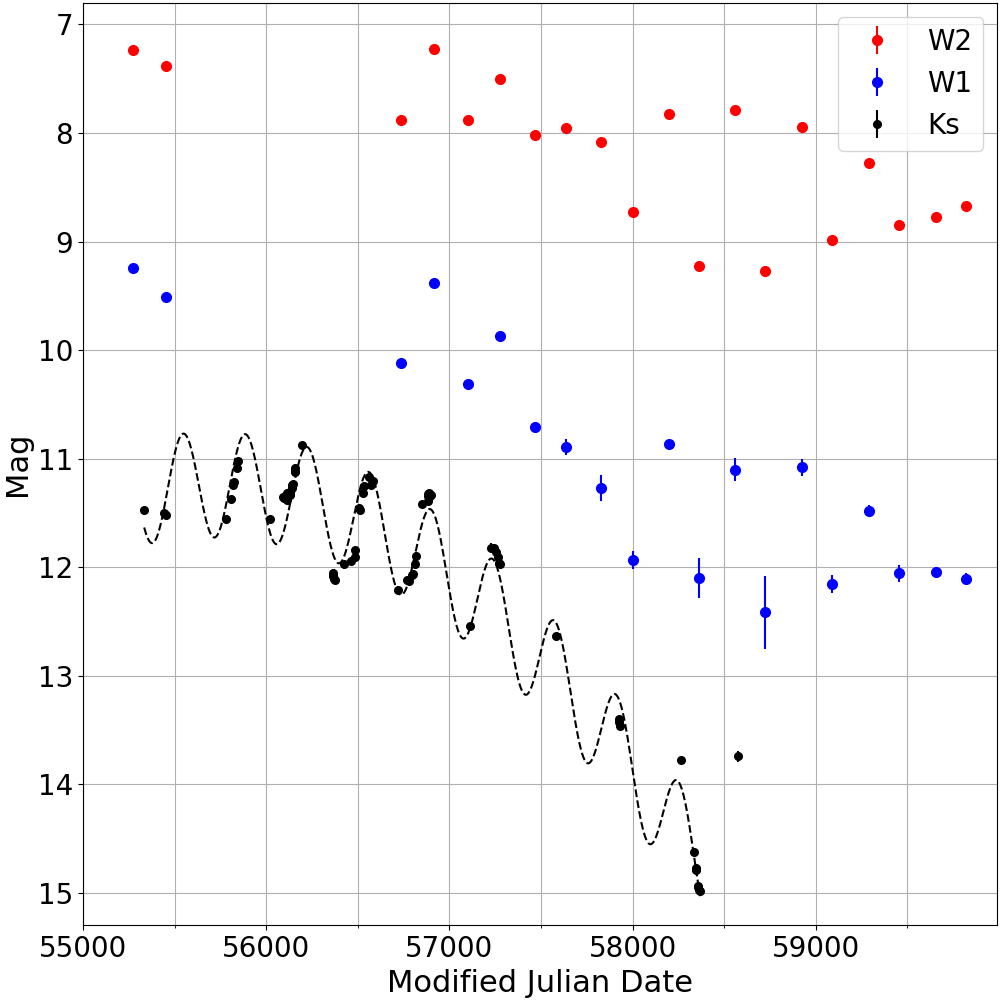}    
        \vspace{-4mm}
    \caption{VVV and WISE light curves of source 163. These appears to be a combination of pulsation and a slower, extinction-driven dip. The dashed line shows a simple model fit, see main text.
    The background grid aids comparison of the amplitude of pulsation in the three filters.}
    \label{fig:s163}
\end{figure}

Many of the 35 LPVs have slow trends in brightness that can rival the amplitude of the pulsation. The most extreme cases are source 163 and source 147 (see Figure \ref{fig:s163} and \ref{fig:3LPVs}, upper panel)
where the aperiodic trends have much higher amplitude than the pulsations. Both these sources are projected in or on the border of the Nuclear Disc so they are interesting in the context our discussion of aperiodic 
dipping giants in that region (see $\S$\ref{sec:dipgiants}). (Source 147 is near the Galactic centre, at $l=0.13^{\circ}$, $b=-0.14^{\circ}$, whereas source 163 is at $l=0.49^{\circ}$, $b=-0.69^{\circ}$, about two
scale heights below the mid-plane of the Nuclear Disc). Only one other member of the sample, source 135, is projected in the Nuclear Disc. Notably, this star shows a 2~mag fading trend over the 2010 to 2019
time series. Other cases not in the Nuclear Disc include sources 184 and 203, each with a 2~mag fading trend and source 52, with a $\sim$3~mag fading trend.

Source 147 is listed by \citet{glass01} as Mira variable 12-228 (=V5084~Sgr). In that work (with a four year time series in $K$), a perfectly regular period of 570~d was found and a 1.5 mag peak to trough amplitude was reported,
although inspection of the published light curve shows a value closer to 1~mag, consistent with the pulsations seen in Figure \ref{fig:3LPVs}). Without this prior work, the VVV light curve of source 147 might be interpreted as
showing signs of a long secondary period with a 4 mag peak to trough amplitude and a possible period 
$P \sim 2100$~d. However, this seems unlikely in view of the previous lack of any trend over fours years, coupled with the fact that the flux in the VVV light curve remains near maximum after recovering from the dip.
A 2100~d period is also a bit longer than the 200 to 1600~d range usually seen in such systems \citep{soszynski21}. \citet{glass01} noted that many of the Miras are ``quite irregular" but did not discuss
aperiodic trends further.

\begin{table}
	\caption{Periods of the LPVs}
	\label{tab:LPVs}
	\begin{tabular}{lcc} \hline
Source & Other Name & Period \\
      &   & (d)  \\ \hline
2 & OH~259.8-1.8 & 1452 \\
24 & VVVv588 & 990 \\
27 &	[RMB2008] G323.4702-00.1421 & 769 \\
30 &	PDS 143 & 903 \\
35 & 	VVVv229 & 996 \\
52 &	[RMB2008] G338.1117-00.3902 & 761\\
56 &	IRAS 16462-4414 \\
57 &	VVVv345 & 940 \\
58 &	IRAS 16491-4336 & 1265 \\
62 &	& 1065 \\
63 &	VVVv382 & 1436 \\
64 & OH 344.905+00.115 & 1480 \\
65 &	IRAS 17002-4248 & 920 \\
71 &	VVVv796 & 894 \\
76 &	VVVv812 & 864 \\
80 &	OH 350.29+00.06 & 1390 \\
85 &	IRAS 17219-2756 & 717 \\
94 &	& 664 \\
98 &	OH 354.529+00.038 & 1458\\ 
104 & OH 354.884-00.539 & 1547 \\
107 & OH 355.815-00.226 & 2263\\
113 & IRAS 17363-2802 & 504 \\
129 & V1185 Sco & 2182 \\
135 & OH 359.836+00.119 & 572 \\
147 & V5084 Sgr & 570 \\
163 & & 338 \\
176 & IRAS 17495-2534 & 1069 \\
178 & & 777 \\
179 & & 1472 \\
181 & & 721 \\
184 & & 480 \\
186 & IRAS 17535-2421 & 888 \\
203 & IRAS 18005-2244 & 705 \\
205 & IRAS 18016-2540 & 945 \\
218 & IRAS 18195-2804 & 779 \\
\end{tabular}
\end{table}

Source 163 shows a long term fading trend after 2012, with sinusoidal oscillations superimposed upon it (period $P\approx$1~yr, peak to trough 
amplitude $\sim$1 mag), see Figure \ref{fig:s163}. 2MASS measured $K_s =11.39$ in 1998, similar to VVV measurements before the decline began. The VIRAC2-$\beta$ light curve is moderately well
fit by the sum of sine wave with $P=338 \pm 2$~d and a quadratic function (plotted as a dashed curve), until the fading trend appears to bottom out near MJD = 58500. The system is classified in the VIRAC2-based 
VIVACE catalogue \citep{molnar22} as an LPV with $P=692.8~d$, approximately double the true period, after fitting and subtracting a quadratic long term trend (which was standard procedure in VIVACE). Similarly, 
\citet{nikzat22} identified it as an LPV within a list of 1013 VVV LPVs with imperfectly determined periods near the Galactic Centre.

The periodic variations and the overall decline are fairly well sampled by NEOWISE in $W1$ and $W2$ (better than the 
unTimely data in this case), with similar behaviour to that seen in VVV and no significant blending issues. Within the individual cycles, the magnitude changes between MJD = 58000 and MJD = 59000 are very similar 
in $K_s$, $W1$ and $W2$. This is consistent with Mira-like pulsations: \citet{iwanek21} found that the amplitude of such pulsations declines with increasing wavelength from the optical to the mid-infrared but 
there was no significant change between the $K_s$, $W1$ and $W2$ passbands. (There is no useful WISE light curve for source 147: the VVV $K_s$ images show it is heavily blended in both WISE and 
{\it Spitzer}/GLIMPSE).

Considering the longer term fading, we can compare the flux levels in similar (faint) parts of the periodic cycle in 2014 (MJD~$\approx 56700$) and 2018 (MJD~$\approx 58350$). Source 163 faded by 2.8 mag, 2.1 mag 
and 1.3 mag in $K_s$, $W1$ and $W2$. The change in $K_s$ is smaller than expected from the 1: 0.52: 0.33 ratio predicted by the \citet{wang19} extinction law but the $\Delta W1$ to  $\Delta W2$ ratio is as expected. 
This could be explained by a model where the mid-infrared fluxes decline as extinction rises but the $K_s$ flux declines more slowly due to a contribution by scattered light. VVV provides only two epochs of 
contemporaneous photometry in $H$ and $K_s$, the
source being undetected in the $Z$, $Y$ and $J$ passbands. The system is very red, with colour changing from $(H-K_s) = 3.91$ at $K_s=11.40$ in 2010 to $(H-K_s) = 4.26$ at $K_s=11.96$ in 2015. This colour change 
is consistent with increased extinction, though the measurements were taken in different parts of the cycle so they cannot be interpreted in a simple way.

Dusty Mira variables often show aperiodic variability on a range of time scales \citep{iwanek21}, slow trends typically being attributed to mass loss and variable extinction.
Whilst sources 163 and 147 could simply be seen as extreme examples of the variable extinction seen in many dusty Miras in the Milky Way, their locations in the Nuclear Disc mean that they could plausibly have super-solar metallicity,
as proposed for the 21 dipping giants discussed earlier, which may enhance dust production. 
Such aperiodic events have been observed, albeit with considerably lower amplitude in $K_s$ than source 163, in the Mira variables R For \citep{whitelock97}, II Lup \citep{feast03} and IRAS~17103-0559 \citep{jimenez-esteban06}.
We should consider whether a long secondary period can explain the event in source 163. Such periods are quite common in LPVs, though less common in the bulge \citep{soszynski13} and they have been 
convincingly attributed to eclipses by either a circumsecondary 
disc around a very low mass companion or by matter spread around the orbit of such a companion \citep{soszynski21}, see also \citet{wood99}. It is proposed by \citep{soszynski21} that the companion is a former 
giant planet that has gained mass from the stellar wind of the pulsating Mira variable and become a brown dwarf. A secondary period is not yet established in source 163 and it would have to be rather longer than 
the 200~d to 1600~d periods usually observed \citep[see Figure 2 of][]{soszynski21}. However, a few rare cases of very long secondary periods are known, such as the C-rich Mira R Lep \citep{whitelock97} 
and V Hya \citep{knapp99}, the latter recently revealed to have an expanding spiral circumstellar disc \citep{sahai22}.

Source 163 is projected in a bright \ion{H}{II} region seen in the WISE mid-infrared images, apparently corresponding to the giant \ion{H}{II} region G000.5-00.7 identified by \citet{kuchar97} and \citet{conti04}. 
It also has a large infrared excess and the {\it Spitzer} colour $[8]-[24] =  3.3$ is more typical of YSOs than dusty AGB stars \citep{robitaille08}. For these reasons it was initially thought that source 163 might well be a 
dipping YSO of the AA Tau type. However, those systems usually have much shorter periods \citep[$P<15$d,][]{rice15}, with the longer-term dip in AA Tau and some other systems attributed to occultation by 
non-axisymmetric matter in a much wider orbit \citep{bouvier13}. The VIRAC2 proper motion helps to reject the possibility of a dipping YSO: the values
are $\mu_{\alpha cos \delta} = -5.04 \pm 0.42$~mas/yr and $\mu_{\delta}=-7.41 \pm 0.46$~mas/yr, which corresponds to $\mu_l = -8.03 \pm 0.43$~mas/yr and $\mu_b = -3.97 \pm 0.45$~mas/yr in Galactic coordinates.
For a source projected very close to the Galactic centre at heliocentric distance $d$/kpc, these relatively large proper motions correspond to tangential velocity components, $V=38 d$~km/s and $W=-19 d$ km/s, parallel
and perpendicular to the Galactic disc, respectively. This is incompatible with a YSO in the Galactic disc, unless it were located within $\sim$1~kpc of the sun. That is highly unlikely, given that no such nearby star 
formation region is known on the line of sight to the Galactic centre.

\section{Cataclysmic Variables and other transients}
\label{sec:CVs}

The great majority of these 70 sources are expected to be classical novae. In fact, 43 of them have been previously classified as classical novae, recurrent novae (i.e. novae 
for which more than one explosion has been observed) or nova candidates. These 43 were discovered either in the VVV-based searches listed in $\S$ \ref{overview}, by optical surveys,
e.g. OGLE in 12 cases, the All-Sky Survey for Supernovae \citep[ASAS,][]{shappee14} in one case, the {\it Gaia} Photometric Science Alerts system \citep{hodgkin21} in one case, by 
amateur astronomers, or, by the mid-IR WISE satellite in three cases \citep[e.g.][]{zuckerman23}. (Individual references and identifications are given in Table \ref{tab:table1a}).

Some of the remaining 27 nova candidates are poorly sampled so their classification is tentative. While most of these 27 light curves display a typical nova-like monotonic fade, there are a few cases 
where a smooth-topped secondary peak attributable to dust condensation \citep{gehrz88, gehrz99, hachisu06} is observed, with varying degrees of confidence. Source 84 shows such a peak and
there are hints of one in sources 45 and 222, though source 45 has noisy data due to saturation and source 222 is poorly sampled. Sources 20 and 109 show a modest rise from the initial detection up to the 
maximum flux: we have chosen to classify these two as nova candidates where the initial outburst maximum was missed. Sources 160 and 183 show a similar behaviour but with 
noteworthy differences (see below).

\begin{figure*}
        \includegraphics[width=\textwidth]{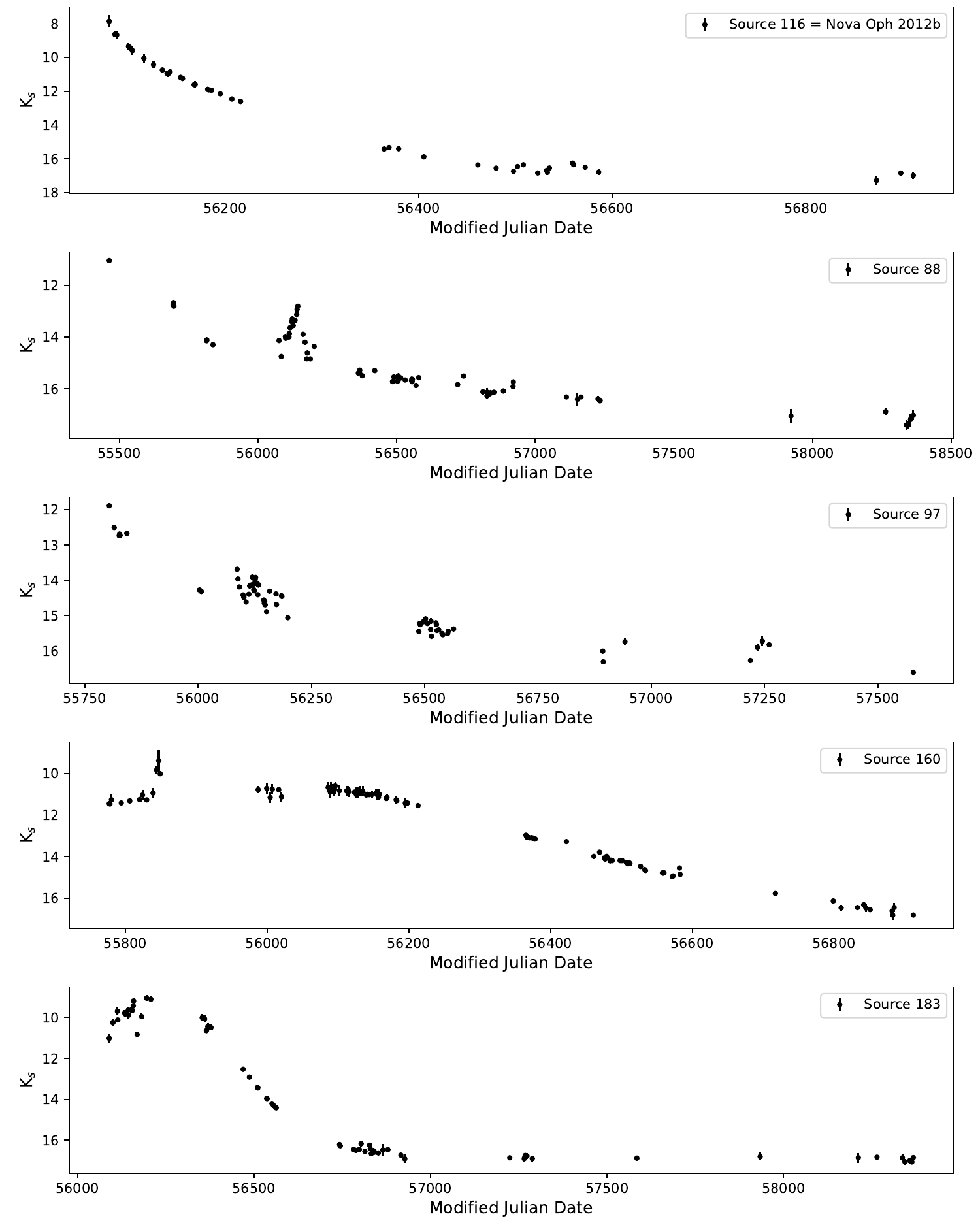}
    \caption{Light curves of novae and some unusual nova candidates. Source 116 is Nova Oph 2012b \citep{walter12} with a typical monotonic light curve. The other four have unusual 
    light curves. Source 88 shows a "cusp`` similar to those seen in a small number of optical nova light curves. Source 98 shows quasi-periodic oscillations near MJD=56150. Sources 160 
    and 183 each show a rise to a maximum that we interpret as a secondary dust peak but there are atypical features in each case, see main text.}
    \label{fig:novaLCs}
\end{figure*}

Four of the 29 new nova candidates have rather unusual light curves, discussed briefly below and illustrated in Figure \ref{fig:novaLCs}.

\begin{enumerate}
\item Source 88 has a sharp secondary peak in its light curve 680 days after the first observation (see Figure \ref{fig:novaLCs}), rising almost 2 mag above the overall 
declining trend. This feature, distinctly not a rounded secondary dust peak, lasted for at least 100 days and it resembles the "cusp`` seen in the rare C-class of novae 
defined by \citet{strope10} in optical light curves. That work stated that only three examples of such novae were known at the time ($\sim$1 per cent of novae) and the 
origin of the cusp features is not yet clearly determined, though two theories are noted in that work. A more recent example is found in \citet{wee20}.

\item Source 97 was first detected on 2011 August 30. The flux declined rapidly but quasi-periodic oscillations within the overall decline were seen in the 2012 observing season (data at Modified Julian Dates, 
MJD = 56086 to 56198). The period was $\sim$45 days. In subsequent years the photometry began to suffer from added noise due to blending with adjacent stars 
so it is difficult to determine whether the oscillations continued. Oscillatory behaviour in optical light curves is described as the O-class in \citet{strope10} but only two examples were mentioned. We have not seen 
this behaviour in other infrared light curves of novae in the literature, such as the numerous examples available in the SMARTS database \citep{walter12}.

\item Source 160 was first detected as a bright star ($K_s \approx 11.4$ on 2011 August 4) but it then brightened by at least 2~mag over the following 70 days until observations 
were interrupted by the end of the observing season. From February 2012 the light curve was initially flat at $K_s \approx 10.8$ for about 4 months but it then began to decline
steadily, fading from view after 2014. This source is projected close to the Galactic centre ($l=0.74$, $b=-0.36$) and detections in the $Y$ filter on two dates in August 2011 
allow us to calculate a colour $Y - K_s \approx 7.2$. If we employ the extinction law of \citet{cardelli89} and assume an unreddened colour $Y-K_s = 0$ for a hot source, this 
implies $A_V \approx 27$ mag, consistent with a location in the Nuclear Disc of the Milky Way and the CMZ. The initial brightening behaviour can probably be accommodated within
the framework of a classical nova with a secondary dust peak, from which we infer that the unobserved primary peak occurred between the last pre-outburst observation on 2010 
September 12 and the initial detection almost a year later. A concern is that the proposed secondary peak has a rather long duration for a nova, given the event took 400 days to fade 
below the level of the initial detection. Several examples of novae with a secondary peak are shown in the SMARTS database (e.g. N Mus 2018) but 
in every case the peaks have a shorter duration than is seen in this source, by factor of two or more. However, the timescales over which nova shells expand and fade 
have a very wide range, determined by the energetics of the explosion \citep[see e.g.][]{strope10}, so our initial classification is that this was a slow nova event corresponding to 
a low energy outburst.

\item Source 183 rose in brightness by $\sim$2 mag over $\sim100$ days up to the time of maximum flux, suggesting that we are seeing a secondary dust peak. While this is not especially remarkable, there is a 
also a dip in the light curve just before maximum flux, having a duration of approximately two to four weeks, though measured on only
two nights in the relevant interval. This might be caused by high optical depth in the ejecta.

\end{enumerate}

\section{Microlensing Events}
\label{micro}

We classify 38 sources from Table \ref{tab:table1a} and a further 15 sources from Table C1 as microlensing events, with varying degrees of confidence owing to the sparse sampling of the VVV survey. The 
classifications are based on visual inspection of the light curves for the characteristic morphology and a timescale of a few weeks. Almost all events appear to be single star lenses, as opposed to 
binary lens events. Three of the events (sources 180, 189 and 214) were previously reported by the OGLE early warning system \citep{udalski03} and two additional events with unusually good sampling, 
sources 215 and DR4\_v51 from Table C1 , were verified by co-author Navarro using a microlens event-fitting code. This gives us some confidence that most of these events are correctly classified. These 
sources are mostly located in the inner bulge as one would expect, see Figure \ref{fig:galdistrib}. The spatial distribution of infrared microlensing events discovered with VVV data was characterised in detail by 
\citep{navarro18, navarro20a, navarro20b} using a much larger sample of lower amplitude events, not including any of the events reported here. The events in the present sample do not add 
significantly to those works, despite their higher amplitudes, so we do not discuss them further.

\section{Unusual sources}
\label{unusual}
\begin{itemize} 
\item Source 175 is V4334~Sgr (= Sakurai's star), thought to be a born-again giant star undergoing a very late thermal pulse after entering the white dwarf cooling
sequence. The ongoing near infrared photometric brightening is described by \citet{evans20}, drawing on the VVV data among other data sets. Here we note briefly that the VIRAC2 \textsc{dophot}
light curve (not shown) displays a smoother long-term brightening trend than the aperture photometry presented in that work. 

\item Source 16 is MAXI~J1348-630, a black hole X-ray binary candidate 
\citep{tominaga20}. This source has been extensively investigated by the X-ray binary community so we do not analyse it here, except to note that the VIRAC2 light curve shows an approximately
constant brightness from 2010 to 2016, before starting to rise in 2017 and becoming much brighter in 2019, at the time of its discovery in the X-ray waveband.
\end{itemize} 

\begin{figure*}
        \includegraphics[width=\textwidth]{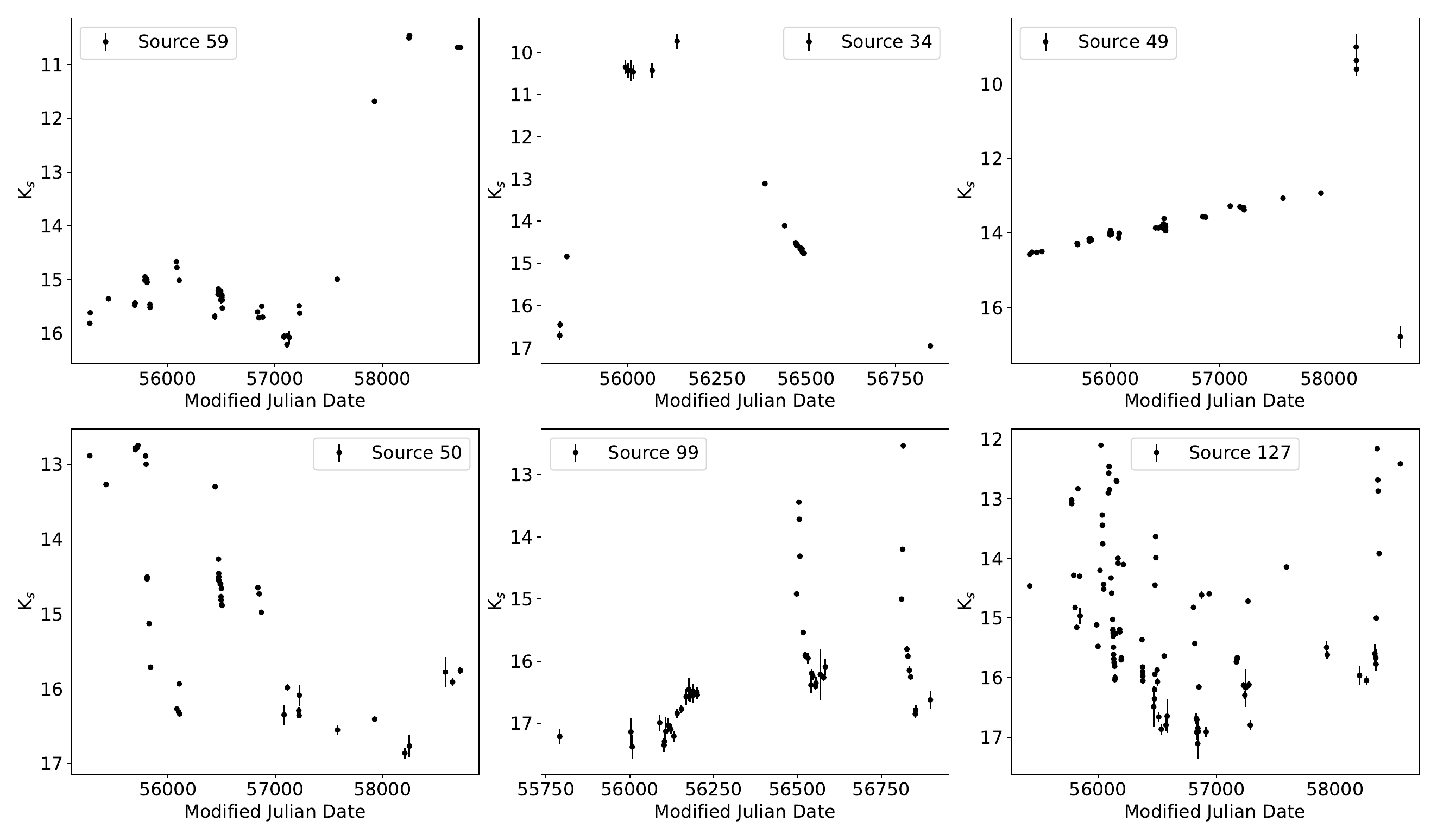}
    \caption{VIRAC2 light curves of six of the unusual variable stars that are poorly understood at present, see main text.}
    \label{fig:unusLCs}
\end{figure*}

The remaining eight ``unusual" sources are as follows. Light curves of the six previously unreported sources in this set are shown in Fig. \ref{fig:unusLCs}.
\begin{itemize} 
\item Source 59 was listed by CP17a as an eruptive YSO candidate (VVVv746) having a relatively modest (1.3 mag) amplitude in the 2010 to 2014 VVV time
series but not discussed further. \citet{lucas20b} also listed it as an eruptive YSO candidate with a very high mid-infrared amplitude (5.3 mag in $W1$), having recovered it in a WISE/NEOWISE search for 
variable stars and transients projected in Infrared Dark Clouds. However, the companion paper by GLK finds that its infrared spectrum does not resemble that of known eruptive YSOs 
and shows almost unique molecular absorption features, discussed in that work. The nature of this system is currently under investigation.

\item Sources 199 and 101 are VVV-WIT-08 and VVV-WIT-10 respectively, see \citet{smith21}. These are noted in section \ref{sec:dipgiants} as giant stars that underwent dimming events that, at least in the 
case of source 199, have a different nature to most of the other aperiodic dipping giant stars in this work. The amplitude and characteristics of the dimming event of VVV-WIT-10 are unclear since
it was undetected throughout the 2015 observing season and had no detectable variability in other years. Uniquely among sources in this work, VVV-WIT-10 was detected as a highly variable star
despite there being no bona fide detection of variability save for an upper limit when it dropped below the detection threshold. The high amplitude in the VIRAC2 database is due to a few spurious 
\textsc{dophot} detections in 2015, caused by blending with adjacent star.

\item Source 34 is a slow-rising transient 
that was first detected in September 2011 at $K_s=16.7$, clearly on a rising trend (brightening by $\sim$2 mag over 19 days in that month). 
It was next observed at a maximum of $K_s \sim 10$ from March to July of 2012. By the following year it was on a fading trend: observations from April 2013 to July 2014 show it fading smoothly down to $K_s = 17$ 
and it was subsequently undetected. Given the initial slow rise, this was not a conventional nova event. Unfortunately there no observations in filters other than $K_s$ in the years from 2011 to 2014. According to the 
Vizier database it was undetected in any other surveys so no colour information is available. The source is located in the mid-plane of the Galactic disc at ($l$, $b$) = (328.9, -0.2), which tends to suggest a distant 
source that would be expected to suffer significant foreground interstellar extinction. The nature of this transient is unknown.

\item Source 49 had a slow, linear rising trend over 7 years from $K_s \approx14.5$ in 2010 to $K_s \approx 13$ in 2017. In May 2018 it was seen at $K_s \approx 9$, fading by 0.6 mag over 3 days. No other observations
were made that year and it was next seen at a much fainter magnitude, $K_s = 16.8$, in June 2019, near the end of the VIRAC2 time series, before fading from view. The colours during the rising stage 
were very red: $H-K_s=3.68$ in 2010 and $H-K_s=3.98$ in 2015 (with no detections in $J$, $Y$ or $Z$.) The {\it Spitzer}/GLIMPSE survey data from 2004 also detected it as a red star: [4.5]=10.0, [3.6-[4.5]=1.09, 
[4.5]-[5.8]=0.68, [4.5]-[8.0]=0.69.
The source is located in the mid-plane of the Galactic disc at ($l$, $b$) = (337.9, -0.1) and there are numerous submillimetre clumps within a few arcminutes on the sky, detected by the ATLASGAL survey \citep{contreras13,
urquhart18}. However, the light curve does not resemble that of any known eruptive YSO.   
Unless spectroscopy can be undertaken, the nature of this source will probably remain unknown. 

\item Source 50 has relatively blue colours and is located in the Galactic disc at ($l$, $b$) = (339.6, 1.2). It shows an irregular fading trend from 2010 to 2019, with large intra-year variability and a secondary maximum in 2013.
Changes in flux were almost achromatic:  the colours and magnitudes in March 2010 were $K_s=12.89$, $H-K_s=0.59$ and $J-H=0.86$, compared to $K_s=16.29$, $H-K_s=0.50$ and $J-H=0.92$ in July 2015.
The 2MASS PSC gives $K_s=13.38$, slightly fainter than the initial VVV measurements but with broadly similar colours. It is listed as a blue source in catWISE but the unTimely time series data are 
too heavily blended to be very informative. The nature of this source is unknown.

\item Source 99 is in the Galactic bulge at ($l$, $b$) = (1.2, 4.2). It was seen as a faint star ($H=18.0$) in 2010, not detected in $K_s$ or $J$ that year. In $K_s$ it was first seen in 2011 and last seen in 2014. In 2011 to 2012 
it was at $K_s \approx 17$, brightening slightly towards the end of the 2012 observations. In 2013 and 2014 it showed two separate brief outbursts that each brightened and faded on a timescale of order 1 week, somewhat 
resembling microlensing events in morphology. In 2015 it was detected in $J$ and $H$ (only) with a relatively blue colour: $J=18.8$, $H=18.0$, $J-H=0.8$. This source was near the sensitivity limit of VVV in most years so the non-detection in $K_s$ in 2010 and 2015 to 2019 is unsurprising, given that there were few observations of the field in those years. The Vizier database shows no detections in surveys other than VVV. The nature of this source is unknown.

\item Source 127 is projected in the Nuclear Disc of the Milky Way and the CMZ, at ($l$, $b$) = (359.300, -0.005). It shows signs of periodicity: the VIVACE catalogue \citep{molnar22} found a best-fit period, $P \approx 64.8$~d 
and our period fit has a very similar value ($P \approx 64.65$~d) but there is considerable scatter in the phase-folded light curve. The star has red mid-infrared colours in the {\it Spitzer}/GLIMPSE-II dataset and 
a separate {\it Spitzer} study of the Galactic centre region \citep{ramirez08}. The SPICY catalogue \citep{kuhn21} lists it as a YSO candidate of uncertain nature, based on detections in three mid-infrared filters in GLIMPSE-II. However, \citet{yoshikawa14} list it as a candidate class II YSO with a little more confidence, based on their detection of intrinsic polarisation and detections in four mid-infrared filters in the study of \citet{ramirez08}. However, 
the VIRAC2 light curve is very unusual so the nature of this source remains in doubt. 

\end{itemize}


\bsp	
\label{lastpage}
\end{document}